\documentclass[a4paper, 12pt]{article}

\usepackage{bm, jheppub,xcolor}
\makeatletter
\gdef\@fpheader{}
\makeatother
\usepackage{comment}  
\usepackage{tikz}
\newcommand{\vev}[1]{{\left\langle #1 \right\rangle}}
\usetikzlibrary{decorations.pathreplacing}
\usepackage{amsmath,amssymb,graphicx,hyperref,slashed,subcaption}
\usepackage{psfrag}
\usepackage{subcaption}
\usepackage[dvipsnames]{xcolor}

\usepackage{graphicx} 
\newcommand{\be}{\begin{equation}}
\newcommand{\ee}{\end{equation}}
\usepackage{float} 
\newcommand{\ba}{\begin{aligned}}
\newcommand{\ea}{\end{aligned}}\newcommand{\re}{{\rm e}}

\def\CO{{\mathcal{O}}}

\def\rd{{\rm d}}
\def\ri{{\rm i}}
\def\vn{{\vec{n}}}

\newcommand{\CE}{{\cal E}}
\newcommand{\CF}{{\cal F}}

\newcommand{\CK}{{\cal K}}

\newcommand{\CN}{{\cal N}}

\newcommand{\cita}{{\textcolor{red}{[cite]}}}

\preprint{CERN-TH-2026-245}
\title{\Large Large Charge Jets in ${\cal N}=4$ SYM Energy Correlators}
\author{Alba Grassi,$^{a,b}$
Cristoforo Iossa,$^{b}$
Luigi Tizzano,$^{a}$
and Alexander Zhiboedov$^{a}$}

\affiliation[a]{Theoretical Physics Department, CERN, 1211 Geneva 23, Switzerland }
\affiliation[b]{Section de Math\'ematiques, Universit\'e de Gen\`eve, 1211 Gen\`eve 4, Suisse}
\abstract{
We compute the energy-energy correlator in a state created by a heavy canonical operator of scaling dimension $\Delta_H$ in $\mathcal{N}=4$ $SU(N)$ SYM, at fixed $N$ and arbitrary values of the large charge 't Hooft coupling $\lambda \sim \Delta_H g_{\rm YM}^2$. The state created by the canonical operator triggers spontaneous breaking of conformal symmetry and the Higgs mechanism,  with squared W-boson masses proportional to $\lambda$.
We start with the known finite-coupling result for the underlying four-point function of local operators, perform the detector and Fourier transforms and express the resulting energy-energy correlator as an infinite OPE sum in the source-detector channel, which we then analyze both numerically and analytically.
We show that the finite-coupling energy-energy correlator exhibits a hierarchy of distinct physical regimes as a function of $z=(1-\cos\theta)/2$, where $\theta$ is the angle between the detectors. A small-angle vacuum-dominated behavior at $z\ll\lambda^{-1}$ is followed by a non-monotonic medium-induced transient behavior at $\lambda z \sim1$.
A large charge jet regime follows, where the energy-energy correlator displays a power-law behavior $\sim1/(\lambda z^3)$ for $\lambda^{-1}\lesssim z \lesssim \lambda^{-1/3}$, before approaching an angle-independent plateau at larger angles.
}

\begin{document}

\maketitle

\section{Introduction}

Large quantum numbers can enhance interaction effects,
leading to effective couplings that need not be small even in weakly coupled quantum field theories. This observation motivates
double-scaling limits in which the coupling tends to zero while
the quantum number grows, with their appropriate product held fixed. A familiar example of this type relevant for the present paper
is high-multiplicity particle production, where multiplicity
enhancements can compensate for the smallness of the microscopic
coupling \cite{Cornwall:1990hh,Goldberg:1990qk,Libanov:1994ug}.

The same double-scaling structure was subsequently found in the study of the large charge operator $\phi^n$ in the $O(2)$ Wilson--Fisher CFT in $d=4-\epsilon$ \cite{Badel:2019oxl}, where one keeps $n \epsilon$ fixed as $n \to \infty$ and $\epsilon \to 0$. 
In supersymmetric theories, an analogous limit was first explored for extremal correlators in four-dimensional $\mathcal N=2$ theories \cite{Bourget:2018obm} and further established in \cite{Grassi:2019txd}. It was later also applied to study higher-rank theories and integrated correlators in $\mathcal N=4$ super-Yang–Mills (SYM) theory in \cite{Paul:2023rka,Brown:2023why, Grassi:2024bwl, Brown:2024yvt}. For gauge group $SU(2)$, a semiclassical description of this limit was developed in \cite{Caetano:2023zwe}, leading to a finite-coupling expression for heavy-heavy-light-light correlators (HHLL) in $\mathcal{N} = 4$ SYM. This result was subsequently extended to general fixed $N$ in \cite{Brown:2025cbz} for a distinguished class of half-BPS heavy operators, known as ``canonical'' operators. The resulting four-point function plays a key role in our analysis.

We consider canonical half-BPS operators $\mathcal{O}_K^m$, obtained by orthogonalizing the $m$-th power of a single-trace half-BPS operator of dimension $K$, with $3\leq K\leq N$. Their protected scaling dimension is $\Delta_H=mK$. We take the large-charge double-scaling limit by sending:
\begin{equation}
    m\to\infty\,,
    \qquad
    g_{\rm YM}\to 0\,,
    \qquad
    \lambda
    \equiv
    \frac{g_{\rm YM}^2 m}{16\pi^2}
    \quad\text{fixed}\,,
    \label{eq:limitLCH}
\end{equation}
with $K$ and the rank of the gauge group $N-1$ held fixed.

In this paper, we explore the limit \eqref{eq:limitLCH} in the context of a conformal collider experiment \cite{Hofman:2008ar}. More precisely, we calculate the energy-energy correlator in a state created by a canonical operator $\mathcal{O}_K^m$ in the limit \eqref{eq:limitLCH} as a function of $\lambda$.
As $\Delta_{H} \to \infty$, we can think of the operator $\mathcal{O}_K^m$ creating a multi-particle state that admits an emergent semi-classical description \cite{Brown:2025cbz}, where the scalar fields acquire a nontrivial profile $\langle \Phi_I \rangle \neq 0$ leading to the following symmetry breaking pattern
\be
SU(N) \rightarrow U(1)^{K-1} \times U(N-K) \ 
\ee
in addition to spontaneous breaking of conformal symmetry.
The fluctuations on top of this background consist of $(K-1)+(N-K)^2$ massless modes, and $K(2N-K-1)$ massive W-bosons with different masses $M_s^2 \sim \lambda$, $s=1,\cdots,K$. It was found in \cite{Brown:2025cbz} that in the large charge 't Hooft limit the massless modes remain effectively free, whereas the interaction effects are captured by massive W-bosons which therefore will play a central role in our analysis.

One prominent example of an emergent semiclassical description of the large charge dynamics is provided by the study of the Wilson--Fisher $O(2)$ CFT \cite{Hellerman:2015nra, Monin:2016jmo}. In this case, the effective description is provided by a superfluid for a spontaneously broken $U(1)$ symmetry, and the energy-energy correlator in this case has been recently calculated in \cite{Cuomo:2025pjp}, where an enhanced sound-jet behavior in the collinear limit has been observed. An important difference of the present work compared to \cite{Cuomo:2025pjp} is that, after taking the large charge limit \eqref{eq:limitLCH}, we are still left with a nontrivial coupling dependence understanding which is the principal subject of this work.

We consider the heavy state in the center-of-mass frame with the four-momentum $q^\mu=(Q,\vec 0)$. The energy-energy correlator (EEC) takes the form 
\be
\vev{\CE(\vn_1)\CE(\vn_2)}_H
=
\left(\frac{Q}{4 \pi}\right)^2
\Big( \CF_{\text{free}}(z) + {1 \over \Delta_H^2} \CF_{\text{int}}(\lambda, z) \Big), ~~~ z = {1 - \cos \theta \over 2} \ ,
\label{eq:eec_intro_sup}
\ee
where $\theta$ is the angle between detectors $\vec n_1 \cdot \vec n_2 \equiv \cos \theta$. In the formula above, we have separated the contribution of the known free theory result $\CF_{\text{free}}(z)$ 
\be
\CF_{\text{free}}(z)
=
1+
\frac{18z-12+3\delta(z)}{2\Delta_H}
+\frac{144z^2-162z+36-3\delta(z)}{2\Delta_H^2} +
O(\Delta_H^{-3}),
\ee
from the interacting part $\CF_{\text{int}}(\lambda, z)$, where $\CF_{\text{int}}(0, z)=0$. The prefactor ${\Delta_H^{-2}}$ multiplying the interaction-dependent part can be motivated as follows: the connected part of the EEC is suppressed by ${1 \over \Delta_H}$, which, combined with the coupling constant factor $g_{\text{YM}}^2$, produces $g_{\rm YM}^2/\Delta_H\sim\lambda/\Delta_H^2$ in the limit \eqref{eq:limitLCH}.
Understanding $\CF_{\text{int}}(\lambda, z)$ is the main goal of the present paper.

\begin{figure}[t]
\centering 
\includegraphics[width=1\linewidth]{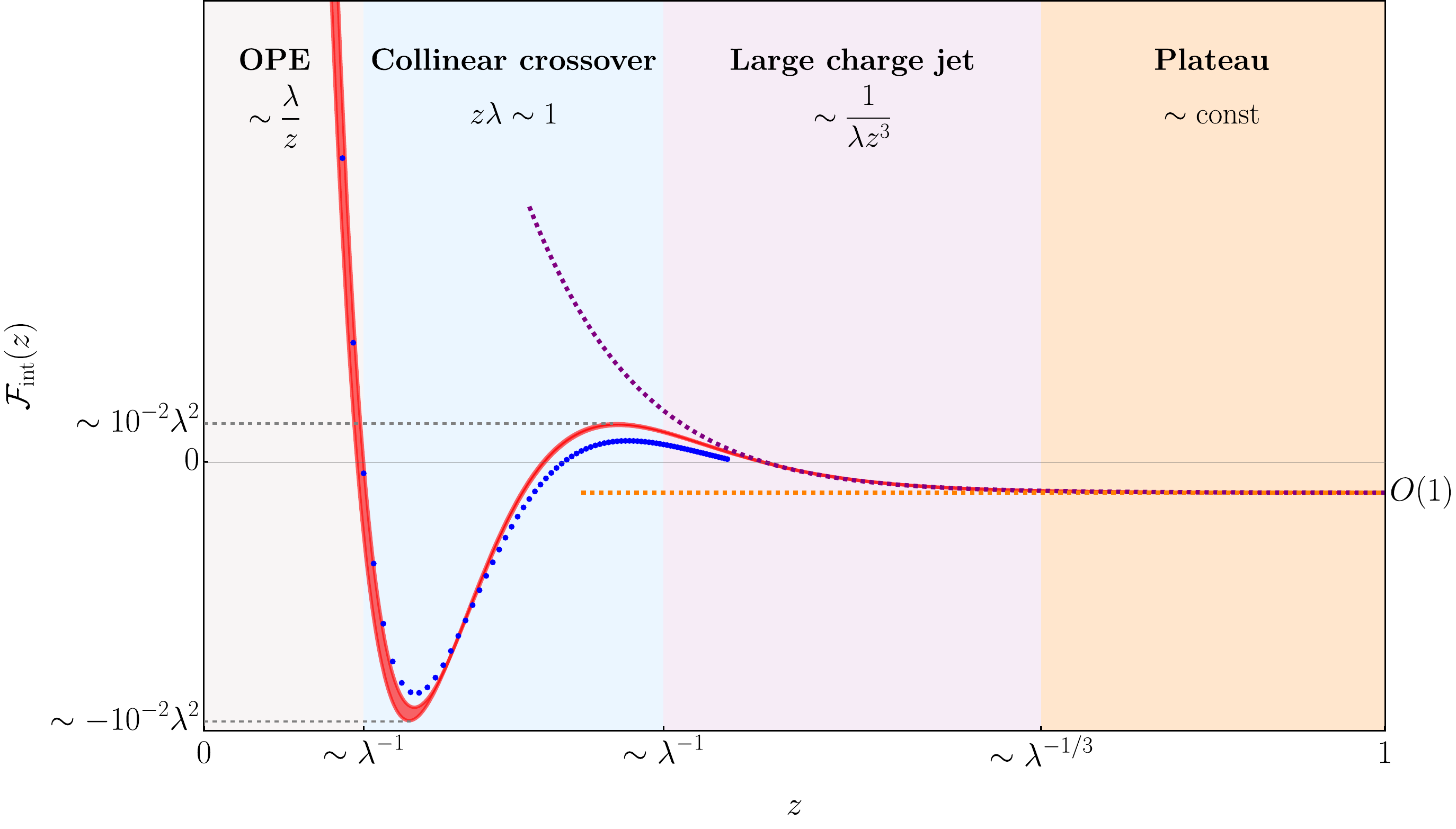}
\caption{Characteristic regions of the EEC as a function of the angle $z$ at $\lambda=10$ in SU(3) SYM. The horizontal axis is plotted on a logarithmic scale. The red curve is obtained by truncating the spin sum in eq \eqref{eq:Ffc}. Its width represents the numerical uncertainty, estimated by comparing truncations with cutoffs $J_{\mathrm{max}}=1000$ and $1200$. The blue dots are obtained using the double-scaling limit calculation $\lambda \to \infty$, $z\,\lambda \sim 1$  described in Section \ref{sec:crossover}.}
\label{fig:intro}
\end{figure}

It is convenient first to write the salient features of $\CF_{\text{int}}(\lambda, z)$ at weak $\lambda \ll 1$ and strong $\lambda \gg 1$ coupling. At fixed $0<z<1$, we have 
\begin{equation}
\CF_{\text{int}}(\lambda, z) \simeq \begin{cases}
     KN \lambda
\left(
{6\over z}
+4\operatorname{Li}_2(1-z)
-12\log z
- {2 \pi ^2 \over 3}
-13
\right) \,,
    & \lambda \ll 1 ~, \\[6pt]
   {1\over 24 \pi^2}  \left(\frac{K(K^2-1)}{12}+2K(N-K)\right) \frac{1}{\lambda \,z^3}- \frac{K(2N-K-1)}{2}~,
    & \lambda \gg 1~.
\end{cases}
\end{equation}
The dependence on $K$ and $N$ has a simple interpretation in terms of the
massive W-boson spectrum: the weak-coupling pole is proportional
to the sum of squared masses, the strong-coupling cubic tail to
the sum of inverse squared masses, and the constant term to  minus the total number of massive states, with each sum weighted by the  degeneracy of a given mass state.
At weak coupling we have separated the leading small-angle behavior.
At large but finite $\lambda$, the energy correlator exhibits the following characteristic regions:
\begin{itemize}
    \item \emph{Vacuum core}: for $z \ll {1 \over \lambda}$, $\CF_{\text{int}}(\lambda, z)$ is well-approximated by the weak-coupling vacuum scaling exponent $\sim {\lambda \over z}$. This behavior is  reproduced by perturbative collinear splitting of partons in vacuum.
    \item \emph{Collinear crossover}: for $ z \sim {1 \over \lambda}$ the medium effects become important. Microscopically, this is related to the fact that at this angle detectors start resolving the characteristic length scales of excitations propagating in the medium. In this regime, $\CF_{\text{int}}(\lambda, z) \sim \lambda^2$ and its angular profile can develop a pronounced dip followed by a peak.
    \item \emph{Large charge jet}:\footnote{In this paper, we use the term ``jet'' to refer to a collimated flux of energy \cite{Sterman:1975xv,Sterman:1977wj}.} for ${1 \over \lambda} \lesssim z \lesssim {1 \over \lambda^{1/3}}$, the $\CF_{\text{int}}(\lambda, z)$ decays with a characteristic power law $\CF_{\text{int}}(\lambda, z) \sim {1 \over \lambda z^3}$.\footnote{In QCD, the $\Lambda_{\text{QCD}}/Q$-suppressed nonperturbative correction
to the EEC with angular dependence $[z(1-z)]^{-3/2}$, were studied, for example, in \cite{Korchemsky:1999kt,Schindler:2023cww}. Also, the sound-jet mode of  \cite{Cuomo:2025pjp} has small-angle behavior $z^{-3/2}$ within the regime of validity of the conformal superfluid EFT.} 
     This regime is governed by massive W-bosons, each contributing a term proportional to ${1 \over M_s^2} {1 \over z^3}$ to $\CF_{\text{int}}(\lambda, z)$. As the detector angle decreases, the interacting contribution
grows from $O(1)$ near $z\sim\lambda^{-1/3}$ to $O(\lambda^2)$
near the crossover at $z\sim\lambda^{-1}$.
This angular enhancement is one of the most striking
features of our result.
     
    \item \emph{Plateau}: for ${1 \over \lambda^{1/3}} \lesssim z \leq 1$, the inverse-cubic term becomes subleading and $\CF_{\rm int}(z,\lambda)$ approaches a negative, angle-independent plateau,
$\CF_{\text{int}}(\lambda,z)\simeq-K(2N-K-1)/2$. This $O(1)$ value depends only on the total number of massive
W-bosons, each contributing $-1/2$ in our normalization.
\end{itemize}
Figure~\ref{fig:intro} illustrates the four angular regimes for $\lambda=10$ in $SU(3)$ SYM. We emphasize that all these interesting features are not visible at weak coupling $\lambda \lesssim  O(1)$.

In the formulas above we assumed $0 < z \leq 1$. In the main text we also discuss the detailed structure of the distributional contact terms at $z=0$, which are important for example to have a detailed control over the properties of the energy multipoles.

One motivation for this work is to provide a controlled, non-holographic strongly-coupled toy model of a high-multiplicity collider experiment.
As explained above, we see in the explicit example of ${\cal N}=4$ SYM that finite-coupling effects could introduce an interesting hierarchy of angular scales in the energy-energy correlator. It would be very interesting to understand if there are other situations in which similar phenomenon happens. The most natural arena for large multiplicity collider experiments is provided by heavy-ion collisions and energy correlators have received a lot of attention in this context recently \cite{Andres:2022ovj,Barata:2023zqg,Yang:2023dwc,Barata:2023bhh,Bossi:2024qho,Duan:2026icj,Barata:2026pgh}. This has been accompanied by recent experimental measurements of energy correlators in heavy-ion and small collision systems \cite{CMS:2025ydi,ALICE:2026giw,ALICE:2026zyx}.

The paper is organized as follows. Our starting point is the finite-$\lambda$ result for the leading
$1/\Delta_H$ correction to the heavy-heavy-light-light (HHLL)
four-point function obtained in \cite{Brown:2025cbz}:

\begin{enumerate}

    \item In Section~\ref{sec:basics}, we derive a
    source-detector OPE representation \cite{Kologlu:2019bco}
    of the leading, order-$\Delta_H^{-2}$ interaction-dependent
    correction to the EEC, using the OPE data of
    \cite{Brown:2025cbz}. The resulting representation,
    \eqref{eq:Ffc}, together with \eqref{eq:ECC_final},
    is valid at arbitrary $\lambda$.
    We also discuss the unresolved angular region
    $z\lesssim\Delta_H^{-1}$, where the fixed-angle
    large-charge expansion ceases to be uniform.

    \item In Section~\ref{sec:weakcoupling}, we calculate the
    EEC at weak coupling. We obtain the full one-loop result
    and compute the two-loop collinear double logarithm
    proportional to $\lambda^2\log^2z$, whose coefficient
    is given in \eqref{eq:weak_phi2_small_z}.

    \item In Section~\ref{sec:scoupl}, we analyze the
    strong-coupling limit at fixed angle. We derive the
    large-charge jet cubic scaling law and identify its
    semiclassical origin in regions of the detector integrals
    where both detectors approach the same endpoint.
    We then study the collinear crossover at $z\sim\lambda^{-1}$
    by taking $\lambda\to\infty$ and $z\to0$ with $\lambda z$
    fixed. The resulting scaling function,
    \eqref{eq:double_scaling_full_EEC}, is determined by the
    large-spin OPE data and interpolates between the collinear
    and inverse-cubic angular regimes.

    \item In Section~\ref{sec:finitecoupling}, we resolve all the features of the EEC at finite coupling. We first analyze the collinear limit, recovering the $\lambda/z$ behavior expected
    from the light-ray OPE \cite{Kologlu:2019mfz}. We show that the leading collinear pole is one-loop exact and that the coefficient
    of $\lambda^2\log^2z$ is two-loop exact, consistently
    with the explicit weak-coupling calculations.
    We then evaluate the OPE sum numerically over a range
    of couplings, resolving the hierarchy of angular regimes
    illustrated in Figure~\ref{fig:intro}.

    \item In Section~\ref{sec:conclusions}, we summarize our
    results and discuss their implications. In particular,
    we hint at a relation between strong-coupling energy
    correlators and half-light transforms of stress tensors
    on the Coulomb branch in flat spacetime. We also discuss
    possible extensions to higher-point energy correlators, and possible future research directions.

\end{enumerate}
Additional material includes the Ward identities, positivity
constraints, and endpoint contact terms discussed in
Section~\ref{sec:basics}, and the analysis of finite-coupling
Legendre moments in Section~\ref{sec:finitecoupling}.
A series of appendices contains technical details and extensions.

\vspace{0.25cm}

\noindent \textbf{Note added:} We have coordinated our submission with the work by Max Jackson, Alessandro Georgoudis, and Congkao Wen who have independently explored the same problem \cite{JacksonGeorgoudisWen}.

\section{Basic Properties of the EEC}\label{sec:basics}

In this section, we review the basic properties of the energy correlators and known results about the four-point function computed in \cite{Brown:2025cbz} that we will use to calculate the EEC.  

\subsection{Generalities on EEC}\label{sec:generalities}
\begin{figure}[t]
    \centering
    \includegraphics[width=0.6\textwidth]{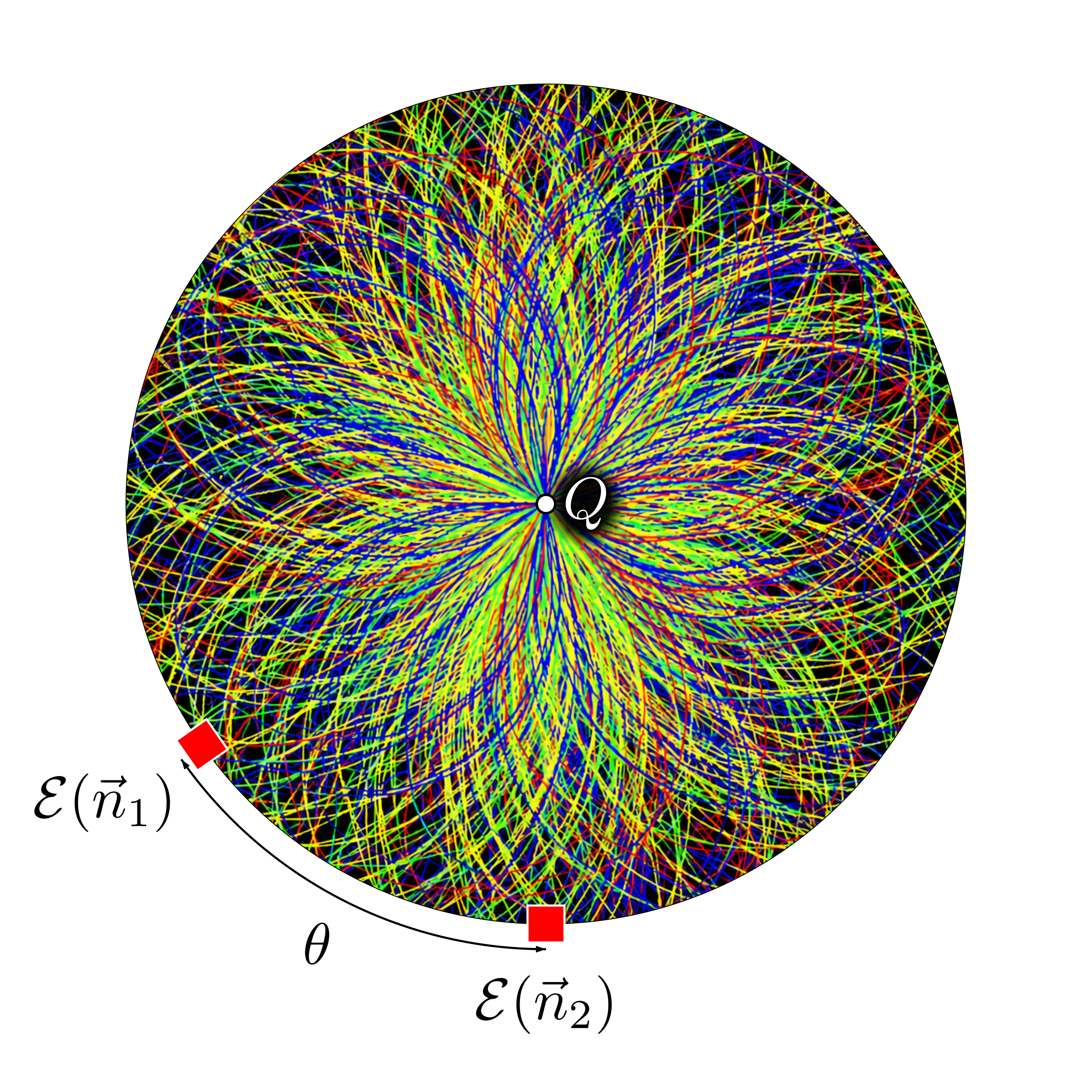}
    \caption{Two energy detectors which we place at future null infinity, $\mathcal{E}(\vec{n}_1)$ and
    $\mathcal{E}(\vec{n}_2)$, separated by an angle $\theta$,
    with $\cos\theta=\vec{n}_1\cdot\vec{n}_2$ and $z={1 - \cos \theta \over 2}$, in the rest frame
    of the source with total energy $Q$ and zero spatial momentum. In this paper, we consider the limit in which the state contains infinitely many particles, $\Delta_H \to \infty$. The illustration of the multiparticle state shown above is taken from
Figure 56 of \cite{Moult:2025nhu}.}
    \label{fig:detector-geometry}
\end{figure}

Let us begin by recalling some basic ingredients of conformal collider physics that will be needed below. The energy detector  is defined by placing the stress-energy tensor 
$T_{\mu\nu}(t,\vec{x})$ at null infinity and integrating it along retarded time  $u=t-r$ \cite{Hofman:2008ar,Sveshnikov:1995vi, Korchemsky:1997sy, Korchemsky:1999kt}:
\begin{equation}\label{eq:energy_detector}
\CE(\vec{n})
=\frac{1}{4}\int_{-\infty}^{\infty}\rd u\,
\lim_{r\to\infty}r^2
T_{\mu\nu}(u+r,r\vec n)\,\bar n^\mu\bar n^\nu~,
\end{equation}
where $\vec{n}$, with $\vec n^{\,2}=1$ is a unit vector on $\mathbb{R}^3$ pointing in the 
direction in which energy flux is measured, $n^\mu=(1,\vn)$ and $\bar n^\mu=(1,-\vec n)$. The above 
operator detects the total energy per unit solid angle reaching null infinity along the direction $\vec{n}$.
Our goal in this work will be to analyze some detailed properties of the correlations between 
two such detectors labeled respectively by $\vec{n}_1$ and $\vec{n}_2$ that are 
separated by an angle $\theta$ as in Figure~\ref{fig:detector-geometry}.

The state in which we measure the energy correlations will be prepared by a local
operator denoted by $\mathcal O_H(x)$ which is a scalar primary of scaling
dimension $\Delta_H$. The definition of such a state is given by acting with $\mathcal{O}_H$ on the vacuum
and taking its appropriate Fourier mode:
\begin{equation}
    |H(q)\rangle \equiv \int d^4x e^{-iq\cdot x}\overline{\mathcal{O}}_H(x)|0\rangle~,
\end{equation} 
where $q^\mu$ is a given four-momentum which we assume to be timelike and we use mostly-minus signature. All Lorentzian
correlation functions in this paper are understood as Wightman correlators, with the operator ordering displayed explicitly.
The two-point energy correlator in these states is therefore obtained as  
\begin{equation}\label{eq:EEC_generic_definition}
\vev{\CE(\vec{n}_1)\CE(\vec{n}_2)}_{H}
=
\frac{1}{\sigma_H(q)}
\int\rd^4x\,\re^{i q\cdot x}
\vev{\mathcal O_H(x)\CE(\vn_1)\CE(\vn_2)
\overline{\mathcal O}_H(0)},
\end{equation}
where
\begin{equation}
\sigma_H(q)
=
\int\rd^4x\,\re^{i q\cdot x}
\vev{\mathcal O_H(x)\overline{\mathcal O}_H(0)}~,
\end{equation}
is the source spectral density, or equivalently the total cross section for producing an arbitrary number of particles in the final state. Here and below, we refer to the insertion $\mathcal O_H(x)$ as the source and to its conjugate $\overline{\mathcal O}_H(0)$ as the sink.

Since $q^\mu$ is timelike, we will consider the source rest frame $q^\mu = (Q,0)$ in which Lorentz invariance together with the homogeneity of the detector under rescaling of its null direction allows us to fix the form of the correlation 
\begin{equation}\label{eq:EEC_F_definition}
\vev{\CE(\vn_1)\CE(\vn_2)}_{H}
=
\frac{(q^2)^4}{(q\cdot n_1)^3(q\cdot n_2)^3}
\frac{\CF(z)}{(4\pi)^2}~,
\end{equation}
where $\CF(z)$ is an arbitrary function of the angular separation $z$ between the two detectors
\begin{equation}
    z = \frac{q^2(n_1\cdot n_2)}{2(q\cdot n_1)(q\cdot n_2)} =\frac{1}{2}(1-\vec{n}_1\cdot\vec{n_2}) = \frac12 (1-\cos\theta)~, \qquad 0\leq z\leq 1~.
\end{equation}
The collinear limit $z \to 0$ (with all other parameters kept fixed) is controlled by the light-ray OPE \cite{Kologlu:2019mfz,Chang:2020qpj} and it will play an important role in our analysis. The limit $z \to 1$ corresponds to the back-to-back location of detectors.

Note that in the source rest frame, the expectation value of \eqref{eq:energy_detector} in the state $|H(q)\rangle$ is determined by kinematics to be $\langle\CE(\vec{n})\rangle_H=Q/4\pi$. It is also useful to introduce the unit normalized energy operator $\widehat{\CE}(\vn)=\frac{4\pi}{Q}\CE(\vn)$, $\langle\widehat{\CE}(\vn)\rangle_H=1$ such that using rotational invariance and  \eqref{eq:EEC_F_definition} we get
\begin{equation}\label{eq:normalized_flow_F}
\vev{\widehat{\CE}(\vec n_1)\widehat{\CE}(\vec n_2)}_H
=\CF(z).
\end{equation}
The function $\mathcal{F}(z)$ must satisfy some nontrivial conditions imposed by the following sum rules. To derive them, let us first recall that when integrated over the celestial sphere the energy detector reproduces the total four-momentum symmetry generator:
\begin{equation}
    P^\mu = \int_{S^2} d^2\vec{n} ~n^\mu \CE(\vec{n}).
\end{equation}
As a consequence, it also follows that
\begin{equation}
    \int_{S^2} d^2\vec{n}_2 ~n_2^\mu \langle\CE(\vec{n}_1) \CE(\vec{n}_2)\rangle_H = q^\mu\langle\CE(\vec{n}_1)\rangle_H~.
\end{equation}
Combining the above two equations one arrives at the sum rules
\begin{equation}
\begin{aligned}
\int \mathrm{d}^{2}\vn_{1}\int \mathrm{d}^{2}\vn_{2}\,
\left\langle
\mathcal{E}(\vn_{1})\mathcal{E}(\vn_{2})
\right\rangle_{H}
&=Q^{2},
\\
\int \mathrm{d}^{2}\vn_{1}\int \mathrm{d}^{2}\vn_{2}\,
\left(1-\vec n_{1}\cdot\vec n_{2}\right)
\left\langle
\mathcal{E}(\vn_{1})\mathcal{E}(\vn_{2})
\right\rangle_{H}
&=Q^{2}.
\end{aligned}
\end{equation}
Substituting the expression \eqref{eq:EEC_F_definition} and the definition of $z$ we conclude that $\CF(z)$ obeys the following Ward identities:
\begin{equation}\label{eq:WardId}
    \int_{0}^1dz ~z\CF(z) = \int_{0}^1 dz~(1-z)\mathcal{F}(z) = \frac{1}{2}~.
\end{equation}
Since $\CF(z)$ is a function on the celestial sphere depending on the relative angle $\theta$ through a single variable $z$, it can be expanded in Legendre polynomials:
\begin{equation}\label{eq:EEC_Legendre_expansion}
\CF(z)
=
1 +\sum_{\ell=2}^{\infty}(2\ell+1)
a_\ell P_\ell(1-2z)~,\quad
a_\ell
\equiv
\int_0^1\rd z\,P_\ell(1-2z)\CF(z)~,\quad 
\end{equation}
where the first two Legendre moments are fixed by the Ward identities \eqref{eq:WardId}. Up to this point the argument is purely kinematical: any rotationally invariant angular distribution admits such an expansion.\footnote{When the EEC contains endpoint contact terms, these equations are
understood as identities between distributions, namely after integration
against a smooth test function.} However, quite interestingly, it can further be shown that the Legendre moments of the EEC $a_\ell$ are all non-negative \cite{Fox:1978vu,Kologlu:2019bco,Mecaj:2025ecl,Dempsey:2025yiv}. To see this let us consider the standard decomposition in spherical harmonics on the celestial sphere
\begin{equation}\label{eq:spherical_harmonic_conventions}
\int_{S^2}d^2{\vec n}\,
Y_{\ell m}^{*}(\vec n)Y_{\ell' m'}(\vec n)
=
\delta_{\ell\ell'}\delta_{mm'}~,\quad
\sum_{m=-\ell}^{\ell}
Y_{\ell m}^{*}(\vec n_1)Y_{\ell m}(\vec n_2)
=
\frac{2\ell+1}{4\pi}
P_\ell(\vec n_1\cdot\vec n_2)~.
\end{equation}
Combining \eqref{eq:EEC_Legendre_expansion} with the above gives
\be
\left\langle
\widehat{\mathcal E}(\vec n_1)
\widehat{\mathcal E}(\vec n_2)
\right\rangle_H
=
4\pi
\sum_{\ell=0}^{\infty}a_\ell
\sum_{m=-\ell}^{\ell}
Y_{\ell m}^{*}(\vec n_1)
Y_{\ell m}(\vec n_2)~.
\ee
Now, we define the angular-momentum modes of the normalized detector by
\begin{equation}\label{eq:detector_spherical_modes}
\widehat{\CE}_{\ell m}
\equiv
\int_{S^2}d^2\vn\,
Y_{\ell m}^{*}(\vec n)\widehat{\CE}(\vec n),
\qquad
m=-\ell,\ldots,\ell.
\end{equation}
Using spherical-harmonic orthonormality it thus follows that
\begin{equation}
\left\langle\widehat{\CE}^{~\dagger}_{\ell m}\widehat{\CE}_{\ell'm'}\right\rangle_H = 4\pi ~a_\ell~ \delta_{\ell\ell'}\delta_{m m'}~.
\end{equation}
For $\ell'=\ell$ and $m'=m$, the left-hand side is the squared norm of the state obtained by acting with $\widehat{\mathcal E}_{\ell m}$ on the source state; and is therefore non-negative by Hilbert-space positivity
\begin{equation}\label{eq:Legendre_positivity}
    a_\ell
=
\frac{1}{4\pi}
\left\langle
\widehat{\mathcal E}_{\ell m}^{\dagger}
\widehat{\mathcal E}_{\ell m}
\right\rangle_H
\geq0~.
\end{equation}
Thus every Legendre moment of the full physical EEC is non-negative. Note that the sum rules obtained in \eqref{eq:WardId} are indeed consistent with this since they imply that $a_0=1$ and $a_1=0$. The
$\ell=0$ mode measures the total energy and hence gives $a_0=1$,
whereas the three $\ell=1$ modes are proportional to the total spatial
momentum and therefore vanish in the source rest frame. Positivity is
thus a nontrivial constraint on the moments with $\ell\geq2$.

Positivity of \eqref{eq:Legendre_positivity} is an important consistency check that, for the heavy EEC considered here, will be further analyzed in Section \ref{sec:positivityprop}.

\subsection{EEC in Heavy States}\label{sec:cumulantsheavyexp}
We now consider families of scalar states for which the dimension of the source becomes large, $\Delta_H\rightarrow\infty$. As discussed in the introduction, such states are expected to admit a semiclassical description. At the level of energy correlators, this expectation is expressed through the factorization of normalized detector insertions. In the rest frame of a scalar state, the two-detector correlator can therefore be decomposed as
\be
\left\langle
\widehat{\mathcal E}(\vn_1)\widehat{\mathcal E}(\vn_2)
\right\rangle_H
=
1+
\left\langle
\widehat{\mathcal E}(\vn_1)\widehat{\mathcal E}(\vn_2)
\right\rangle_{H,c}~.
\ee
The subscript $c$ denotes the connected contribution. It is not fixed by symmetry and contains state-dependent information, including correlations already present at zero coupling. 

More generally, the cumulant expansion expresses a $k$-detector correlator as a sum of products of connected correlators. For fixed $k$ and fixed noncoincident detector directions, it has been conjectured that \cite{Chicherin:2023gxt,Firat:2023lbp}
\begin{equation}\label{eq:cumulants}
    \left\langle
\widehat{\mathcal E}(\vn_1)\cdots
\widehat{\mathcal E}(\vn_k)
\right\rangle_{H,c}
=
 O \left(\Delta_H^{-h_k}\right)~,
\end{equation}
where the exponents $h_k>0$ may depend on the theory and on the particular family of heavy states. We will use this scaling as an organizing principle rather than as a universal theorem determining the powers of $\Delta_H$. In the normalization introduced in \eqref{eq:normalized_flow_F} we thus have:
\begin{equation}
    \CF(z) = 1+ \left\langle
\widehat{\mathcal E}(\vn_1)
\widehat{\mathcal E}(\vn_2)
\right\rangle_{H,c}~.
\end{equation}
The first constant term is the factorized semiclassical limit, while the second term measures connected fluctuations of the energy flux.

To organize these corrections, it is important to distinguish the expansion in $1/\Delta_H$ from the expansion in the microscopic coupling. We define the free event shape $\CF_{\rm free}(z)$ as the result obtained at zero microscopic coupling, and separate the full event shape according to
\begin{equation}
    \label{eq:splitting}
\mathcal F(z)
=
\mathcal F_{\mathrm{free}}(z)
+
{1 \over \Delta_{H}^2}\mathcal F_{\mathrm{int}}(z)~.
\end{equation}
The free event shape is determined by free-field Wick contractions of the source and detector operators, we will present this computation in the next section. We emphasize that the free event shape is not purely disconnected: even at zero coupling, a state containing a large but finite number of constituents exhibits connected energy correlations.
Indeed, for the heavy operators considered in this work, the general structure of the  free contribution can be organized as
\begin{equation}\label{eq:Borngeneral}
    \mathcal F_{\mathrm{free}}(z)
=
1+
\frac{1}{\Delta_H}
\left[
\mathcal F_{\mathrm{free,reg}}^{(1)}(z)
+
\mathcal F_{\mathrm{free,ct}}^{(1)}(z)
\right]
+
 O \left(\Delta_H^{-2}\right)~.
\end{equation}
Here the superscript $ (1)$ labels the coefficient of the first subleading term in the large-$\Delta_H$ expansion, it does not denote a perturbative loop order. The regular term $\mathcal F_{\mathrm{free,reg}}^{(1)}(z)$ describes correlations between detectors at separated angles, while the contact term $\mathcal F_{\mathrm{free,ct}}^{(1)}(z)$ is supported when the detectors coincide. Together, they give the leading failure of semiclassical factorization already present at zero coupling.

Our main interest in this work is the interaction-dependent contribution $\CF_{\rm int}(z)$. For the particular heavy limit considered in this work, the interaction-dependent contribution is further suppressed and begins at order $\Delta_H^{-2}$, see \eqref{eq:eec_intro_sup}.
We stress that this suppression is a property of the specific family of heavy states and scaling limit introduced in the next section. It does not follow from the general cumulant conjecture \eqref{eq:cumulants} alone.

Before concluding, let us record a consequence of the Ward identities derived in the previous subsection. The leading disconnected contribution already saturates both Ward identities \eqref{eq:WardId}, since 
\begin{equation}
    \int_0^1 dz\,z\,1
=
\int_0^1 dz\,(1-z)\,1
=
{1 \over 2}~.
\end{equation}
Consequently, every subleading coefficient in the large-$\Delta_H$ expansion must satisfy the corresponding homogeneous sum rules. Since the full and free event shapes separately satisfy the same Ward identities, their difference obeys homogeneous sum rules. In particular,\footnote{These equations are understood as distributional identities and therefore include all endpoint contributions.}
\begin{equation}\label{eq:sumrulesint}
    \int_0^1 dz\,z\,
\mathcal F_{\mathrm{int}}(z)
=
\int_0^1 dz\,(1-z)\,
\mathcal F_{\mathrm{int}}(z)
=
0~.
\end{equation}

\subsection{Large-Charge 't Hooft Limit}
We now specialize to $\CN=4$ super Yang--Mills theory with gauge group $SU(N)$ and identify the heavy source $\mathcal{O}_H$ introduced above. In particular, we will be interested in a special class of half-BPS scalar primaries. Note that at fixed scaling dimension these operators are highly degenerate: many different multi-trace operators have the same scaling dimension and $SO(6)_R$ quantum numbers. A special set of operators dubbed  as ``canonical" operators \cite{Brown:2024yvt,Grassi:2024bwl,Brown:2025cbz} provide a distinguished orthogonal basis in this degenerate space which will prove extremely useful for our purpose.

The half-BPS chiral ring of $\CN=4$ SYM is described starting from single-trace generators given by
\begin{equation}\label{eq:halfBPS}
    T_p(x,Y)
=
\left(\frac{1}{g^2_{\rm YM}}\right)^{p/2}
Y_{I_1}\cdots Y_{I_p}\,
\operatorname{Tr} \left(
\Phi^{I_1}(x)\cdots\Phi^{I_p}(x)
\right),
\qquad
p=2,\ldots,N,
\end{equation}
where $Y^2=0$ is an $SO(6)_R$ null polarization vector and $\big\langle \Phi_I^a(x)\Phi_J^b(0)\big\rangle_0
=\frac{g_{\mathrm{YM}}^2}{4\pi^2 x^2}\,\delta^{ab}\delta_{IJ}$. A basis of multi-trace half-BPS operators can be labelled by non-increasing partitions $\textbf{p}=(p_1,\dots, p_\ell)$, with $p_i \in \{2,\dots,N\}$:
\begin{equation}
    T_{\mathbf{p}}(x,Y) = T_{p_1}(x,Y)\cdots T_{p_\ell}(x,Y)~, \quad \Delta_{\mathbf{p}} = \sum^\ell_{i} p_i~.
\end{equation}
At fixed dimension, the trace basis is not orthogonal. Its protected two-point-function matrix takes the form
\be
\left\langle
T_{\boldsymbol p}(x_1,Y_1)
T_{\boldsymbol q}(x_2,Y_2)
\right\rangle
=
\mathcal N_{\boldsymbol p\boldsymbol q}(N)
\left(\frac{d_{12}}{4\pi^2}\right)^{\Delta_{\boldsymbol p}},
\qquad
d_{12}=\frac{Y_1\cdot Y_2}{x_{12}^2},
\ee
where $\Delta_{\boldsymbol p}=\Delta_{\boldsymbol q}$. The matrix
$\mathcal N_{\boldsymbol p\boldsymbol q}(N)$ defines the inner product on the space of half-BPS operators.

To specify an orthogonal basis, one must supplement this inner product with an ordering prescription. Among partitions of a fixed dimension, we first compare the number of factors $T_2$, placing operators with fewer such factors first. Ties are resolved by comparing successively the numbers of $T_3,T_4,\ldots,T_N$, see \cite{Brown:2025cbz}. Once the operators of a fixed dimension have been ordered, we perform Gram--Schmidt starting from the rightmost operator and moving to the left allowing us to define a new basis of operators:
\begin{equation}
\mathcal O_{\boldsymbol p}(x,Y)
=
T_{\boldsymbol p}(x,Y)
+
\sum_{\substack{
\Delta_{\boldsymbol q}=\Delta_{\boldsymbol p}
}}
c_{\boldsymbol q}^{(\boldsymbol p)}
T_{\boldsymbol q}(x,Y)~,\quad \langle\CO_{\mathbf{p}}\CO_{\mathbf{q}}\rangle \propto \delta_{\mathbf{p}\mathbf{q}}~.
\end{equation}
Thus, $\mathcal O_{\boldsymbol p}$ is the unique operator whose coefficient in front of $T_{\boldsymbol p}$ is fixed to one, which contains only operators following $T_{\boldsymbol p}$ in the prescribed ordering, and which is orthogonal to those operators. The ordering is therefore part of the definition. Since the two-point-function matrix is protected, the coefficients $c_{\boldsymbol q}^{(\boldsymbol p)}$ depend on $N$ and on the partitions, but not on the Yang--Mills coupling.

The canonical operators considered in this work correspond to partitions containing $m$ identical entries, all equal to $K$. We define
\begin{equation}
    \mathcal{O}^m_K(x,Y) \equiv \mathcal{O}_{(K,\dots,K)}(x,Y)~, \quad 3\leq K\leq N~.
\end{equation}
In words, $\mathcal{O}^m_K$ is the orthogonalized half-BPS operator whose unorthogonalized seed is the product $T^m_K$. 

As a simple example, consider $SU(3)$, $K=3$ and $m=2$. The relevant partition is $(K,K)=(3,3)$, and hence $T_{(3,3)}=T_3^2$ and $\Delta_{(3,3)}=6$.
In $SU(3)$, the other partition of dimension six is $(2,2,2)$, corresponding to $T_2^3$. The canonical operator therefore has the schematic form
\begin{equation}
\mathcal O_3^2
=
\mathcal O_{(3,3)}
=
T_3^2+c\,T_2^3~,
\end{equation}
where $c=- { \left\langle
 T_3^2\,T_2^3
\right\rangle\over \left\langle
 T_2^3\,T_2^3
\right\rangle}=-{1\over 24}$ is fixed by the protected two-point function matrix. This simple example illustrates the general definition: $\mathcal O_K^m$ is not the raw monomial $T_K^m$, but the definite state obtained after subtracting its overlaps with the other half-BPS states of the same dimension.

In this work, we will identify the local operator $\mathcal{O}_H$ introduced in  Section \ref{sec:generalities}, to prepare the state in which we measure energy correlations, with 
\begin{equation}
    \mathcal{O}_H(x) = \CO^m_K(x,Y)~, \quad \Delta_H = mK~.
\end{equation}
The reason for such choice is as follows. In the last few years, it has become clear that the limit $\Delta_H\to \infty$ can be studied very efficiently by considering the following double-scaling limit:
\begin{equation}\label{eq:dstooft}
    m\to\infty~,\quad g^2_{\rm YM}\to 0~,\quad K~\textrm{fixed}~,\quad \lambda=\frac{mg^2_{\rm YM}}{16\pi^2}~\textrm{fixed}~.
\end{equation}
Although the microscopic coupling vanishes, the interaction does not disappear. At $L$ loops, the suppression $g_{\rm YM}^{2L}\sim m^{-L}$ can be compensated by the order-$m^L$ number of ways of attaching the interaction vertices to the heavy operator. The surviving perturbative expansion is consequently organized by powers of the collective coupling $\lambda$. This is the large-charge analogue of a ’t Hooft limit, but it is not a planar limit: the rank $N$ remains fixed while the charge of the external state becomes large.

Double-scaled limits of this type were first introduced for extremal large-$R$-charge correlators in four-dimensional $\mathcal N=2$ theories \cite{Bourget:2018obm}, inspired by the emerging large-charge effective-field-theory framework
\cite{Hellerman:2015nra, Monin:2016jmo, Alvarez-Gaume:2016vff,  Hellerman:2017sur, Hellerman:2018xpi}, and were proven to exist to all orders in $1/m$ in \cite{Grassi:2019txd}. Starting with the work \cite{Paul:2023rka,Brown:2023why,Caetano:2023zwe}, it has also been understood that the $SU(2)$ 4-point function HHLL, involving two dimension-two operators, can be explicitly computed in this regime as a resummation of conformal ladder integrals.

At higher rank, to leading order in $1/m$, insertions of $\mathcal O_K^m$ were argued, and extensively checked, to source a Coulomb-branch-like scalar background.
In this background, the HHLL correlator can be expressed again in terms of massive scalar propagators, or equivalently as an all-loop resummation of conformal ladder integrals \cite{Brown:2025cbz}. This is an important dynamical input that will be used in the next subsection to determine the interacting contribution to the EEC.

\paragraph{Free Contribution.} 
Before turning to the interaction-dependent part, let us determine the free event shape introduced in \eqref{eq:splitting}. Here free means that $g_{\rm YM}$ is set to zero at fixed $m$, before taking the double-scaling limit. It should not be confused with $g_{\rm YM}\to0$ at fixed nonzero $\lambda$, where the large-charge combinatorial enhancement keeps interaction effects finite.

Every trace structure entering $\mathcal O_K^m$ contains exactly $\Delta_H$ scalar fields and no derivatives. Its tree-level form factor therefore creates $\Delta_H$ massless scalar particles and is independent of their momenta. After summing inclusively over color and $R$-symmetry labels, the internal structure of the form factor reduces to a single quadratic norm. The same norm multiplies the free spectral density $\sigma_H(q)$, and hence cancels from the normalized EEC \eqref{eq:EEC_generic_definition}. The free event shape is therefore determined entirely by the universal phase space of $\Delta_H$ massless particles and depends on $m$ and $K$ only through $\Delta_H=mK$.

On a fixed final state, the product of two detectors decomposes as
\begin{equation}
    \begin{aligned}
\mathcal E(\vec n_1)\mathcal E(\vec n_2)
={}&
\sum_{a\neq b}
E_aE_b\,
\delta^{(2)} \left(\vec n_1-\widehat{\vec p}_a\right)
\delta^{(2)} \left(\vec n_2-\widehat{\vec p}_b\right)
\\
&+
\delta^{(2)}(\vec n_1-\vec n_2)
\sum_a E_a^2\,
\delta^{(2)} \left(\vec n_1-\widehat{\vec p}_a\right)~.
\end{aligned}
\end{equation}
where $E_a=p_a^0$ and $\widehat{\vec p}_a=\vec p_a/E_a$. The first line corresponds to detecting two distinct particles and gives a regular function for noncoincident detector directions. The second line corresponds to measuring the same particle twice and produces a contact term at $z=0$.
The required phase-space integral was evaluated in \cite{Chicherin:2023gxt}. Replacing the source weight used there by the total number of constituents $\Delta_H$, one obtains, for $\Delta_H\geq3$,
\begin{equation}
\mathcal F_{\mathrm{free}}(z)
=
\frac{
(\Delta_H-1)(\Delta_H-2)
}{
(\Delta_H+1)(\Delta_H+2)
}
\,{}_2F_1
 \left(
3,3;\Delta_H+3;z
\right)
+
\frac{3}{2(\Delta_H+1)}\delta(z)
~.
\end{equation}
Expanding the exact result at large $\Delta_H$, we obtain
\begin{equation}\label{eq:EECBorn}
  \mathcal F_{\mathrm{free}}(z)
=
1+
\frac{18z-12+3\delta(z)}{2\Delta_H}
+\frac{144z^2-162z+36-3\delta(z)}{2\Delta_H^2} +
 O(\Delta_H^{-3}).
\end{equation}
Comparing with the general expansion \eqref{eq:Borngeneral}, we identify
\begin{equation}
    \mathcal F_{\mathrm{free,reg}}^{(1)}(z)
=
9z-6,
\qquad
\mathcal F_{\mathrm{free,ct}}^{(1)}(z)
=
\frac{3}{2}\delta(z)~.
\end{equation}
The constant $1$ is the factorized heavy-state contribution. The regular correction $9z-6$ originates from detecting two different particles whose momenta are correlated by fixed total four-momentum, whereas the contact term arises from detecting the same particle twice. All dependence on the microscopic interactions is contained instead in ${\mathcal F_{\mathrm{int}}}(z)$, to which we turn next.

\subsection{EEC for Canonical Operators}\label{sec:eecforcanop}

In this section we will present the main technical result of this work: an explicit formula for the interaction-dependent EEC in the large-charge double scaling limit defined in \eqref{eq:dstooft},  valid at arbitrary coupling $\lambda$. The calculation is based on a method developed in \cite{Belitsky:2013xxa,Belitsky:2013bja,Belitsky:2013ofa} based on the Mellin representation of CFT four-point functions \cite{Costa:2012cb}. 

By definition, the energy correlation \eqref{eq:EEC_generic_definition} is related to the four-point function 
\begin{equation}\label{eq:HHTT}\langle\CO_H(x_1)T_{\mu\nu}(x_2)T_{\rho\sigma}(x_3)\overline{\CO}_H(x_4)\rangle~,
\end{equation}
involving two heavy-source operators and two stress-energy tensors. Here and below, the $R$-symmetry polarizations are chosen so that the operator inserted at $x_4$ is the conjugate of the operator inserted at $x_1$. In $\mathcal N=4$ SYM, the stress tensor is a descendant of the dimension-two half-BPS superconformal primary $T_2$. The relevant stress-tensor correlator is therefore determined by superconformal Ward identities from the HHLL correlator
\begin{equation}\label{eq:HHLL}
\langle \CO_H(x_1)T_2(x_2)T_2(x_3)\overline{\CO}_H(x_4)\rangle~.
\end{equation}
\iffalse
Let us consider the light transform of the half-BPS primary $ T_2$,
\begin{equation}
\mathcal J_0(\vec{n})
=
\int_{-\infty}^{+\infty}du\,
\lim_{r\to\infty}
r^2\,
 T_2(u+r,r\vec{n})~,
\label{eq:scalar_detector}
\end{equation}
where the relevant $R$-symmetry channel and normalization conventions are summarized in Appendix~\ref{app:fourpoint_mellin}.
Then the superconformal Ward identities give \cita~
\begin{equation}
\begin{aligned}
&
\left\langle
\mathcal O_H(x)\,
\mathcal E(\vec{n}_1)\mathcal E(\vec{n}_2)\,
\overline{\mathcal O}_H(0)
\right\rangle
\\
&\qquad =
\frac{4}
{(n_1 \cdot n_2)^2(x^2)^2}
\mathcal D_\chi
\left\langle
\mathcal O_H(x)\,
\mathcal J_0(\vec{n}_1)\mathcal J_0(\vec{n}_2)\,
\overline{\mathcal O}_H(0)
\right\rangle ,
\end{aligned}
\label{eq:scalar_to_energy_detector}
\end{equation}
where
\begin{equation}
\chi=
\frac{2(x \cdot n_1)(x \cdot n_2)}
{x^2(n_1 \cdot n_2)},
\qquad
\mathcal D_\chi
=
\frac{d^2}{d\chi^2}
(1-\chi)^2\chi^2
\frac{d^2}{d\chi^2}.
\label{eq:scalar_to_energy_operator}
\end{equation}
\fi
A key feature of the HHLL correlator \eqref{eq:HHLL} is that,
once its kinematic and $R$-symmetry structures have been factored out,
the entire coupling dependence is encoded in a single function
 \begin{equation}
    \Phi_{\Delta_H}(u,v;\lambda)
=
\frac{v}{2\Delta_H}
\mathcal T_{m,K}(u,v;\lambda)~,
\end{equation}
of the conformal cross-ratios:
\begin{equation}
    u=\frac{x_{12}^2 x_{34}^2}{x_{13}^2 x_{24}^2} \,, \quad v=\frac{x_{14}^2 x_{23}^2}{x_{13}^2x_{24}^2} ~,
\end{equation}
see \eqref{eq:app_HHLL_split}.
For canonical operators, the large-charge result of \cite{Brown:2025cbz} gives
\begin{equation}\label{eq:resummedHHLL}
    \mathcal T_{m,K}(u,v;\lambda)
=
\frac{1}{2v}
\sum_{s=1}^{K}
\nu_s
\left[
t_s(u,v;\lambda)^2-1
\right]
~.
\end{equation}
This result follows from supersymmetric localization~\cite{Binder:2019jwn}
combined with the large-charge diagrammatic analysis~\cite{Brown:2024yvt}. 
Here $\nu_s$ denotes the multiplicity of a given massive sector $s$ while $t_s$ is its scalar propagator normalized as follows. We place heavy insertions at the past and future ends of the Euclidean cylinder $\mathbb{R}\times S^3$, with unit sphere radius. The light insertions have time separation $\sigma$ and angular separation $\phi$ related to the cross-ratios by
\begin{equation}
    \sigma = -\frac12 \log u~,\quad  \cos\phi
=
\frac{1+u-v}{2\sqrt u}~,\quad \frac{v}{\sqrt{u}} = 2(\mathrm{cosh} \sigma -\cos \phi)~.
\end{equation}
In these variables, 
the relevant equation satisfied by the Green function takes the form
\begin{align}
\left[-\partial_\sigma^2-\Delta_{S^3}+1+4M_s^2\right]G_s(\sigma,\phi)
&=\delta(\sigma)\,\delta_{S^3}(n,n'),
\end{align}
where $n\cdot n'=\cos\phi$ and 
\be
G_s(\sigma,\phi)
=\frac{t_s(u,v;\lambda)}{8\pi^2(\cosh\sigma-\cos\phi)}
=\frac{\sqrt{u}}{4\pi^2v}\,t_s(u,v;\lambda).
\ee
The solution to this equation can be written in the partial wave expansion as follows
\cite{Giombi:2020enj,Caetano:2023zwe}
\begin{equation}\label{eq:propagatorts}
    t_s(u,v,\lambda) = \frac{v}{\sqrt{u}}\sum_{r=1}^{\infty} \frac{r}{\sqrt{r^2+4M^2_s}}u^{\frac{1}{2}\sqrt{r^2+4M^2_s}}\mathcal G_{r-1}^1(\cos\phi)~,\quad \mathcal G_r^1(\cos\phi)
=
\frac{\sin((r+1)\phi)}{\sin\phi}~,
\end{equation}
where the mass parameters $M_s$ are given below. The function $\mathcal{G}^1_r$ is the four-dimensional Gegenbauer polynomial, written in a normalization convenient for the large-charge conformal blocks.\footnote{
Remarkably, the coefficient of $(-M_s^2)^\ell$ in the Taylor expansion of
$t_s(u,v;\lambda)$ is the $\ell$-loop conformal ladder
integral~\cite{Isaev:2003tk,Giombi:2020enj,Caetano:2023zwe}.}

Let us briefly recall the semiclassical origin of this expression.\footnote{As explained in \cite{Brown:2025cbz}, the semiclassical description is inferred by combining  localization techniques from \cite{Binder:2019jwn} with the large-charge diagrammatic analysis~\cite{Brown:2024yvt}. A direct saddle point derivation from the path integral remains to be established.} The heavy canonical insertions source a Coulomb-branch-like scalar background. Due to this profile, off-diagonal scalar fluctuations acquire spacetime dependent masses, while the unbroken Cartan modes remain massless. A canonical operator of scaling dimension $\Delta_H=mK$ in $SU(N)$ $\mathcal{N}=4$ Higgs the gauge group to
\begin{equation}
    SU(N)\to U(1)^{K-1}\times U(N-K) \,.
\end{equation}
 After mapping the configuration to the cylinder $\mathbb{R}\times S^3$, with the radius of $S^3$ set to one, the position-dependent masses in flat space become dimensionless cylinder masses $M_s$. The propagator $t_s$ is normalized so that $t_s(u,v;0)=1$ for generic $u,v$. Therefore, the explicit subtraction term $-1$ appearing in \eqref{eq:resummedHHLL} is part of the canonical correlator and ensures that its interaction-dependent contribution vanishes at $\lambda=0$.

The perturbative expansion in $M_s^2$ can be resummed. The square $t_s^2$ in $\eqref{eq:resummedHHLL}$ arises from the two scalar propagators connecting the dimension-two light insertions. The massive sectors and their multiplicities are
\begin{equation}\label{eq:masses}
\begin{aligned}
\nu_s &= K~,
&\qquad
M_s &= 2\sqrt{\lambda}\,
\sin \left(\frac{\pi s}{K}\right),
&\qquad
s &= 1,\ldots,K-1~,
\\
\nu_K &= 2K(N-K)~,
&
M_K &= \sqrt{\lambda}~.
\end{aligned}
\end{equation}
Upon squaring the propagator \eqref{eq:propagatorts} and decomposing into total spin $J$ we obtain the heavy-light OPE decomposition
\begin{equation}
t_s(u,v;\lambda)^2
={}
\frac{v^2}{u}
\sum_{J\geq0}
\sum_{n\geq0}
\sum_{r=1}^{J+1}
C_{J,n,r}^{(s)}
u^{\frac{\Delta_{J,n,r}^{(s)}-\Delta_H}{2}}
\mathcal G_J^1(\cos\phi)~.
\end{equation}
The positive heavy-light OPE weights are
\begin{equation}\label{eq:Cs}
    C_{J,n,r}^{(s)}
=
\frac{
(r+n)(J+2+n-r)
}{
\sqrt{(r+n)^2+4M_s^2}\,
\sqrt{(J+2+n-r)^2+4M_s^2}
}~,
\end{equation}
while the corresponding exchanged dimensions are
\begin{equation}
\Delta_{J,n,r}^{(s)}
=
\Delta_H + \sqrt{(r+n)^2+4M_s^2} +\sqrt{(J+2+n-r)^2+4M_s^2}~.
\end{equation}
where the spectral labels range over $J\geq 0$, $n\geq 0$ and $r=1,\dots,J+1$.
It is convenient to define the shifted heavy-light gap
\begin{equation}\label{eq:gammaHL}
    \begin{aligned}
\gamma_{J,n,r}^{(s)}
&\equiv
\Delta_{J,n,r}^{(s)}-\Delta_H-2
\\
&=
\sqrt{(r+n)^2+4M_s^2}
+
\sqrt{(J+2+n-r)^2+4M_s^2}
-2~.
\end{aligned}
\end{equation}
In what follows, we will often denote this quantity by $\gamma_{\mathrm{HL}}$. Substituting all these results into $\Phi_{\Delta_H}$, we obtain
\begin{equation}
\Phi_{\Delta_H}(u,v;\lambda)
=
\frac{1}{2\Delta_H}
\bigg[
\frac12v^2
\sum_{s=1}^{K}\nu_s
\sum_{J,n,r}
C_{J,n,r}^{(s)}
u^{\frac{\gamma_{\rm HL}}{2}}
\mathcal G_J^1(\cos\phi)
-
\frac12
\sum_{s=1}^{K}\nu_s
\bigg]~.
\end{equation}

The superconformal Ward identities imply that the interaction-dependent EEC can be computed directly from $\Phi_{\Delta_H}$. An efficient way to do this is via the Mellin transform  of the function $\Phi_{\Delta_H}$, that is 
\begin{equation}
   \Phi_{\Delta_H}(u,v;\lambda) =
\int{dj_1\,dj_2\over(2\pi i)^2}\,
M(j_1,j_2;\lambda)\,U^{j_1}V^{j_2}~,
\end{equation}
where $U=u/v$ and $V=1/v$. The Mellin amplitude $M(j_1,j_2;\lambda)$ contains the coupling-dependent information. Its relation to the reduced Mellin amplitude used in the OPE analysis is given in Appendix~\ref{app:fourpoint_mellin}.
More precisely, the interaction-dependent EEC is obtained from $M(j_1,j_2;\lambda)$  through the universal detector
kernel of \cite{Chicherin:2023gxt}
\begin{equation}\label{eq:EEC_Mellin_master}
\CF_{\rm int}(z,\lambda)
=\frac{\Delta_H^2}{2}
\int {dj_1\,dj_2\over(2\pi i)^2}
\left[
{\Gamma(1-j_1-j_2)\over
\Gamma(1-j_1)\Gamma(1-j_2)}
\right]^2
M(j_1,j_2;\lambda)\,\CK_{\Delta_H}(j_1+j_2,z)~.
\end{equation}
At large charge,
\begin{equation}\label{eq:large_charge_kernel}
\CK_{\Delta_H}(j,z)
=
{1\over\Delta_H}\kappa(j,z)+O(\Delta_H^{-2})~,
\end{equation}
where
\begin{equation}\label{eq:kappa_general}
\kappa(j,z)
=
(j-2)(j-1)z^{-j-2}
\left[
j(j+1)+2j(1-j)z+(j-1)(j-2)z^2
\right]~.
\end{equation}
For later use, it is convenient to express $\kappa(j,z)$ as the action of a fourth-order differential operator on a monomial:
\begin{equation}\label{eq:calD}
    \kappa(j,z) = \mathcal{D}z^{-j+1}~,\qquad \mathcal{D}\equiv \frac{1}{z}\frac{d}{dz}z^2\frac{d}{dz}\frac{(1-z)^2}{z^2}\frac{d}{dz}z^2\frac{d}{dz}~.
\end{equation}
We will make repeated use of this differential representation throughout the paper. 

To expose the action of the detector transform on the heavy-light OPE,
consider an ordinary exchanged primary of spin $J$ and dimension
$\Delta_{\rm ord}=\Delta_H+\gamma_{\rm HL}$. This is the dimension
entering the reduced scalar conformal block; the corresponding physical
superconformal primary has dimension
$\Delta_{\rm phys}=\Delta_{\rm ord}+2
=\Delta_H+\gamma_{\rm HL}+2$.
The poles of the reduced Mellin amplitude associated with this primary
and its descendants can be evaluated by residues. Following the
derivation in Appendix~\ref{app:fourpoint_mellin}, one obtains the
ordinary detector block
\begin{equation}\label{eq:ordinary_detector_block}
\begin{aligned}
\mathcal I_{J,\gamma_{\rm HL}}(z)
={}&
\sum_{a=0}^{J-1}\sum_{p=0}^{J-a-1}
\frac{(J-a-p)!(J-p)!}{a!}\binom{J-a}{p}\,
\frac{2^{1-J}(-1)^{a+p}\left(-\frac{J+\gamma_{\rm HL}}{2}\right)_p^2}
{\Gamma \left(\frac{2+J-2a-\gamma_{\rm HL}}{2}\right)^2
\Gamma \left(\frac{2+J+\gamma_{\rm HL}}{2}\right)^2}
\\
&\times
\left[\kappa(a+p-J,z)+6\delta_{1,J-a-p}\delta(z)\right].
\end{aligned}
\end{equation}
Here $(x)_p$ denotes the Pochhammer symbol. The indices $a$ and $p$ originate from the descendant sum and the large-charge expansion of the Mellin residues; they are unrelated to the physical labels $n$ and $r$.

The ordinary detector blocks $\mathcal I_{J,\gamma_{\rm HL}}$ must be recombined into complete
superconformal multiplets. Applying the Gegenbauer decomposition derived
in Appendix~\ref{app:fourpoint_mellin}, including its $J=0$
prescription, gives the detector superblock
\begin{equation}\label{eq:detector_superblock}
\begin{aligned}
\mathcal S_{J,\gamma_{\rm HL}}(z)
&=
(-1)^J2^J
\bigg[
{1\over2}\mathcal I_{J,\gamma_{\rm HL}}(z)
+2\mathcal I_{J,\gamma_{\rm HL}+2}(z)
+{1\over2}\mathcal I_{J,\gamma_{\rm HL}+4}(z)
\\
&\hspace{2.7cm}
+\sum_{\sigma=\pm1}2^\sigma
\left(
\mathcal I_{J+\sigma,\gamma_{\rm HL}+1}(z)
+\mathcal I_{J+\sigma,\gamma_{\rm HL}+3}(z)
\right)
+\sum_{\sigma=\pm1}2^{2\sigma-1}
\mathcal I_{J+2\sigma,\gamma_{\rm HL}+2}(z)
\bigg]~,
\end{aligned}
\end{equation}
Terms with negative spin are understood to vanish. Notice that the delta-function contribution appearing in the individual detector blocks cancel in this superblock combination. Summing over the spectral labels of the physical heavy-light superblocks leads to
\begin{equation}\label{eq:Ffc}
\mathcal F(z,\lambda)
\equiv
\sum_{s=1}^K\nu_s
\sum_{J\geq0}\sum_{n\geq0}\sum_{r=1}^{J+1}
C_{J,n,r}^{(s)}
\mathcal S_{J,\gamma_{J,n,r}^{(s)}}(z)~.
\end{equation}
Looking at \eqref{eq:gammaHL} and \eqref{eq:Cs},  at zero coupling we have 
$\gamma_{J,n,r}^{(s)}=J+2n$ and
$C_{J,n,r}^{(s)}=1$. Furthermore, the detector superblocks obey
\begin{equation}
    \mathcal S_{J,J+2n}(z)
=
\delta_{n0}\mathcal S_{J,J}(z)~.
\end{equation}
As a result, only the $n=0$ tower contributes, and the sum over the $J+1$ possible values of $r$ in \eqref{eq:Ffc} gives
\begin{equation}\label{eq:F_zero_coupling}
\mathcal F(z,0)
=
\sum_{s=1}^K\nu_s
\sum_{J\geq0}(J+1)\mathcal S_{J,J}(z)~.
\end{equation}
This is the detector-space image of $t_s(u,v;0)^2=1$. The explicit $-1$ in \eqref{eq:resummedHHLL} therefore subtracts the coupling independent contribution $\mathcal F(z,0)$. Combining this subtraction with the normalization of $\Phi_{\Delta_H}$, the interaction-dependent event shape is
\begin{equation}\label{eq:ECC_final}
\CF_{\rm int}(z,\lambda)
=
\frac{
\mathcal F(z,\lambda)-\mathcal F(z,0)
}{
8
}~.
\end{equation}
with $\CF(z,\lambda)$ defined above.\footnote{Let us reiterate: ${\cal F}(z)$ is defined in \eqref{eq:EEC_F_definition}; ${\cal F}(z,\lambda)$ is defined in \eqref{eq:Ffc}; finally, $\CF_{\rm int}(z,\lambda)$ is defined in \eqref{eq:ECC_final}. We hope that these distinctions clarify the notation.}

\subsection{Positivity Properties}\label{sec:positivityprop}

Let us now discuss the positivity properties of the result obtained above. We recall that, Hilbert-space positivity constrains the Legendre moments of the full physical EEC, as shown in $\eqref{eq:Legendre_positivity}$. The interaction-dependent event shape \eqref{eq:ECC_final}, however, is defined as the difference between the finite-coupling and zero-coupling superblock sums and therefore need not be positive moment by moment.

At fixed exchanged spin $J$, the detector superblock $\mathcal{S}_{J,\gamma_{\rm HL}}(z)$ defined in \eqref{eq:detector_superblock} is a finite
polynomial in $z$ of degree at most $J+2$, and therefore admits the finite
Legendre expansion
\begin{equation}\label{eq:blockLegendre}
    \mathcal{S}_{J,\gamma_{\rm HL}}(z) = \sum^{J+2}_{\ell=2}(2\ell + 1)a_{\ell,J}(\gamma_{\rm HL})P_\ell(1-2z)~,
\end{equation}
where
\begin{equation}
    a_{\ell,J}(\gamma_{\rm HL}) = \int^1_0 dz P_\ell(1-2z)\mathcal{S}_{J,\gamma_{\rm HL}}(z)~.
\end{equation}
Importantly, the sum in \eqref{eq:blockLegendre} starts from $\ell=2$, indeed 
by direct inspection one can show that the first two Legendre moments $\ell=0,1$ vanish:
\begin{equation}
    a_{0,J}(\gamma_{\rm HL})=a_{1,J}(\gamma_{\rm HL})=0~.
\end{equation}
The absence of the $\ell=0,1$ components shows that the homogeneous sum rules
are satisfied block by block. By similar arguments,  it can also be shown that the non-trivial positivity constraints on the Legendre moments coming from \eqref{eq:Legendre_positivity} are indeed satisfied
\begin{equation}
    a_{\ell\geq 2,J}(\gamma_{\rm HL}) \geq 0~.
\end{equation}
Together with the positivity of the multiplicities $\nu_s$ and OPE coefficients $C_{J,n,r}^{(s)}$, this implies that the unsubtracted superblock sum \eqref{eq:Ffc} has non-negative Legendre moments:
\begin{equation}
\sum_{s=1}^{K}\nu_s
\sum_{J\geq0}
\sum_{n\geq0}
\sum_{r=1}^{J+1}
C_{J,n,r}^{(s)}
a_{\ell,J} \left(\gamma_{\rm HL}\right)
\geq0~.
\end{equation}
The infinite sum over the exchanged spin $J$ nevertheless requires some care. This is particularly transparent at zero coupling in \eqref{eq:F_zero_coupling}. To see this, let us first consider the region $0<z<1$. Introducing a finite spin cutoff $J_{\rm max}$, the sum can be performed explicitly and one has
\begin{equation}
    \lim_{J_{\rm max}\to\infty}\sum^{J_{\rm max}}_{J= 0}(J+1)\mathcal{S}_{J,J}(z) = 4~,\qquad 0<z<1~.
\end{equation}
The constant $4$ alone, however, cannot satisfy the Ward identities \eqref{eq:sumrulesint}. It follows that the limit $J_{\max}\to\infty$ cannot be interchanged with the angular integrations. The convergence of the spin sum is non-uniform near an endpoint. Therefore we expect contributions from distributional contact terms localized at the endpoint $z = 0$ and $z = 1$.

Notice that endpoint contact terms are intrinsically tied to the infinite sum and, in particular, to its large-$J$ behavior. No individual superblock and no finite truncation of the spin sum can produce such terms. The analysis of the large-$J$ tail in  Section \ref{sec:smallz} will show that there is no contribution supported at $z=1$. Anticipating that result, the zero-coupling sum is completed as\footnote{Here we adopt the conventions
\be
\int_0^1\rd z\,\delta(z)\varphi(z)=\varphi(0)~,
\qquad
\int_0^1\rd z\,\delta'(z)\varphi(z)=-\varphi'(0)~,
\ee
where $\varphi(z)$ is a test function.}
\begin{equation}
    \sum_{J\ge0}(J+1)S_{J,J}(z)=4 -4\,\delta(z)+2\delta'(z) ~.
\end{equation}
After summing over the different $s$-sectors with $\nu_s$ as in \eqref{eq:masses}, it follows that the zero-coupling subtraction in \eqref{eq:ECC_final} is given by
\begin{equation}\label{eq:subtractiondeltaprime}
 -\frac{K(2N-K-1)}{8}\left(4 -4\,\delta(z)+2\delta'(z)\right)~.
\end{equation}
We emphasize that the term proportional to $\delta'(z)$ should not be interpreted as an exact contact term in the finite-$\Delta_H$ event shape. Its presence is required by the $\ell=1$ Ward identity. However, at finite-$\Delta_H$ this term is in fact prohibited by the light-ray OPE. This will be further discussed in Section \ref{sec:lightrayOPE}. Rather, it encodes the first subleading moment of an unresolved boundary layer of width $z\sim\Delta_H^{-1}$. In other words $ \lim_{\Delta_H \to \infty} \sum_{\ell} \neq \sum_{\ell} \lim_{\Delta_H \to \infty}$. The former is what produces the full energy correlator, the latter is what is done in the present paper.

The Legendre moments of the zero-coupling complete sum follow directly from the identity
\begin{equation}\label{eq:zeromoments}
    \int_{0}^1 dz~ P_\ell(1-2z)[4-4\delta(z)+2\delta'(z)] = 
\begin{cases}
0~,\quad  \ell=0,1~,\\[2mm]
2(\ell-1)(\ell+2)~,  \quad \ell\geq2~.
\end{cases}
\end{equation}
Let us now make compatibility with positivity of the full EEC \eqref{eq:splitting} explicit. The Legendre moments derived from the free contribution \eqref{eq:EECBorn} obey
\begin{equation}\label{eq:BornLegendre}
    a_{{\rm free},0} =1~,\quad a_{{\rm free},1} =0~,\quad a_{{\rm free},\ell \geq 2
} = \frac{3}{2\Delta_H} + O(\Delta^{-2}_H)~.
\end{equation}

To analyze the contribution from $\CF_{\rm int}(z,\lambda)$, we use \eqref{eq:zeromoments} and rewrite the result as
\begin{equation}
  \frac{1}{8\Delta_H^2}
\sum_{s=1}^{K}\nu_s
\sum_{J\geq0}
\sum_{n\geq0}
\sum_{r=1}^{J+1}
C_{J,n,r}^{(s)}
a_{\ell,J} \left(\gamma_{\rm HL}\right)
-
\frac{K(2N-K-1)}{4\Delta_H^2}
(\ell-1)(\ell+2)
~.
\end{equation}
The first term is non-negative by positivity of the detector-superblock moments, whereas the second term is the negative zero-coupling subtraction. Their difference therefore need not have a definite sign. Combining this result with the free contribution, the full physical moment takes the form
\begin{equation}
    \begin{aligned}
a_\ell(\lambda)
={}&
a_{\text{free},\ell}
-
\frac{K(2N-K-1)}{4\Delta_H^2}
(\ell-1)(\ell+2)
\\
&+
\frac{1}{8\Delta_H^2}
\sum_{s=1}^{K}\nu_s
\sum_{J\geq0}
\sum_{n\geq0}
\sum_{r=1}^{J+1}
C_{J,n,r}^{(s)}
a_{\ell,J} \left(\gamma_{\rm HL}\right)~,
\end{aligned}
\end{equation}
with fixed $\ell \geq 2$. Comparing with the behavior of the free contribution from \eqref{eq:BornLegendre} we see that the potentially negative zero-coupling subtraction begins at order $\Delta_H^{-2}$. A negative interaction-dependent moment is therefore a subleading correction to the positive free result and is fully compatible with positivity of the complete EEC in the large-charge expansion. This provides an important consistency check of our result.

As already alluded above, the comparison is not uniform in $\ell$. When $\ell\sim\sqrt{\Delta_H}$, the subtraction becomes comparable to the free contribution. In this regime, the Legendre polynomial $P_\ell(1-2z)$ resolves angular distances $z\sim\ell^{-2}\sim\Delta_H^{-1},$ precisely the scale of the boundary layer represented by the derivative contact term. The fixed-$\ell$ large-charge expansion then ceases to be uniform, and the truncated result cannot by itself be used to test positivity.

\subsection{Light-Ray OPE Constraints}\label{sec:lightrayOPE}

The small-angle behavior of the EEC, with every other physical parameters of the problem kept fixed, is governed by the light-ray OPE \cite{Dixon:2019uzg,Korchemsky:2019nzm,Kologlu:2019mfz,Chicherin:2023gxt}. In the normalization introduced in $\eqref{eq:normalized_flow_F}$, its leading contribution takes the form
\begin{equation}\label{eq:OPEpred}
    \left\langle
\widehat{\mathcal E}(\vec{n}_1)
\widehat{\mathcal E}(\vec{n}_2)
\right\rangle_H
\underset{z\to0}{\sim}
\frac{
\left\langle\mathbb O_3^+\right\rangle_H
}{
z^{1-\gamma_3^+/2}
}
+\ldots~,
\qquad
0<z\ll1~.
\end{equation}
Here $\mathbb O_3^+$ is the positive-signature spin-three light-ray operator on the stress-tensor Regge trajectory, while $\gamma_3^+$ is its anomalous dimension. The normalization of $\mathbb O_3^+$ is understood to include the corresponding light-ray OPE coefficient.\footnote{Notice that $\gamma_3^+$ should not be confused with the heavy-light gaps $\gamma_{J,n,r}^{(s)}$ entering the detector superblocks.} For $g^2_{\rm YM} \ll 1$, we have:
\begin{equation}
    \gamma^+_3 = \frac{g_{\rm YM}^2 N}{2\pi^2} + O(g^4_{\rm YM})~.
\end{equation}
Therefore, in the large-charge double scaling limit \eqref{eq:dstooft} considered here:
\begin{equation}
    \gamma_3^+
=
\frac{8KN\lambda}{\Delta_H}
+
 O(\Delta_H^{-2})~,
\end{equation}
where $N$, $K$, and the collective coupling $\lambda$ are held fixed. Thus, although $\lambda$ remains finite, the anomalous dimension controlling the light-ray OPE approaches its free-theory value,
\begin{equation}
    \gamma^+_3 = O(\Delta^{-1}_H)~.
 \end{equation}
The normalization of the leading collinear contribution can be fixed by matching it to the free contact term using the identity
\begin{equation}
    \frac{\gamma_3^+}{2}
z^{-1+\gamma_3^+/2}
=
\delta(z)
+
\frac{\gamma_3^+}{2}
\left[\frac1z\right]_+
+
\frac{(\gamma_3^+)^2}{4}
\left[\frac{\log z}{z}\right]_+
+\ldots,
   \end{equation}
where $[\cdot]_{+}$ denotes a plus-distribution.\footnote{We follow \cite{GelfandShilov1964}, where one defines \begin{equation}\nonumber
    \int_{0}^1 dz~\left[\frac{1}{z}\right]_+\varphi(z) \equiv \int_{0}^1~dz~\frac{\varphi(z)-\varphi(0)}{z}~,\qquad \int_{0}^1 dz~\left[\frac{\log^k(z)}{z}\right]_+\varphi(z) \equiv \int_{0}^1~dz~\log^k(z)\frac{\varphi(z)-\varphi(0)}{z}~.
\end{equation}\label{footnote}}~Since the free event shape contains
\begin{equation}
    \mathcal F_{\rm free}(z)
\supset
\frac{3}{2(\Delta_H+1)}\,\delta(z)~,
\end{equation}
matching the delta function fixes the leading light-ray contribution to be
\be
\mathcal F(z)=
\frac{3}{2(\Delta_H+1)}
\frac{\gamma_3^+}{2}
z^{-1+\gamma_3^+/2} + \textrm{subleading}~.
\ee
Expanding at one-loop gives the prediction
\be
\mathcal F(z)=
\frac{3}{2(\Delta_H+1)}
\left[
\delta(z)
+
\frac{g_{\rm YM}^2N}{4\pi^2}
\left[\frac1z\right]_+
\right]~+\dots~,
\ee
up to terms which are less singular for $z\to 0$. After subtracting the free contact term and using the double-scaling relation for $g^2_{\rm YM}$, this becomes 
\be\label{eq:F_col_exp} 
{1 \over \Delta_{H}^2}{\mathcal F_{\mathrm{int}}}(z,\lambda)
\underset{z\to 0}{\sim}
\frac{6KN\lambda}
{\Delta_H(\Delta_H+1)}
\left[\frac1z\right]_+
=
\frac{6KN\lambda}{\Delta_H^2}
\left[\frac1z\right]_+
+
 O \left(\Delta_H^{-3}\right).
\ee
Notice that this implies that the ${1 \over z}$ term is one-loop exact.

Equivalently, the heavy-state matrix element appearing in the light-ray OPE satisfies, at this order, 
\be
\left\langle\mathbb O_3^+\right\rangle_H
=
\frac{3\gamma_3^+}{4(\Delta_H+1)}
+\ldots
=
\frac{6KN\lambda}{\Delta_H^2}
+
 O \left(\Delta_H^{-3}\right)~.
\ee
Notice that this is stronger than the bound
\be
\left\langle\mathbb O_3^+\right\rangle_H
\lesssim
\Delta_H^{-1}~,
\ee
suggested by the constituent estimate of \cite{Chicherin:2023gxt} for planar $\CN=4$ SYM. As already mentioned in Section \ref{sec:cumulantsheavyexp} that bound is not saturated in the large-charge double-scaling limit: the matrix element acquires an additional suppression because the microscopic coupling, and hence $\gamma_3^+$, vanishes as $1/\Delta_H$.

\section{Weak Coupling}\label{sec:weakcoupling}
At small $\lambda$ the EEC admits the expansion
\begin{equation}\label{eq:weak_coupling_EEC_expansion}
    \CF_{\rm int}(z,\lambda) = \frac{1}{8}\sum_{k\geq 1} \varphi^{(k)}(z) \lambda^k
    + O(\Delta_H^{-1}) \,.
\end{equation}
The sum starts with $k=1$ since by construction $\CF^{\rm int}(z,0)=0$, see for example equation \eqref{eq:ECC_final}. As $M_s\sim \sqrt{\lambda}$, the weak-coupling expansion is obtained by
expanding the OPE data at small $M_s$ and performing the sums over $J,r,n$ order by order. In what follows, we perform the full one loop computation and discuss the main features of the two loop result.

\subsection{One-loop Correction}
Let us compute the one loop correction to the interacting EEC. Expanding to one loop the OPE coefficient and scaling dimensions we find
\begin{equation}\label{eq:weak_small_M_data}
\begin{aligned}
C_{J,n,r}^{(s)}
&\simeq
1
-2M_s^2
\left[
{1\over (J+n+2-r)^2}
+{1\over (n+r)^2}
\right] \\ \gamma_{J,n,r}^{(s)}&\simeq J+2n+2M_s^2\left[{1\over J+n+2-r}+{1\over n+r}\right]
\end{aligned}
\end{equation}
Expanding the whole summand we get
\begin{equation}\label{eq:weak_one_loop_block_expansion}
\begin{aligned}
C_{J,0,r}^{(s)}
\mathcal S_{J,\gamma_{J,0,r}^{(s)}}(z)
-\mathcal S_{J,J}(z)
&=
2M_s^2
\left[
\left(
{1\over J+2-r}
+
{1\over r}
\right)
 \partial_\gamma \mathcal{S}_{J,\gamma}(z)|_{\gamma=J}\, +
\right.
\\
&\hspace{1.6cm}
\left.
-
\left(
{1\over (J+2-r)^2}
+
{1\over r^2}
\right)
\mathcal S_{J,J}(z)
\right]
 +O(M_s^4) \,,
\end{aligned}
\end{equation}
where we have used the fact that only the $n=0$ tower contributes at this order (see equation \eqref{eq:sublargJSn0}). The sum over $r$ can be performed explicitly in terms of harmonic numbers
\begin{equation}\label{eq:weak_one_loop_spin_sum}
\sum_{r=1}^{J+1}
\left[
C_{J,0,r}^{(s)}
\mathcal S_{J,\gamma_{J,0,r}^{(s)}}(z)
-\mathcal S_{J,J}(z)
\right]
\simeq
4M_s^2
\left[
H_{J+1} \partial_\gamma \mathcal{S}_{J,\gamma}(z)|_{\gamma=J}\
-H_{J+1}^{(2)}\mathcal S_{J,J}(z)
\right] \,.
\end{equation}
Using the expression for $\mathcal{S}_{J,J}$ and its derivative with respect to the scaling dimension given in \eqref{eq:1LSS}, the sum over $J$ then reads
\begin{equation}\label{eq:weak_phi1_exact}
\varphi^{(1)}(z)
=
16KN
\left[
{3\over z}
+2\operatorname{Li}_2(1-z)
-6\log z
-2\zeta(2)
-{13\over2}
\right].
\end{equation}
The function of $z$ in square brackets coincides with the planar
large-charge one-loop EEC found in
\cite{Chicherin:2023gxt}.  This agreement is expected from the
large-charge diagrammatic analysis of \cite{Brown:2024yvt}: at one loop the
surviving diagrams are the same as  in the planar
large-$N$ computation of \cite{Chicherin:2023gxt}. This equivalence does
not persist beyond one loop.

Note that we have reproduced the one loop exact $1/z$ term predicted from the light ray OPE analysis, see equation  \eqref{eq:F_col_exp}. In fact in the collinear limit we get
\begin{equation}\label{eq:weak_phi1_small_z}
\varphi^{(1)}(z)
\simeq
16KN\left({3\over z}-6\log z+{O}(1)\right),
\qquad z\to0.
\end{equation}
It is instructive to reproduce \eqref{eq:weak_phi1_small_z} directly from the large-spin tail
of \eqref{eq:weak_one_loop_spin_sum}. As we will further explain in  Section \ref{sec:smallz} the relevant scaling limit is
\begin{equation}
z\to0,\qquad J\to\infty,\qquad x=Jz\quad {\rm fixed}.
\end{equation}
In this limit, the sum over spin can be replaced by
\begin{equation}\label{eq:J_int}
\sum_{J\geq0}\ \simeq\ {1\over z}\int_0^\infty dx\,.
\end{equation}
Using  equations \eqref{eq:xderi}, at leading order we need to expand the harmonic numbers in \eqref{eq:weak_one_loop_spin_sum} and evaluate
\begin{equation}\label{eq:small1z1lint}
\begin{aligned}
{1\over z}\int_0^\infty dx\,\bigg[
\left(\log{x\over z}+\gamma_E\right)
\mathcal \partial_x^4\left(
x^4\int_0^1dt\,{t^2e^{-tx}-e^{-x}\over1-t}
\right)
-{\pi^2\over6}\mathcal \partial_x^4 \, \left(x^4 e^{-x} \right)
\bigg]={6\over z} \,,
\end{aligned}
\end{equation}
where again since the superblock integrates to zero, the only term that contributes is the $\log J$ divergence. In order to reproduce the $\log z$ singularity we expand both the harmonic numbers and the superblock at subleading order. The leading superblock multiplying $1/J$ terms in the large $J$ expansion of the harmonic numbers will produce a $\log z$ divergence inherited from small $x$ singularities, as discussed at the end of Section \ref{sec:smallz}. There is an extra term coming from the subleading superblock multiplying the $\log z$ appearing already in \eqref{eq:small1z1lint}. We then need the subleading expression in
\begin{equation}
\begin{aligned}
    \partial_\gamma \mathcal{S}_{J,\gamma}(z)|_{\gamma=J} \simeq& \partial_x^4\left(
x^4\int_0^1dt\,{t^2e^{-tx}-e^{-x}\over1-t}
\right)+ \\ +&
\frac{1}{2J}\,
\partial_x^3\left(\int_0^1 dt
\frac{x^4}{t-1}
\left(
e^{-x}(x^2-2x-12)
-
t^2 e^{-tx}\left(t^3x^2-6t^2x+4tx-12\right)
\right)
\right)
\end{aligned}
\end{equation}

Adding up all together we get
\begin{equation}
\begin{aligned}
    &\log \frac{1}{z} \int_0^\infty\frac{1}{2x}\,
\partial_x^3\left(\int_0^1 dt
\frac{x^4}{t-1}
\left(
e^{-x}(x^2-2x-12)
-
t^2 e^{-tx}\left(t^3x^2-6t^2x+4tx-12\right)
\right)
\right)+\\&+\int_{J_0 z}^\infty\frac{1}{x}\,\partial_x^4\left(\frac{3}{2}
x^4\int_0^1dt\,{t^2e^{-tx}-e^{-x}\over1-t} 
+x^4e^{-x}\right)\simeq-12\log z
\end{aligned}
\end{equation}
Summing over $s$ we recover \eqref{eq:weak_phi1_small_z}.

\begin{figure}[t]
\centering 
\includegraphics[width=0.8\linewidth]{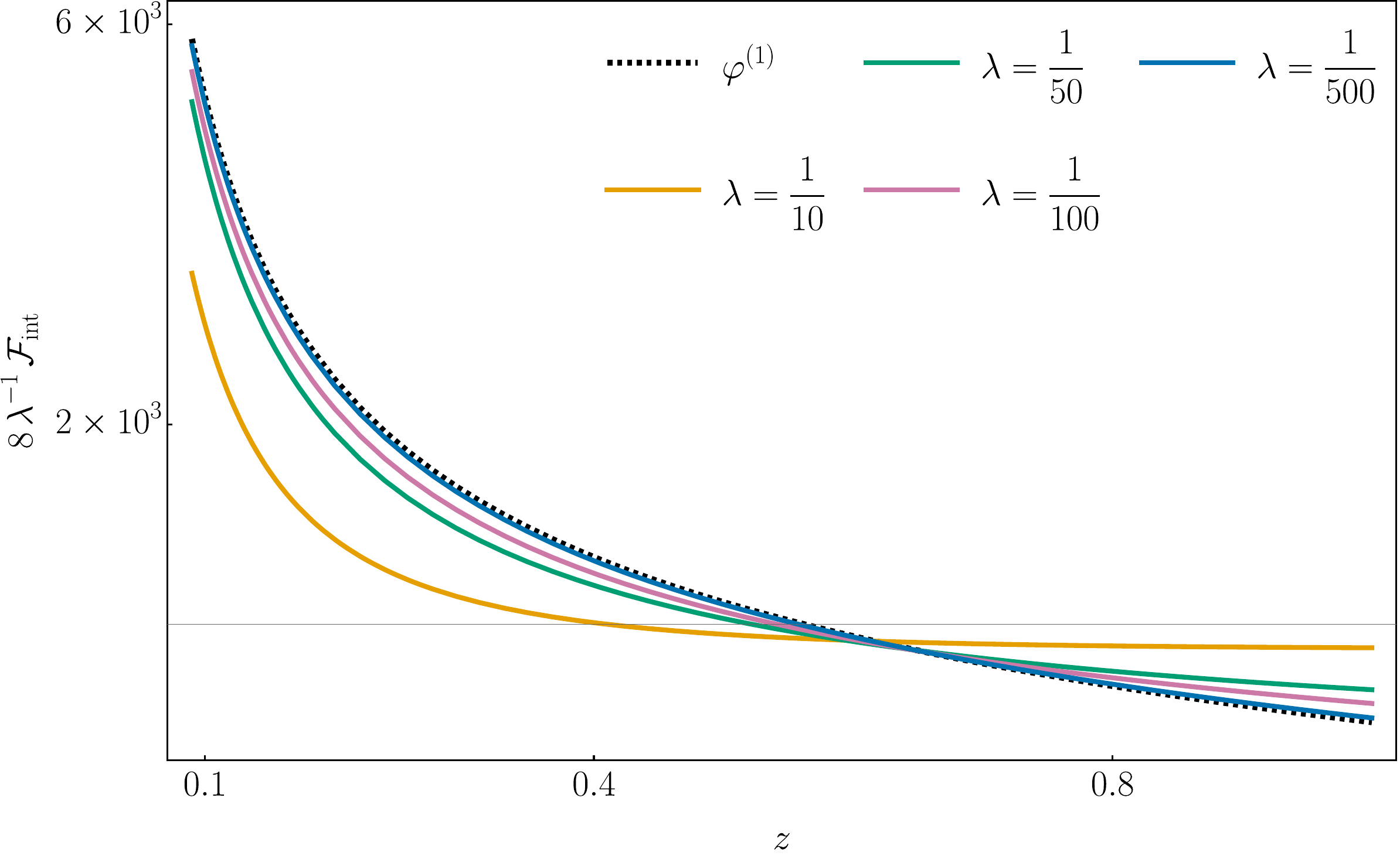}
\caption{$8\CF^{\rm int}(z,\lambda)/\lambda$ at weak coupling for $N=K=3$. The dashed black one loop prediction is nicely approached as we decrease the coupling. Here $J_{\text{max}}=n_{\text{max}}=700$.}
\label{fig:1L3}
\end{figure}

The one loop result is nicely approached by numerics at weak coupling as shown in figure \ref{fig:1L3} in the case $N=K=3$.

\subsection{Two-loop Correction}\label{sec:twoloopss}
At two loops order we need to expand at order $O(M_s^4)$ OPE coefficients, scaling dimensions and superblocks.
Adding up all terms together and performing the sums over $r$, the two loops correction can be written as
\begin{equation}\label{eq:weak_two_loop_spin_sum}
\begin{aligned}
{\varphi^{(2)}(z)\over 2K(2K+N)}
&=
\sum_{J\ge0}
\left[
12H_{J+1}^{(4)}
+{4\over (J+2)^2}
\left(
2H_{J+1}^{(2)}
+{4H_{J+1}\over J+2}
\right)
\right]
\mathcal S_{J,J}(z)
\\
&\quad
+\sum_{J\ge0}
\left[
-12H_{J+1}^{(3)}
-{4\over J+2}
\left(
2H_{J+1}^{(2)}
+{4H_{J+1}\over J+2}
\right)
\right]
\mathcal \partial_\gamma S_{J,\gamma}(z)|_{\gamma=J}
\\
&\quad
+\sum_{J\ge0}
\left[
4H_{J+1}^{(2)}
+{8H_{J+1}\over J+2}
\right]
\partial_\gamma^2\mathcal S_{J,\gamma}(z)\big|_{\gamma=J}
\\
&\quad
+\sum_{\substack{J\ge0\\ n\ge1}}
\left[
4\left(H_{J+n+1}^{(2)}-H_n^{(2)}\right)
+8{H_{J+n+1}-H_n\over J+2n+2}
\right]
\partial_\gamma^2\mathcal S_{J,\gamma}(z)\big|_{\gamma=J+2n}.
\end{aligned}
\end{equation}
The first line can be resummed explicitly to
\begin{equation}
\begin{aligned}
&\sum_{J\ge0}
\left[
12H_{J+1}^{(4)}
+{4\over (J+2)^2}
\left(
2H_{J+1}^{(2)}
+{4H_{J+1}\over J+2}
\right)
\right]\mathcal S_{J,J}(z)
\\
&\qquad
=
\mathcal D_z
\left[
{4z^3\over (1-z)^2}\left[\operatorname{Li}_2(1-z)\right]^2
+{12z^2\over 1-z}\operatorname{Li}_4(1-z)
\right],
\end{aligned}
\end{equation}
which is regular as $z\to0$.  The terms involving one $\gamma$-derivative
can also be resummed:
\begin{equation}\label{eq_one_derivative}
\begin{aligned}
&\sum_{J\ge0}
\left[
-12H_{J+1}^{(3)}
-{4\over J+2}
\left(
2H_{J+1}^{(2)}
+{4H_{J+1}\over J+2}
\right)
\right]\mathcal \partial_\gamma S_{J,J}(z)
\\
&\qquad
=
\mathcal D
\left[
4z^3
\int_0^1dt\,{t^2f(tz)-f(z)\over 1-t}
\right]=-48\pi^2\log z+{O}(1),
\end{aligned}
\end{equation}
where
\begin{equation}
f(z)
=
{2\log z\over(1-z)^2}\operatorname{Li}_2(1-z)
-{3\over z(1-z)}\operatorname{Li}_3(1-z).
\end{equation}
The first two lines of \eqref{eq:weak_two_loop_spin_sum} therefore contribute
with a $\log z$ as $z\to 0$.

Let us now focus on the second-derivative terms. The large $J$ expansion of the harmonic numbers
\begin{equation}
4\left(H_{J+n+1}^{(2)}-H_n^{(2)}\right)
+8{H_{J+n+1}-H_n\over J+2n+2} \sim {O}(1)+ 8 \frac{\log J}{J}+{O}(1/J) 
\end{equation}
contains a $\log J/J$ term that can source a $\log^2 z$ singularity as we will see around equation \eqref{eq:logjj}.
Again in the regime $J\to\infty$ with $J z = x$ fixed we can replace the sum over $J$ with an integral, and the $\log J/J$ produces the leading two loops collinear divergence
\begin{equation}
    8 \, \sum_{n=0}^{\infty}\int_{J_0 z}dx \,\log \frac{x}{z} \frac{\partial_\gamma^2S_{J,\gamma}\left(\frac{x}{J}\right)|_{\gamma=J+2n}}{x} \simeq 4\log^2 z\sum_{n=0}^\infty \partial_\gamma^2S_{J,\gamma}\left(0\right)|_{\gamma=J+2n} \,,
\end{equation}
where as shown in equation \eqref{eq:S0laJ} we have
\begin{equation}
  \partial_\gamma^2\mathcal S_{J,\gamma}(0)\big|_{\gamma=J}\simeq 8(\pi^2-6) \,,\quad \partial_\gamma^2\mathcal S_{J,\gamma}(0)\big|_{\gamma=J+2n}\simeq24\left(\frac{\pi^2}{2}-3H_n^{(2)}-\frac{(2n+1)(3n^2+3n-1)}{2n^2(n+1)^2}\right)\,.
\end{equation}
The sum over $n$ can be evaluated explicitly and we get 
\begin{equation}\label{eq:weak_phi2_small_z}
{\varphi^{(2)}(z)\over 2K(2K+N)}
\simeq
96\left(3-{\pi^2\over6}\right)\log^2 z
+{O}(\log z)~,
\quad \text{as }z\to0~.
\end{equation}
We have independently checked the result \eqref{eq:weak_phi2_small_z} by calculating the OPE sum \eqref{eq:weak_two_loop_spin_sum} exactly in terms of multiple polylogarithms.\footnote{We evaluated the relevant integrals using \texttt{HyperInt}~\cite{Panzer:2014caa}.
The integrals and the resulting expression are similar in structure to those
in (5.21) of \cite{Chicherin:2023gxt}.
We do not present the explicit result here, since we have not attempted to simplify it fully. We are happy to provide it upon request.}

\section{Strong Coupling}\label{sec:scoupl}
We now study the strong coupling limit $\lambda\to\infty$ at fixed angle $0<z<1$. The finite coupling superblock sum in \eqref{eq:Ffc} is controlled by states whose quantum numbers scale with cylinder masses $J, n, r \sim M_s\sim \sqrt{\lambda}\to \infty$. Its leading strong coupling behavior is 
\begin{equation}\label{eq:cubicl}
\CF_{\text{int}}(\lambda, z) \simeq 
   {1\over 24 \pi^2}  \left(\frac{K(K^2-1)}{12}+2K(N-K)\right) \frac{1}{\lambda \,z^3}- \frac{K(2N-K-1)}{2}~.
\end{equation}
We will refer to the inverse-cubic dependence on $z$ as the ``cubic law". In what follows, we first derive the cubic law from the source-detector decomposition and then show how this result admits a nice semiclassical description.

We further show that when $z \sim \lambda^{-1}$,  $\mathcal{F}_\text{int}$ enters the collinear crossover regime introduced in Figure \ref{fig:intro}. This will be the subject of Subsection \ref{sec:crossover}.

\subsection{Cubic Law}\label{sec:cubic_law}
We begin with the source-detector decomposition of the superblock sum in equation \eqref{eq:Ffc}. Consider first a fixed massive sector $s$. The OPE coefficients \eqref{eq:Cs} and heavy-light dimensions \eqref{eq:gammaHL} show that no finite set of low-lying states can control the large-$M_s$ limit. The dominant contribution comes from a continuum of states such that
\begin{equation}
    J, n, r \sim M_s~.
\end{equation}
We therefore introduce the rescaled variables
\begin{equation}
    J=M_s k \,, \quad \gamma_{J,n,r}^{(s)} = M_s \alpha \,, \quad r+n=M_s \,x \,, \quad J+2+n-r=M_s y \,.
\end{equation}
The large-$M_s$ detector superblock is derived in appendix \ref{app:strong} and it takes the form
\begin{equation}\label{eq:strong_superblock}
\begin{aligned}
    &\mathcal S_{J,\gamma_{J,n,r}^{(s)}}(z)\simeq \frac{\sin^2\left(M_s\pi \frac{k-\alpha}{2}\right)}{\pi^2 M_s^5} \mathcal{D}_z\left(\partial_k^2-\partial_\alpha^2\right)^2 h(\alpha,k,z) \,, \\ &h(\alpha,k,z)=\left(\frac{-4zk}{\alpha^2-k^2}+\frac{2\sqrt{z}\,\text{arctanh}\left(\frac{2k\sqrt{z(\alpha^2-(1-z)k^2)}}{\alpha^2-(1-2z)k^2}\right)}{\sqrt{\alpha^2-(1-z)k^2}} \right) \,.
\end{aligned}
\end{equation}
Some comments on the above expression are in order. First, an individual superblock is suppressed as $M^{-5}_s$. Second, the sine factor which can be rewritten as
\begin{equation}
    \sin^2 \left[\frac{\pi M_s}{2}(k-\alpha)\right]
=
\sin^2 \left[\frac{\pi}{2}(\gamma_{\mathrm{HL}}-J)\right]~,
\end{equation}
becomes rapidly oscillatory as $M_s\to \infty$. Once the discrete sums have been replaced by continuum integrals, these oscillations may be averaged at leading order. The three sums supply a phase-space factor of order $M^3_s$, leaving an overall suppression
\be
M^3_s \times M^{-5}_s = M^{-2}_s \sim \lambda^{-1}~.
\ee
Thus the leading strong coupling effect is the collective effect of parametrically many heavy-light states contributing to the sum. At the same order, the OPE coefficients and dimensions become
\begin{equation}
C_{J,n,r}^{(s)} = \frac{x y}{\sqrt{(x^2+4)(y^2+4)}} \,, \quad \alpha= \sqrt{x^2+4}+\sqrt{y^2+4} \,.
\end{equation}
The sum over $J,n,r$ may now be replaced by integrals. It is convenient to trade $x,y$ for
\begin{equation}
    \alpha= \sqrt{x^2+4}+\sqrt{y^2+4}~, \qquad \delta=\sqrt{x^2+4}-\sqrt{y^2+4}~.
\end{equation}
Note that the integral is $\delta-$independent, therefore one finds
\begin{equation}
\sum_{J,n,r} C_{J,n,r}^{(s)} \longrightarrow \frac{M_s^3}{2} \int_0^\infty dk~k \, \int_{\sqrt{k^2+16}}^\infty d \alpha \sqrt{\frac{\alpha^2-k^2-16}{\alpha^2-k^2}} \,, 
\end{equation}
where the square root in the integrand comes from the integral over $\delta$. Averaging the wildly oscillating sine squared gives
\begin{equation}\label{eq:scint}
    \sum_{J,n,r} C_{J,n,r}^{(s)}  \mathcal S_{J,\gamma_{J,n,r}^{(s)}}(z) \simeq \frac{1}{4\pi^2M_s^2}\mathcal{D}_z\int_0^\infty dk~k \, \int_{\sqrt{k^2+16}}^\infty d \alpha \sqrt{\frac{\alpha^2-k^2-16}{\alpha^2-k^2}}\left(\partial_k^2-\partial_\alpha^2\right)^2h(\alpha,k,z) \,.
\end{equation}
We now evaluate the integrals in \eqref{eq:scint}. It is convenient to introduce two new variables $\rho,w$ such that
\begin{equation}
    \alpha=\frac{\rho}{\sqrt{1-w^2}} \,, \quad k = \frac{\rho \, w}{\sqrt{1-w^2}} \,, \quad \text{with}\,\, \rho \in [4,\infty) \,, \quad w \in [0,1] \,.
\end{equation}
In these variables the integrand takes a simpler form and one has
\begin{equation}
    dk \, d\alpha ~k\sqrt{\frac{\alpha^2-k^2-16}{\alpha^2-k^2}}\mathcal{D}_z\left(\partial_k^2-\partial_\alpha^2\right)^2h(\alpha,k,z) =  d\rho\,d w~\frac{\sqrt{\rho^2-16} }{\rho^{4}} \,  \,\mathcal{K}_4(w,z)  \,.
\end{equation}
Therefore, \eqref{eq:scint} can be rewritten as
\begin{equation}\label{eq:intfactorized}
    \sum_{J,n,r} C_{J,n,r}^{(s)}  \mathcal S_{J,\gamma_{J,n,r}^{(s)}}(z) \simeq \frac{1}{4\pi^2M^2_s}\int_{4}^{\infty}d\rho \frac{\sqrt{\rho^2-16}}{\rho^{4}}\int_{0}^1 dw ~\mathcal{K}_4(w, z)~,
\end{equation}
where
\begin{equation}
\begin{aligned}
    &\mathcal{K}_4(w,z)= \frac{w}{(1-w^2)^\frac{3}{2}} \left(\left((1-w^2)\partial_w\right)^2-9\right)\left(\left((1-w^2)\partial_w\right)^2-1\right) \times \\ &\times \mathcal{D}_z\, 4\sqrt z \left[ \sqrt{\frac{1-w^2}{1-(1-z)w^2}}\,
\text{arcsinh}\left(\frac{\sqrt z\,w}{\sqrt{1-w^2}}\right)-\frac{\sqrt z\,w}{\sqrt{1-w^2}}\right] \,.
\end{aligned}
\end{equation}
The above equation defines a special angular kernel first introduced in \cite{Chen:2024iuv}. 

The $\rho$-integral in \eqref{eq:intfactorized} factorizes and can be evaluated immediately. All the remaining angular dependence is encoded in $\CK_4(w,z)$ and the superblock sum reduces to 
\begin{equation}\label{eq:nophsh}
    \sum_{J,n,r} C_{J,n,r}^{(s)}  S_{J,\gamma_{J,n,r}^{(s)}}(z) \simeq \frac{1}{4\pi^2 M_s^2} \frac{1}{48} \int_0^1 \mathcal{K}_4(w,z) \, dw \,.
\end{equation}

Finally, since the kernel $\mathcal{K}_4(w,z)$ is a total derivative with respect to $w$, the integral in \eqref{eq:nophsh} can be evaluated exactly and we get the simple result
\begin{equation}
\sum_{J,n,r} C_{J,n,r}^{(s)}  S_{J,\gamma_{J,n,r}^{(s)}}(z) \simeq \frac{1}{4\pi^2 M_s^2} \frac{1}{48} \frac{64}{z^3} =\frac{1}{3\pi^2 M_s^2}  \frac{1}{z^3} \,.
\end{equation}
Summing over massive sectors we recover precisely the cubic law presented in equation \ref{eq:cubicl}.

\subsection{Semiclassical Interpretation}\label{sec:semiclassics}

In this section we discuss how the strong coupling behavior found above admits a direct interpretation in the semiclassical description of large-charge states on the cylinder. First, let us recall the dynamical argument that was already used in deriving the canonical HHLL correlator. In the large-charge limit, the path integral in the presence of the two heavy insertions is expected to be dominated by a configuration with a non-trivial scalar profile, see \cite{Brown:2025cbz}. The light operators probe fluctuations around this background, while their backreaction on the leading saddle is suppressed by $1/\Delta_H$. After mapping to the cylinder, the off-diagonal scalar fluctuations corresponding to W-bosons acquire the constant cylinder masses $M_s$ introduced in \eqref{eq:masses}. Since each insertion of the half-BPS light primary $T_2$ is bilinear in the elementary scalar fields, joining two light insertions produces a product of two identical massive propagators. The semiclassical result reviewed around \eqref{eq:resummedHHLL} leads to the main formula
\be\label{eq:Tautots}
\mathcal T_{m,K}(u,v;\lambda)
=
\frac{1}{2v}
\sum_{s=1}^{K}
\nu_s
\left[
t_s(u,v;\lambda)^2-1
\right]
~,
\ee
which we report here again for convenience. Here $t_s$ is the normalized scalar propagator in the $s$-th massive sector. In the previous sections we decomposed this expression into heavy-light superconformal blocks. Here we instead retain its position-space form and perform the detector integrals directly. Since $M_s\sim\sqrt{\lambda}$, the limit $\lambda\to\infty$ becomes a large-mass expansion of the cylinder propagators.

The framework developed by \cite{Cuomo:2025pjp} provides a convenient way of organizing the light-ray calculation. One of their main observations is that the Fourier transform producing the momentum eigenstate \eqref{eq:EEC_generic_definition} can itself be evaluated semiclassically. At leading order in $1/\Delta_H$, the integral over the separation of the heavy insertions localizes at the saddle
\be\label{eq:saddle}
x_*^\mu
=
-2i\Delta_H\frac{q^\mu}{q^2}\equiv -iX^\mu~.
\ee
Equivalently, translating the two heavy insertions so that their midpoint is at the origin, the normalized correlation function becomes
\be\label{eq:saddleFourier}
\frac{
\displaystyle
\int d^4x\,e^{iq\cdot x}
\left\langle
\mathcal O_H(x/2)\,\mathcal{D}_{\mathcal{O}_1}\mathcal{D}_{\mathcal{O}_2}\,
\overline{\mathcal O}_H(-x/2)
\right\rangle
}{
\displaystyle
\int d^4x\,e^{iq\cdot x}
\left\langle
\mathcal O_H(x/2)\overline{\mathcal O}_H(-x/2)
\right\rangle
}
\simeq
\frac{
\left\langle
\mathcal O_H(-iX/2)\,\mathcal{D}_{\mathcal{O}_1}\mathcal{D}_{\mathcal{O}_2}\,
\overline{\mathcal O}_H(iX/2)
\right\rangle
}{
\left\langle
\mathcal O_H(-iX/2)
\overline{\mathcal O}_H(iX/2)
\right\rangle
}~,
\ee
up to relative corrections of order $1/\Delta_H$. Here $\mathcal{D}_{\CO}$ denotes a detector operator obtained by integrating a local operator $\CO$ on a light-ray at null infinity. In the rest frame $q^\mu=(Q,0)$ and $X^\mu=(\frac{2\Delta_H}{Q},0)$, the saddle places $\mathcal O_H$ and $\overline{\mathcal O}_H$ at imaginary times $-iX/2$ and $iX/2$, respectively. Under the conformal map to the cylinder, this pair of insertions prepares the heavy primary state.

We now focus on the case where both
$\mathcal O_1$ and $\mathcal O_2$ are identified with the half-BPS operator $T_2$. The corresponding flat-space detector reads \begin{equation}\label{eq:alt_scalar_detector_flat}
\mathcal D_{T_2}(\vec n)
=\int_{-\infty}^{\infty}du\,
\lim_{r\to\infty}r^2\,
T_2(u+r,r\vec n)\,.
\end{equation}
It is convenient to introduce cylinder coordinates
\begin{equation}\label{eq:alt_ur_map}
u_i=\frac{X_0}{2}\tan\frac{t_i^-}{2},
\qquad
r_i=\frac{X_0}{4}\left(
\tan\frac{t_i^+}{2}-\tan\frac{t_i^-}{2}
\right),
\end{equation}
such that
\begin{equation}
r_i\to\infty\ \Longleftrightarrow\ t_i^+\to\pi
\quad (t_i^-\ \text{fixed})~.
\end{equation}
With these, the detector \eqref{eq:alt_scalar_detector_flat} can be mapped to
\begin{equation}\label{eq:alt_detector_cylinder}
\mathcal D_{T_2}(\vec n_i)
=\frac{X_0}{4}\int_{-\pi}^{\pi}dt_i^-\,
T_2^{\rm cyl}(\pi,t_i^-,\vec n_i)~.
\end{equation}
In what follows, to avoid clutter, we use $T_2^{\rm cyl}(t_i^+,t_i^-,\vec n_i)=T_2(y_i)$ where $y_i=(t^+_i=\pi, t^-_i, \vec{n}_i)$ denotes a point along the $i$-th light-ray contour on the cylinder.\footnote{Here,
$T_2^{\rm cyl}(t_i^+,t_i^-,\vec n_i)
=\Omega(t_i^+,t_i^-)^2
T_2(u_i+r_i,r_i\vec n_i),$ where $\Omega(t_i^+,t_i^-)=\frac{X_0}{4\cos(t_i^+/2)\cos(t_i^-/2)}$ is the Weyl factor. See e.g. \cite{Pappadopulo:2012jk} for the details of the standard plane-to-cylinder map.}
The infinite flat-space detector trajectory becomes the finite interval $-\pi\leq t^-\leq\pi$. The two integrated operators are separated in the interior of their contours, but they collide at the equal endpoints
\be
t_1^-=t_2^-=\pi~,
\qquad
t_1^-=t_2^-=-\pi~.
\ee
Notice that, according to \cite{Cuomo:2025pjp}, since $\Delta + J = 2 < (d+2)/2=3$, the detector \eqref{eq:alt_detector_cylinder} belongs to the endpoint-sensitive class: its equal-endpoint contribution would be UV matching data encoded in the light-ray OPE. In our problem, however, we will use the resummed canonical correlator to provide this matching information explicitly through the massive propagators $t_s$.

The coordinate transformation \eqref{eq:alt_ur_map} is valid for any positive reference length
$X_0$.  As remarked below \eqref{eq:saddleFourier} it
is convenient to choose the conformal frame adapted to the heavy
insertions.  In the rest frame, this amounts to setting
$X_0=2\Delta_H/Q$. All in all,  we get
\begin{equation}\label{eq:alt_cylinder_ratio}
    \begin{aligned}
\left\langle
\mathcal{D}_{T_2}(\vec{n}_1)\mathcal{D}_{T_2}(\vec{n}_2)
\right\rangle_{H}
=
\frac{\Delta_H^2}{4Q^2}
\int_{-\pi}^{\pi}dt_1^-
\int_{-\pi}^{\pi}dt_2^-
~\left\langle H\left|
T_2(y_1)T_2(y_2)
\right|H\right\rangle\left(1+O\left( \Delta_H^{-1}\right)\right)~.
\end{aligned}
\end{equation}

To make the connection between our analysis and the above story more precise, we introduce a general identity involving the scalar-flow operator $\mathcal{D}_{T_2}(\vec{n})$ mentioned above adapted to the $R$-symmetry channel used for the EEC. Conformal symmetry fixes the interaction-dependent two scalar flows correlator to have the form
\begin{equation}\label{eq:sscidentity}
    \left\langle
\mathcal O_H(x)\,
\mathcal D_{T_2}(\vec{n}_1)\mathcal{D}_{T_2}(\vec{n}_2)\,
\overline{\mathcal O}_H(0)
\right\rangle
=
\frac{2}{
(n_1\cdot n_2)(x^2)^{\Delta_H-1}
}
\,G(\chi)~,\qquad \chi \equiv \frac{2(x\cdot n_1)(x\cdot n_2)}{x^2(n_1\cdot n_2)}~.
\end{equation}
The above identity can be conveniently used to evaluate the normalized Fourier transform on the large-$\Delta_H$ saddle $x_*$ as in \eqref{eq:saddleFourier} to obtain
\be\label{eq:SSCWard}
\left\langle
\mathcal{D}_{T_2}(\vec n_1)\mathcal{D}_{T_2}(\vec n_2)
\right\rangle_{H}
=
-\frac{4\Delta_H^2}{
(q\cdot n_1)(q\cdot n_2)
}
\,\chi G(\chi)
\left[1+ O(\Delta_H^{-1})\right].
\ee
In the rest frame $q\cdot n_i=Q$. Moreover, evaluating $\chi$ on the large-$\Delta_H$ saddle $x_*$ gives
\be\label{eq:SSCchiSaddle}
\chi_{*}\equiv\chi\big|_{x=x_*}
=
\frac{
2(q\cdot n_1)(q\cdot n_2)
}{
q^2(n_1\cdot n_2)
}
=
\frac1z~.
\ee
Therefore, comparing \eqref{eq:SSCWard} with
\eqref{eq:alt_cylinder_ratio} we obtain
\be\label{eq:SSCGcylinder}
G(\chi) = -\frac{1}{16\chi} \int_{-\pi}^{\pi} dt^-_1\int_{-\pi}^{\pi} dt^-_2~\left\langle H\left|
T_2(y_1)T_2(y_2)
\right|H\right\rangle +O\left({\Delta_H^{-1}}\right)~.
\ee

This equation will prove very useful for the calculation in the remainder of this section: it expresses the scalar function entering the superconformal Ward identity directly in terms of a two-point function on the cylinder. For the canonical heavy operators $\CO_H$, this cylinder correlator is known semiclassically. Its interaction-dependent part is obtained from the massive propagators appearing in the resummed HHLL correlator. After including the multiplicities of the different massive sectors,  the interaction part of $G$ reads
\be\label{eq:semiclassicsreps}
G_{\rm int}(\chi)
= -\frac{1}{128\pi^4}\sum_s^K\nu_s~G_s(\chi)~,
\ee
where
\be \label{eq:semiEEC2}
G_s(\chi)=\frac{1}{\chi}\int_{-\pi}^{\pi}dt_1^-
\int_{-\pi}^{\pi}dt_2^-\, (t_s^2-1)~.
\ee
Thus, the role played by the Goldstone two-point function in \cite{Cuomo:2025pjp} is played here by the resummed W-boson massive propagators $t_s^2$. This difference is essential: $t_s$ describes the propagation of off-diagonal massive modes on the canonical large-charge background, rather than fluctuations of a conformal-superfluid Goldstone mode.

To reconstruct the full EEC we proceed as follows. In $\mathcal N=4$ SYM, the scalar flow and energy flow belong to the same superconformal multiplet. The superconformal Ward identity (see \cite{Belitsky:2014zha,Korchemsky:2015ssa} for additional details) therefore expresses the energy-flow correlator in terms of the same function $G_{\rm int}(\chi)$,
\begin{equation}\label{eq:EECidentity}
    \left\langle
\mathcal O_H(x)\,
\mathcal {E}(\vec{n}_1)\mathcal{E}(\vec{n}_2)\,
\overline{\mathcal O}_H(0)
\right\rangle_{\rm int}
= \frac{8}{(n_1\cdot n_2)^3(x^2)^{\Delta_H+1}}\mathcal{D}_\chi G_{\rm int}(\chi)~,\qquad \mathcal{D}_\chi \equiv \frac{d^2}{d\chi^2}(1-\chi)^2\chi^2\frac{d^2}{d\chi^2}~.
\end{equation}
Performing the same large-$\Delta_H$ Fourier transform expresses the interaction-dependent EEC as
\be\label{eq:semiEEC}
\CF_{\rm int}(z,\lambda) = \frac{1}{32\pi^2}\sum^K_{s=1}\nu_s \chi^3 \mathcal{D}_\chi~G_s(\chi)\Big|_{\chi=\chi_*}~+O(\Delta^{-1}_H)~.
\ee
Our goal in the remainder of this section thus reduces to evaluate the double light-ray integrals in the limit $M_s\sim\sqrt{\lambda}\to \infty$.

On the cylinder, the massive propagator $t_s$ reads
\be\label{eq:massive_worldline}
t_s = 2(\cosh\sigma -\cos\phi)\sum_{\ell\geq 0}[W_s(2\pi -\phi+2\pi\ell,\sigma)+W_s(\phi+2\pi\ell,\sigma)]~,
\ee
where $W_s(x,\sigma)$ is given in terms of the modified Bessel function as
\be\label{eq:Bessel}
W_s(x,\sigma) = \frac{2 M_sx}{\sin x}\frac{K_1(2 M_s \sqrt{x^2+\sigma^2})}{\sqrt{x^2+\sigma^2}}~.
\ee
The above Euclidean formulas need to be analytically continued to the relevant detector contour by
\begin{equation}
    \sigma\to\frac{i}{2}(t^-_{1}-t^-_{2})~,
\end{equation}
and the angular separation on the cylinder obeys
\begin{equation}
    \cos\phi = \cos\frac{t^-_{1}-t^-_{2}}{2} - \frac{2}{\chi}\cos\frac{t^-_1}{2}\cos\frac{t^-_2}{2}~.
\end{equation}
For generic points in the interior of the integration contour, the argument of $K_1$ is of order $M_s$. In the limit $\lambda\to \infty$ of interest, the propagator is therefore exponentially suppressed. A non-trivial contribution can arise only where the argument of the Bessel function remains of order one. At leading order this occurs near the two same-sign endpoints,\footnote{More precisely, a non-trivial contribution can arise when at least one of the variables $t^-_i$ approaches $\pm \pi$. However, if only one point approaches an endpoint while the other remains at finite distance, the leading contribution is independent of $\chi$ and is therefore annihilated by the superconformal Ward operator in \eqref{eq:semiEEC}. This explains why the first contribution with nontrivial angular dependence comes from the overlap regions in which both points approach the same endpoint,}
\begin{equation}
    (t_1^-,t_2^-)\simeq(\pi,\pi)~,
\qquad
(t_1^-,t_2^-)\simeq(-\pi,-\pi)~.
\end{equation}
Consider first the corner $(\pi,\pi)$ and introduce 
\begin{equation}\label{eq:newvar}
t_i^-
=
\pi-\frac{2\vartheta_i}{M_s}~,
\qquad
\vartheta_i>0~.
\end{equation}
Let us notice that the limit \eqref{eq:newvar} is precisely the macroscopic limit that maps the correlators to the ones on the Coulomb branch of ${\cal N}=4$ SYM. Indeed, keeping the 
physical cylinder-time separation
$R t_{ij}^-/2$ (with $t_{ij}^-=t_i^--t_j^-$ and $R$ the cylinder radius) fixed as $R \to \infty$ corresponds to $R \sim M_s \sim \sqrt{\lambda}$, which is precisely the Coulomb-branch macroscopic limit scaling \cite{Jafferis:2017zna,Caetano:2023zwe,Cuomo:2024fuy}.
Expanding the angular relation gives
\be
\phi^2-\left(\frac{t_{12}^-}{2}\right)^2
=
\frac{4\vartheta_1\vartheta_2}{
\chi M_s^2
}
+
 O(M_s^{-4})~.
\ee
Keeping the leading image of the propagator, one finds
\be\label{eq:tsqqual}
t_s^2
\simeq
16\frac{\vartheta_1\vartheta_2}{\chi}\,
K_1^2 \left(
4\sqrt{\frac{\vartheta_1\vartheta_2}{\chi}}
\right)~.
\ee
The corner at $(-\pi,-\pi)$ gives an identical contribution. It can also be shown that the opposite corners $(\pi,-\pi)$ and $(-\pi,\pi)$ first contribute at order $M_s^{-4}$ and therefore do not affect the leading behavior. We further regulate the endpoint region by restricting $0<\vartheta_i<\vartheta_0$. The matching scale must include the Bessel crossover region $\vartheta_1\vartheta_2\sim\chi$, while remaining within the endpoint expansion. Thus the required hierarchy is
\be
\sqrt{\chi}\ll\vartheta_0\ll M_s~.
\ee
Such a window exists precisely when
\be
zM_s^2\gg1~.
\ee
This is automatically satisfied in the fixed-$z$, large-$\lambda$ limit we are exploring. The above condition also identifies where the expansion breaks down as one approaches the crossover behavior where $zM^2_s$ is kept fixed. This regime will be explored in detail in Section \ref{sec:crossover}.

Let us now consider the double light ray integral in \eqref{eq:semiEEC2}, using the parametrization \eqref{eq:newvar}, this becomes 
\begin{equation}
 G_s(\chi)\simeq   -\frac{4\pi^2}{\chi} +\frac{8}{M^2_s \chi}\int_{0}^{\vartheta_0}d\vartheta_1\int_{0}^{\vartheta_0}d\vartheta_2~\left[16\frac{\vartheta_1\vartheta_2}{\chi}\,
K_1^2 \left(
4\sqrt{\frac{\vartheta_1\vartheta_2}{\chi}}
\right)\right]~.
\end{equation}
The first term is the explicit zero-coupling subtraction. The remaining integral is evaluated by setting
\be
\vartheta_1
=
\frac{\sqrt{\chi}}{4}\rho\,\mathrm{e}^y,
\qquad
\vartheta_2
=
\frac{\sqrt{\chi}}{4}\rho\,\mathrm{e}^{-y},
\qquad
\mathrm{d}\vartheta_1\mathrm{d}\vartheta_2
=
\frac{\chi}{8}\rho\,\mathrm{d}\rho\,\mathrm{d}y~,
\ee
with integration range
\be
|y|
<
\log\frac{4\vartheta_0}{\rho\sqrt{\chi}}~,
\qquad
\rho
<
\frac{4\vartheta_0}{\sqrt{\chi}}~.
\ee
Because of the matching condition, the upper limit of the $\rho$-integral can be extended to infinity up to exponentially small corrections. Using
\be
\int_0^\infty\mathrm{d}\rho\,
\rho^3K_1(\rho)^2
=
\frac{2}{3}~,
\ee
we find
\be
G_s(\chi)
=
-\frac{4\pi^2}{\chi}
+
\frac{2}{3M_s^2}
\log\frac{16\vartheta_0^2}{\chi}
+
\left(
\text{terms annihilated by }\mathcal D_\chi
\right)
+
O(M_s^{-4})~.
\ee
We emphasize that altough the function $G_s(\chi)$ depends explicitly on the matching scale $\vartheta_0$, the EEC \eqref{eq:semiEEC} does not. Indeed, 
\be
\mathcal{D}_\chi \log\frac{16\vartheta^2_0}{\chi} = 2~.
\ee
Putting everything together, we have that
\begin{equation}\label{eq:Fintcubic}
    \CF_{\rm int}(z,\lambda) = -\frac{1}{2}\sum_s^K\nu_s + \frac{1}{24\pi^2 z^3}\sum_s^K\frac{\nu_s}{M_s^2}+O(\lambda^{-2})~.
\end{equation}
Using \eqref{eq:masses}, it follows that there is a non-trivial cubic angular correction to the EEC which we can express as 
\begin{equation}\label{eq:cubicangularlaw}
    \CF_{\rm int}(z,\lambda)|_{\lambda^{-1}} = \frac{K(K^2-1+24(N-K))}{288\pi^2}\frac{1}{\lambda z^3}~.
\end{equation}
The angular power law correction $z^{-3}$ that we observe at strong coupling is produced by a boundary layer of width $M_s^{-1}$ around the two equal light-ray endpoints contributions. This mechanism is qualitatively different from the bulk-contour Goldstone contribution of the conformal-superfluid theory in \cite{Cuomo:2025pjp}. In that theory, the endpoint region is outside the low-energy EFT and is encoded in undetermined local matching data. Here it is calculable because the microscopic large-charge saddle supplies the full massive propagator that resolves this region.

Notice that in \eqref{eq:Fintcubic}, at stricly fixed $z$ the constant term dominates over the power-law behavior for $\lambda\to\infty$. Nevertheless, there is a regime where omitting the constant term is consistent. The cubic term dominates the regular subraction term in the overlap region $$M^{-2}_s\ll z \ll M^{-2/3}_s$$ which is $\lambda^{-1}\ll z \ll \lambda^{-1/3}$ for fixed $K$. This is the regime referred to as the \emph{large-charge jet} in the introduction. As the angle decreases, the EEC sharply grows from $\sim O(1)$ values until reaching the peak $\sim O(\lambda^2)$ at $z \sim \lambda^{-1}$ after which the medium-induced transient effects become important.

Before concluding this section, let us also mention that the light-ray integrals can be evaluated systematically beyond the leading large-mass approximation. We have carried out the explicit calculation through order $M_s^{-4}$. Since the derivation is considerably more involved and does not modify the interpretation, we refer the interested reader to appendix \ref{app:subleading_strong_coupling}. The final result is given by:
\begin{equation}
\label{eq:subleading_strong_coupling_EEC}
\begin{aligned}
\left.
\mathcal F_{\mathrm{int}}(z,\lambda)
\right|_{\lambda^{-2}}
=
\frac1{8}
\sum_{s=1}^{K}\frac{\nu_s}{M_s^4}&
\Bigg[
-\left(\frac1{10\pi^2}+\frac3{5\pi^4}\right)\frac1{z^4}
+\left(\frac1{15\pi^2}+\frac2{5\pi^4}\right)\frac1{z^3}\\
&-\frac{12}{5\pi^4}
\frac{
3(1-z^2)+(1+4z+z^2)\log z
}{
(1-z)^5
}
\Bigg]
~.
\end{aligned}
\end{equation}

\subsection{Collinear Crossover}\label{sec:crossover}

In the collinear limit $z\to 0$ the light ray OPE analysis in Section \ref{sec:lightrayOPE} predicts 
\begin{equation}
\CF_{\rm int}(z,\lambda)\sim{\lambda\over z}~,
\qquad z\to0~.
\end{equation} 
In the strong coupling limit $z\gg 1/\lambda$ on the other hand the analysis in Section \ref{sec:cubic_law} gives 
\begin{equation}
\CF_{\rm int}(z,\lambda)\big|_{\lambda^{-1}}\sim
{1\over\lambda z^3}~,
\qquad \lambda\to\infty~.
\end{equation}
These two asymptotic behaviors of $\CF_{\rm int}$ become comparable for $z\sim\lambda^{-1}$, where both quantities become of order $\lambda^2$.

The emergence of this crossover regime $z \lambda = O(1)$ at strong coupling can be understood as follows. As reviewed in Section  \ref{sec:eecforcanop}, the heavy canonical operators source a semiclassical profile for the scalar fields which Higgs the gauge group. Scalar fluctuations along Higgsed directions acquire a coupling dependent mass $M_s$ on the cylinder, with $M_s\sim \sqrt{\lambda}$. 
As shown in \eqref{eq:newvar} and \eqref{eq:tsqqual}, after Lorentzian continuation, and close to the lightlike configuration relevant for detectors,
the massive propagator is a sum of terms of the form 
\begin{equation}
    W_s(\delta t_1\delta t_2;z) \sim \frac{M_s}{\sqrt{\delta t_1\delta t_2 z}}K_1(2M_s\sqrt{\delta t_1 \delta t_2 z})~,
\end{equation}
up to kinematic prefactors, see for example equation \eqref{eq:massive_worldline}. Here $\delta t_i$ meausres the displacement of the $i$-th insertion from the lightlike separation, so that $\delta t_i = 0$ at the corresponding endpoint of the light-ray contour. Since the Bessel function $K_1(y)$ is exponentially suppressed for $y\gg 1$, the relevant scaling variable is
\be
M_s^2z\sim \lambda z~.
\ee
For $\lambda z \gg 1$, generic points on the detector integration contour are exponentially suppressed, and the integral localizes near the endpoints $\delta t_i=0$ producing the cubic tail discussed in Section \ref{sec:semiclassics}. When $\lambda z \sim O(1)$, this localization breaks down and a finite range of the integration region contributes. In this angular region, labeled \textit{collinear crossover} in Figure \ref{fig:intro}, the separation between detectors is comparable to the characteristic length scales of excitations propagating through the medium. The microscopic structure of the medium becomes relevant and can no longer be neglected.
In this section we zoom in the collinear crossover region by studying the scaling limit
\begin{equation}\label{eq:EEC_double_scaling_xi}
\lambda\to\infty,
\qquad
z\to0,
\qquad
\xi=z\lambda\quad\text{fixed}~.
\end{equation}
As we will see, the resulting function will interpolate between the collinear $1/z$ behavior at small $\xi$ and the inverse-cubic $1/z^3$ behavior of the large-charge jet regime at large $\xi$.

Factoring out the coupling dependence from the effective masses we write
\begin{equation}\label{eq:dimensionless_masses}
M_s=m_s\sqrt{\lambda}~,
\qquad
m_s=2\sin \left(\frac{\pi s}{K}\right)~,\qquad(s=1,\ldots,K-1)~,
\qquad
m_K=1~.
\end{equation}
For each fixed $s$-sector, the natural crossover variable is
\begin{equation}\label{eq:EEC_sector_scaling_xi}
zM_s^2=m_s^2\xi~,
\end{equation}
which is fixed in the limit \eqref{eq:EEC_double_scaling_xi} since $m_s$ is fixed.

The small-$z$ limit is controlled by the large-spin tail of the sum with $x=Jz$ fixed as in Section \ref{sec:weakcoupling}.  In the limit \eqref{eq:EEC_double_scaling_xi}, this requires $J\sim\lambda$ and the dominant contribution comes from the region where we set
\begin{equation}\label{eq:double_scaling_bulk_variables}
Jz=x~,
\qquad
r=\eta J~,
\qquad
0<\eta<1~,
\qquad
n=O(1)~.
\end{equation}
The sums over $J$ and $r$ become
\begin{equation}\label{eq:double_scaling_measure}
\sum_J\sum_r
\longrightarrow
{\lambda^2\over\xi^2}
\int_0^\infty\rd x\,x\int_0^1\rd\eta,
\end{equation}
whereas the sum over $n$ remains finite. 
Expanding the OPE data in this region gives
\begin{equation}\label{eq:double_scaling_OPE_data}
\begin{aligned}
C_{J,n,r}^{(s)}
&=1+ O(\lambda^{-1})~,
\\
\gamma_{J,n,r}^{(s)}
&=J+2\left(n+{m_s^2\xi\over x\eta(1-\eta)}\right)
+ O(\lambda^{-1})~.
\end{aligned}
\end{equation}
In the large-spin limit the detector
superblock becomes (see equation \eqref{eq:S_large_spin_scaling})
\begin{equation}\label{eq:double_scaling_block}
\mathcal S_{J,\gamma_{J,n,r}^{(s)}}(z)
\simeq \mathcal \partial_x^4  x\Phi_{b}(x)|_{b=n+\delta_s}~,
\end{equation}
where
\begin{equation}\label{eq:double_scaling_delta}
\delta_s\equiv{m_s^2\xi\over x\eta(1-\eta)}~.
\end{equation} 
Substituting the above
directly into the superblock sum \eqref{eq:Ffc}, 
gives for the interacting energy correlator defined in \eqref{eq:ECC_final}
\begin{equation}\label{eq:double_scaling_full_EEC}
\CF_{\rm int}\left(z,\lambda\right)
\simeq
{\lambda^2\over8}
\sum_{s=1}^K\nu_s m_s^4\,
\mathfrak F(\xi m_s^2)~,
\end{equation}
with
\begin{equation}\label{eq:double_scaling_integral}
\mathfrak F(\xi m_s^2)=
{1\over(\xi m_s^2)^2}
\sum_{n\geq0}\int_0^\infty\rd x\,x\int_0^1\rd\eta\,
\mathcal \partial_x^4 x\,\Phi_{b}(x)|_{b=n+\delta_s}~,
\end{equation}
where we have used the fact that since in this limit 
\begin{equation}
    \mathcal{F}(z,\lambda) \simeq O(\lambda^2) \,, 
\end{equation}
at leading order one has 
\begin{equation}
    \mathcal{F}_{\text{int}}(z,\lambda) \simeq \frac{1}{8}\mathcal{F}(z,\lambda) \,.
\end{equation}
In other words the plateau region where the constant term dominates in $\mathcal{F}_{\text{int}}(z,\lambda)$ is pushed to infinity since the condition 
\begin{equation}
    z \ll M_s^{-2/3}
\end{equation}
is always satisfied in the collinear crossover region.

The sum over $n$ can be performed analytically using the integral representation of $\Phi_b$ in \eqref{eq:Phi_large_spin_integral}. After a change of variables and evaluation of the remaining $x$- and $\eta$-integrals, one obtains
\begin{equation}\label{eq:double_scaling_final_integral}
\mathfrak F(\xi m_s^2)
={1\over2\pi^2(\xi m_s^2)^2}
\int_0^\infty\rd L\int_0^1\rd\rho\,L
\bigg[
\mathcal H\left(\sqrt{4\,\xi\,m_s^2 Lp}\right)
-\operatorname{Re}\mathcal H\left(
\sqrt{4\,\xi \, m_s^2(L+2\pi\ri)p}\right)
\bigg]~,
\end{equation}
where
\begin{equation}
    p(L,\rho)
=
\frac{
\left(1-\mathrm e^{-\rho L}\right)
\left(1-\mathrm e^{-(1-\rho)L}\right)
}{
1-\mathrm e^{-L}
}~,
\end{equation}
and 
\begin{equation}\label{eq:double_scaling_H}
\mathcal H(a)
={a^4\over2}
\left[
(a^2+12)K_0(a)^2
+(a^2+6)K_1(a)^2
-13aK_0(a)K_1(a)
\right]~.
\end{equation}

\begin{figure}[t]
    \centering
    \begin{subfigure}[b]{0.49\textwidth}
        \centering
        \includegraphics[width=\linewidth]{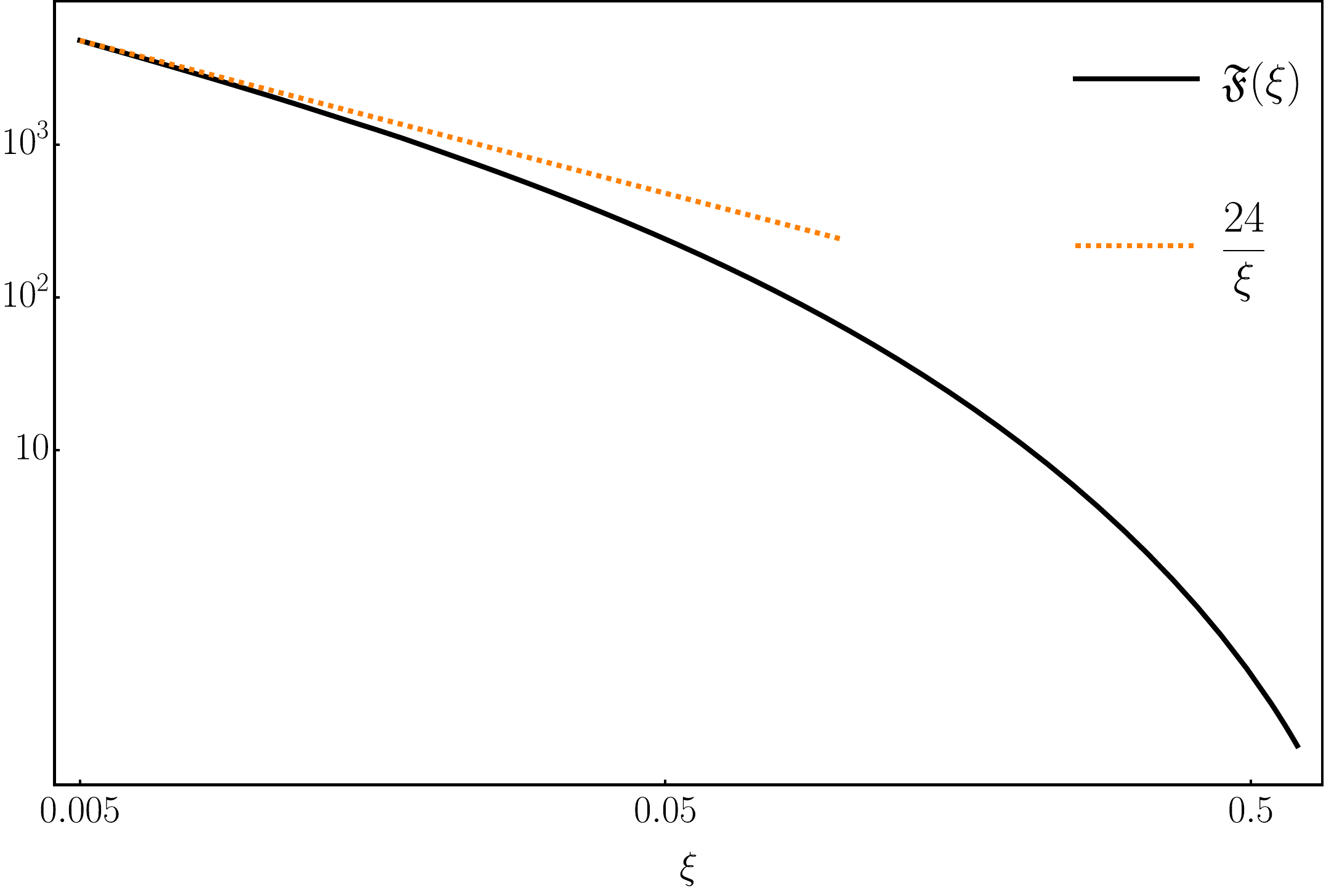}
    \end{subfigure}
    \hfill
    \begin{subfigure}[b]{0.49\textwidth}
        \centering
        \includegraphics[width=\linewidth]{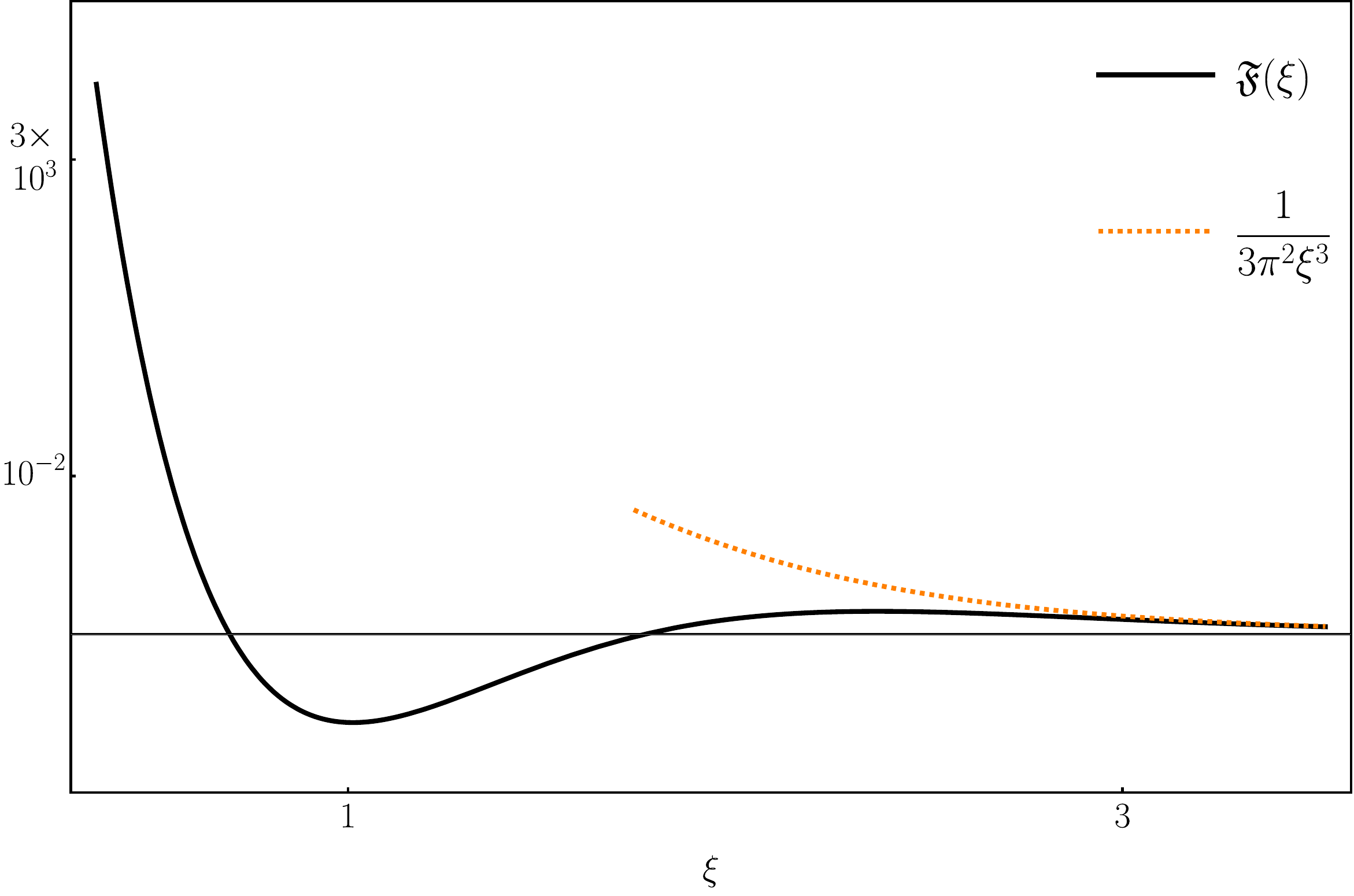}
    \end{subfigure}

    \caption{\textbf{Left panel:} $\mathfrak{F}(\xi)$ in log-log scale at small argument. When $\xi\ll1$ we approach the light-ray OPE prediction. \textbf{Right panel:} $\mathfrak{F}(\xi)$ in linear scale for $\xi=O(1)$. The function develops a global minimum at $\xi_{\text{min}}\simeq 1.0074$ where $\mathfrak{F}(\xi_{\text{min}})=-5.59\times 10^{-3}$, and a local maximum at $\xi_{\text{max}}=2.118$ where $\mathfrak{F}(\xi_{\text{max}})=1.44\times 10^{-3}$. At larger $\xi$ we approach the strong coupling prediction.}
    \label{fig:double_scaling_function}
\end{figure}

The asymptotic behavior of $\mathfrak F(\xi m_s^2)$ for small and large values of $\xi$ is given by
\begin{equation}\label{eq:double_scaling_asymptotics}
\begin{aligned}
\mathfrak F(\xi m_s^2)
&={24\over\xi \, m_s^2}
+16(18-\pi^2)\log^2(m_s^2\xi)
+ O\left(\log(m_s^2\xi)\right),
&&\xi\to0,
\\
\mathfrak F(\xi m_s^2 )
&={1\over\pi^2}
\left[
{1\over3(\xi m_s^2)^3}
-\frac{6+\pi^2}{10\pi^2}{1\over (\xi m_s^2)^4}
+{18(90+\pi^4)\over175\pi^4}{\log(m_s^2\xi)\over (\xi m_s^2)^5}
\right]
+ O \left((\xi m_s^2)^{-5}\right),
&&\xi\to\infty.
\end{aligned}
\end{equation}
The derivation of both limits is given in
appendix \ref{app:double_scaling_asymptotics}.
Notice that the small-$\xi$ limit reproduces the light ray OPE prediction in \eqref{eq:F_col_exp} and the two-loop double logarithm in \eqref{eq:weak_phi2_small_z}, consistently with the fact that these coefficients are exact at one and two loops, respectively. At large $\xi$ we instead match the $1/z^3$ behavior observed in the strong-coupling limit at fixed $z$. Thus \eqref{eq:double_scaling_full_EEC} provides the desired
interpolation between the two asymptotic regimes. When $\xi=O(1)$ on the other hand, the cubic growth saturates to a bump followed to a pronounced dip, as shown in Figure \ref{fig:double_scaling_function}.

For $N=K=3$
\eqref{eq:double_scaling_full_EEC} reduces to
\begin{equation}
\frac{\lambda^{2}}{8}
\sum_{s=1}^{3}\nu_s m_s^{4}\,\mathfrak F(\xi m_s^{2})
=
\frac{27\lambda^{2}}{4}\,\mathfrak F(3\xi).
\end{equation}
hence we have the same behavior as in Figure \ref{fig:double_scaling_function}.

\begin{figure}[t]
\centering
\includegraphics[width=0.8\textwidth]{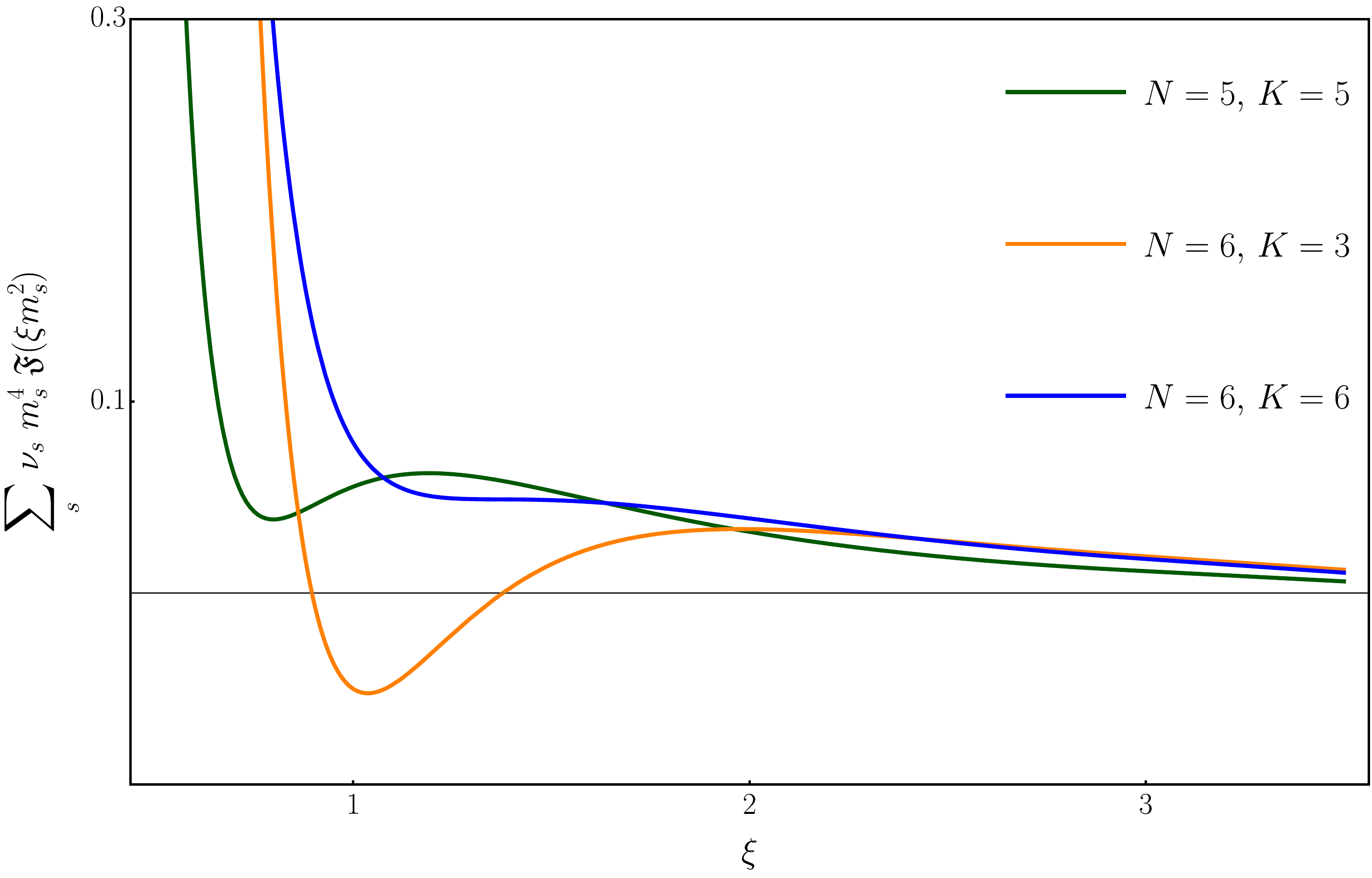}
\caption{Interacting EEC in the collinear crossover regime for different values of $N,K$. The dip goes from being a global minimum to a local one and finally disappears altogether respectively in the \textcolor{orange}{orange}, \textcolor{ForestGreen}{green} and \textcolor{blue}{blue} line.}
\label{fig:double_scaling_full_EEC_N7}
\end{figure}

  For $N>3$ instead the sum over $s$-sectors in
\eqref{eq:double_scaling_full_EEC} produces a richer pattern.  Depending on
$N$ and $K$, the minimum inherited from the universal function may be only
local or disappear altogether due to interference between different mass sectors, as illustrated in Figure \ref{fig:double_scaling_full_EEC_N7}. We will see that this analysis correctly captures the behavior we will encounter at finite coupling, see Section \ref{sec:num_an}.

\section{Finite Coupling}\label{sec:finitecoupling}
In this section we study the EEC at finite coupling $\lambda$ directly from the superblock sum \eqref{eq:Ffc}. We first analyze the the $z\to0$ limit at fixed $\lambda$, showing that it is controlled by the large-spin tail of the sum and agrees with the prediction of the light-ray OPE. We then explore the finite-$z$ region numerically by truncating the sums at $(J_{\text{max}},n_{\text{max}})$. Since the small-$z$ behavior is dominated by large spin, we expect the numerical result to remain reliable down to $z \sim 1/J_{\text{max}}$.

\subsection{Collinear Limit}\label{sec:smallz}
As reviewed in Section \ref{sec:lightrayOPE}, the light-ray OPE predicts a singular behavior of the EEC in the collinear limit $z\to 0$. The superblock $\mathcal{S}_{J,\gamma_{J,n,r}^{(s)}}(z)$ is a polynomial in $z$ which is finite at $z=0$ for any finite value of $J,n,r$. A small-angle singularity can therefore arise only after summing over infinitely many states. The sum over $n$ does not lead to such a singularity. In fact, at large $n$ one has $C_{J,n,r}^{(s)}\sim 1$ and $\gamma_{J,n,r}^{(s)} \sim 2n$, and since at fixed $J$ one can show that
\begin{equation}
\mathcal{S}_{J,2n}(z) =\frac{96 (J+1) P_2(1-2 z) \sin ^2\left(\frac{1}{2} \pi  (2 n-J)\right)}{\pi ^2  n^6}+{O}\left({1\over n^7}\right)\,,
\end{equation}
the infinite $n-$sum is absolutely convergent at $z=0$. 

The small-$z$ singularity is therefore entirely generated by the large-spin tail. The sum over spins in fact does not converge at $z=0$ and accounts for the light-ray OPE singularity. More precisely, the scaling limit that describes the OPE regime is given by $J \sim 1/z$. To argue that this is in fact the case let us consider the simplest case in which $\gamma_{J,n,r}^{(s)}=J$. In this case the detector superblock can be evaluated exactly to (see equation \eqref{eq:1LSS}):~\footnote{As anticipated in Section \ref{sec:positivityprop}, the large-spin tail of the superblock sum remains regular as $z\to1$. This is consistent with the contact terms prescription adopted there.} 
\begin{equation}
    \mathcal{S}_{J,J}(z) = \mathcal{D} \left[z^3(1-z)^J \right] \,.
\end{equation}
As $z\to 0$
\begin{equation}
    (1-z)^J =e^{J \log (1-z)} \simeq e^{-Jz} \,,
\end{equation}
therefore if $J z \gg 1$ the whole superblock is exponentially suppressed and it cannot generate a singularity. In the scaling limit $J z=x$ fixed as $J\to\infty$ the detector superblock has been computed in appendix \ref{app:largesapp} and it reads
\begin{equation}
\mathcal{S}_{J,J+2\beta}(z)|_{2\beta=n-r+\sqrt{4M_s^2+(n+r)^2}}  \simeq \partial_x^4 \left(x \Phi_\beta(x) \right) |_{2\beta=n-r+\sqrt{4M_s^2+(n+r)^2}}~,
\end{equation}
Here $\Phi_\beta(x)$ is the 
large-spin, small-angle behavior of the detector superblock at fixed
$x=Jz$ and $\beta=(\gamma_{\mathrm{HL}}-J)/2$. Its integral
representation is given in \eqref{eq:Phi_large_spin_integral};
the derivatives with respect to $x$ are taken at fixed $\beta$. At large $J$
\begin{equation}
    \gamma_{J,n,r}^{(s)} \simeq J+ n-r+\sqrt{4M_s^2+(n+r)^2} \,,
\end{equation}
therefore, the leading contribution in the collinear limit takes the form
\begin{equation}
\frac{1}{z} \int_0^\infty dx \sum_{r=1}^{J+1} \sum_{n=0}^\infty C_{J,n,r}^{(s)} \partial_x^4 \left(x \Phi_\beta(x) \right) |_{2\beta=n-r+\sqrt{4M_s^2+(n+r)^2}} ~.
\end{equation}
Since the superblock integrates to zero in this scaling regime (see \eqref{eq:x0int})
\begin{equation}
    \int_0^\infty dx \, \partial_x^4 \left(x \Phi_\beta(x) \right) =0 \,,
\end{equation}
a nonzero contribution to the leading collinear singularity requires
residual $J$-dependence from the sum over $r$. In particular, we must
isolate terms whose sum over $r$ diverges as $J\to\infty$. We therefore
examine the large-$r$ behavior, for which
\begin{equation}
    \gamma_{J,n,r}^{(s)} \simeq J+2 n+\frac{2M_s^2}{r} \,.
\end{equation}
Expanding the detector superblock then gives (see \eqref{eq:xderi} and \eqref{eq:no0n})
\begin{equation}
\mathcal{S}_{J,J+2\beta}(z)|_{2\beta=n-r+\sqrt{4M_s^2+(n+r)^2}} \simeq \delta_{n0} \partial_x^4 \left(x^4 e^{-x} \right)+\delta_{n0}\frac{2 M_s^2}{r} \partial_x^4 \left(x^4 \int_0^1 dt\, \frac{t^2e^{-tx}-e^{-x}}{1-t}\right) \,.
\end{equation}
The term proportional to $1/r$, arising from the first derivative, generates a
logarithmic divergence upon summation over $r$ and thus yields a
nonzero leading contribution. By contrast, the OPE coefficients
behave as
\begin{equation}
    C_{J,n,r}^{(s)} \simeq 1-\frac{2M_s^2}{r^2} \,,
\end{equation}
which is summable as $r\to \infty$ and will therefore not contribute at this order. Combining these results, we obtain
\begin{equation}
\frac{2M_s^2}{z}\int_0^\infty dx \, \log J \, \partial_x^4 \left(x^4 \int_0^1 dt\, \frac{t^2e^{-tx}-e^{-x}}{1-t}\right) = \frac{12 M_s^2}{z} \,.
\end{equation}
This is not yet the full leading contribution. The dimensions $\gamma_{J,n,r}^{(s)}$ are invariant under
$r\leftrightarrow J+2-r$, so the opposite end of the $r$-sum gives an identical contribution. Including both contributions and summing over the massive sectors, we obtain
\begin{equation}\label{eq:1zexm}
    \mathcal{F}(z,\lambda) \simeq \sum_s \nu_s \frac{24M_s^2}{z} = \frac{48 K N \lambda}{z} \,,
\end{equation}
in perfect agreement with the light-ray OPE prediction \eqref{eq:F_col_exp} and \eqref{eq:ECC_final}. 
We remark that subleading corrections will start with $J^{-1}=z/x$, therefore subleading singularities in the collinear limit can be at most logarithmic. In other words, the leading collinear singularity in \eqref{eq:1zexm} is one loop exact. 

Let us briefly sketch how to include subleading corrections to the above discussion. Expanding to the next order in $1/J$ we get
\begin{equation}
    \gamma_{J,n,r}^{(s)} \simeq J+n-r+\sqrt{4M_s^2+(n+r)^2}+\frac{2M_s^2}{J} \,.
\end{equation}
Accordingly
\begin{equation}
    \mathcal{S}_{J,\gamma_{J,n,r}^{(s)}}(z)\simeq \partial_x^4 \left(x \Phi_\beta(x)\right)+\frac{M_s^2}{J} \partial_\beta \partial_x^4 \left(x \Phi_\beta(x)\right) \,, \quad 2\beta=n-r+\sqrt{4M_s^2+(n+r)^2} \,.
\end{equation}
Expanding again in the $r\to\infty$ corner the first derivative term contributes with 
\begin{equation}\label{eq:logjj}
    \sum_r\frac{M_s^2}{J} \frac{M_s^2}{r} \partial_\beta^2 \partial_x^4 \left(x \Phi_\beta(x)\right)|_{\beta=n} \simeq \frac{M_s^4 \log J}{J}\partial_\beta^2 \partial_x^4 \left(x \Phi_\beta(x)\right) |_{\beta=n} \,.
\end{equation}
Since the superblock and its derivatives approach a nonvanishing constant at $z=0$ the $1/J=z/x$ term will introduce a divergence and the integral over $x$ must be cut off at $x=x_0=z J_0$. Integrating over $x$ will produce a $\log^2z$ singularity
\begin{equation}\label{eq:2llogex}
2M_s^4\int_{J_0 z}^\infty dx \log \frac{x}{z}\frac{\partial_\beta^2 \partial_x^4 \left(x \Phi_\beta(x)\right) |_{\beta=n}}{x} \simeq  M_s^4 \partial_\beta^2 \partial_x^4 \left(x \Phi_\beta(x)\right) |_{\beta=n,x=0} \log^2 z \,.
\end{equation}
The precise value of $J_0$ is irrelevant as it will only contribute to $O(1)$ terms. Expanding at subleading orders in $J$ will only improve convergence of the $r$ sums. As a result, the coefficient of $\log^2z$ computed in Section \ref{sec:twoloopss} is two-loop exact.

Finally, we emphasize that $\log z$ singularities will appear every time a $1/J$ term arises after summing over $r$. Such terms will then appear at every loop order and get contributions from three-point functions as well.

\subsection{Numerical Analysis}\label{sec:num_an}
\begin{figure}[t]
    \centering
    \begin{subfigure}[b]{0.49\textwidth}
        \centering
        \includegraphics[width=\linewidth]{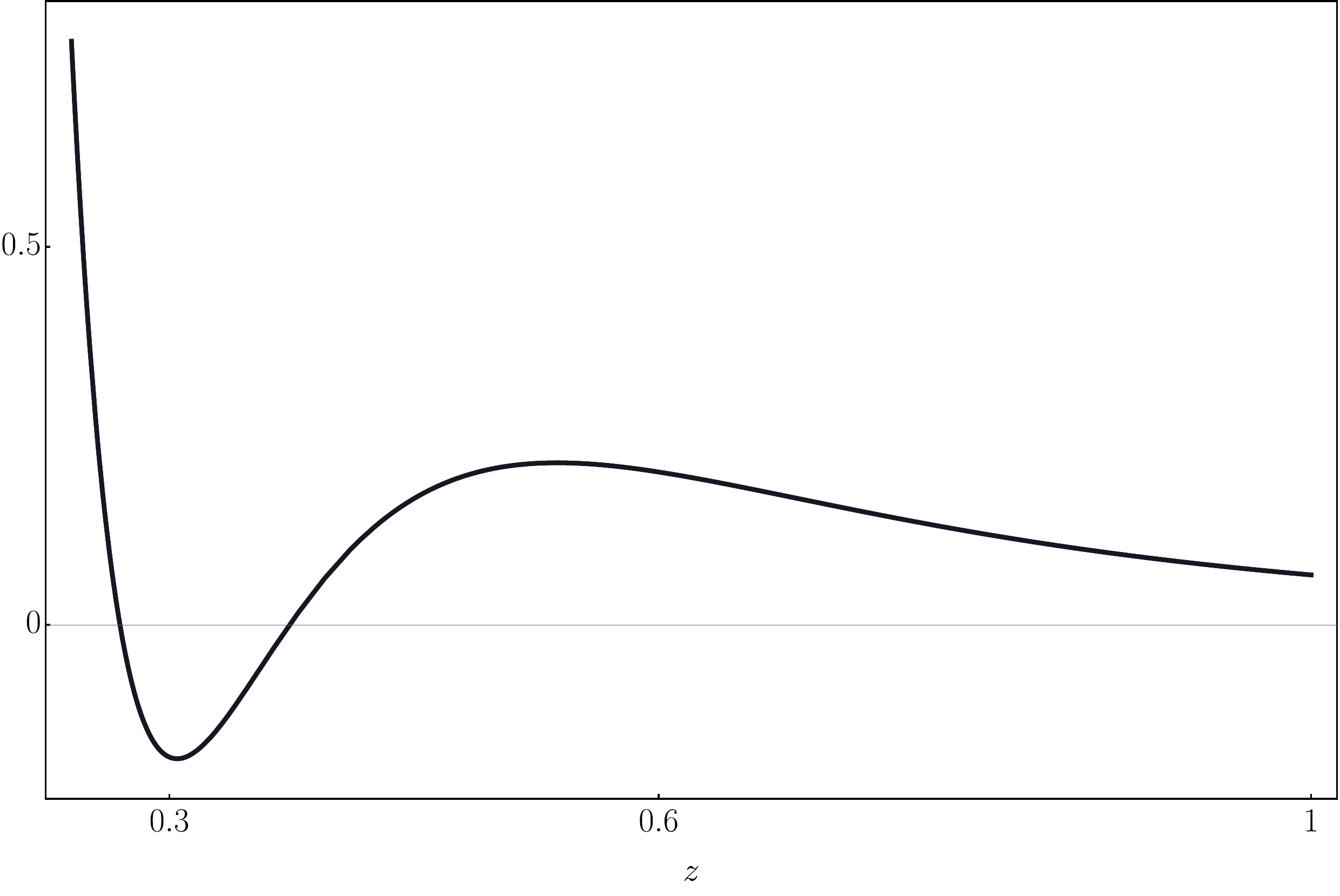}
    \end{subfigure}
    \hfill
    \begin{subfigure}[b]{0.49\textwidth}
        \centering
        \includegraphics[width=\linewidth]{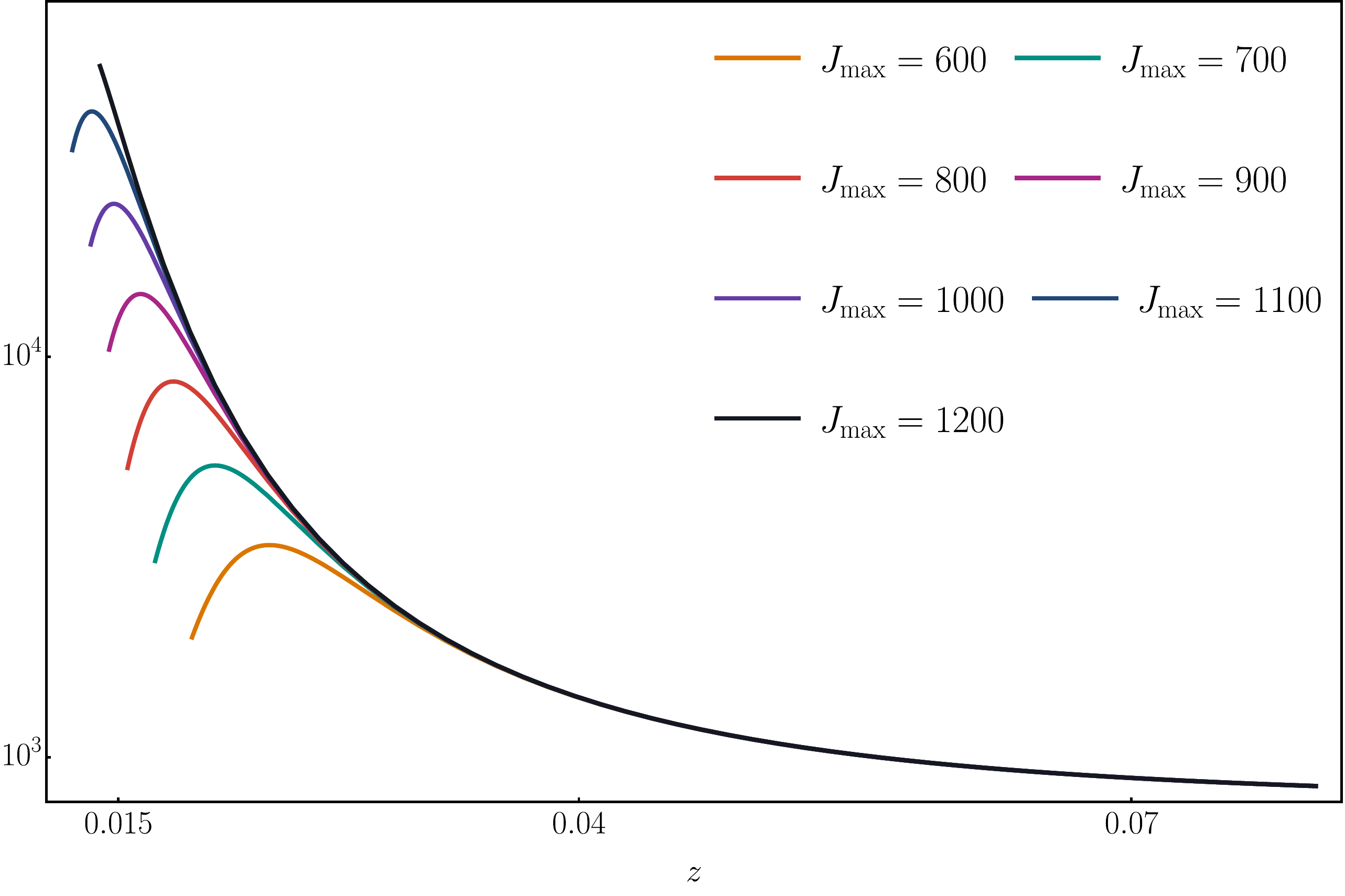}
    \end{subfigure}
    \caption{$\mathcal{F}(z,\lambda=1)$ for $N=K=3$. \textbf{Left panel:} When $z\in[0.24,1]$ curves with $J_{\text{max}}=n_{\text{max}}\in [600,1200]$ sit on top of each other. \textbf{Right panel:} Convergence is much worse close to $z=0$.}
    \label{fig:lambda1z}
\end{figure}

In this section we perform a numerical study of $\mathcal{F}(z,\lambda)$. For simplicity, we set $N=K=3$ so that the sum over $s$ runs over $s=1,2$, and
\begin{equation}
    \nu_1=\nu_2=3~, \qquad M_1^2=M_2^2=3\lambda \,.
\end{equation}
After introducing cutoffs at $(J_{\text{max}},n_{\text{max}})$, $\mathcal{F}(z,\lambda)$ becomes a finite-degree polynomial in  $z$. We compute this polynomial coefficient by coefficient. For each fixed $J$, we first sum over all $n<n_{\max}$ to obtain the corresponding polynomial contribution. Since the different spin sectors are independent, these computations can be performed in parallel. The full truncated result is then reconstructed by summing the polynomial coefficients over $J<J_{\max}$. We report the final result for $\lambda=1$ in  Figure \ref{fig:lambda1z}.
The numerical analysis converges well in the region $z\gtrsim 1/J_{\text{max}}$, while convergence is much slower for $z\lesssim 1/J_{\text{max}}$. 

\begin{figure}[t]
\centering 
\includegraphics[width=0.8\linewidth]{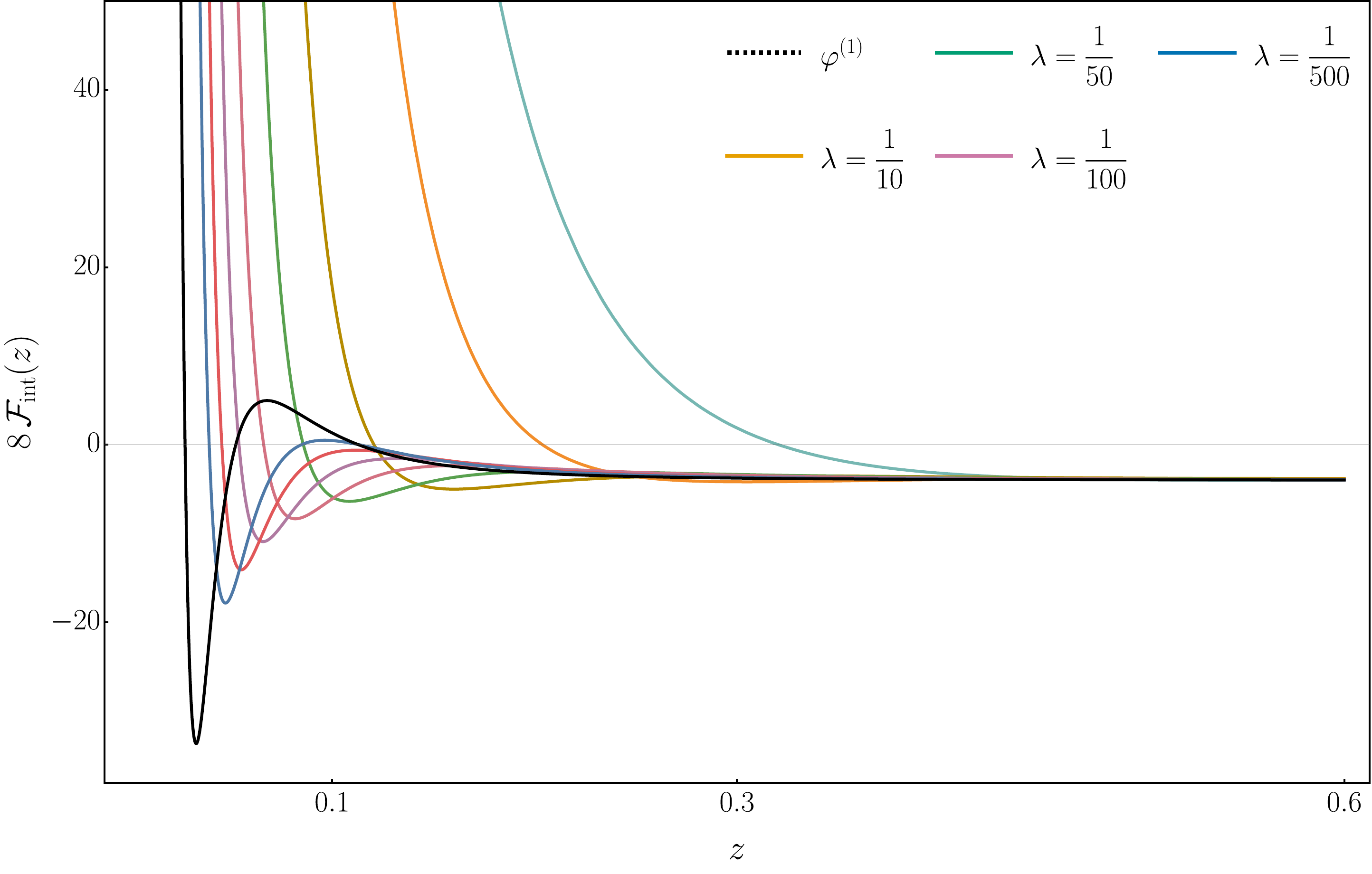}
\caption{$8 \mathcal{F}_{\text{int}}(z,\lambda)$ as a function of $z$ for different values of $\lambda$.}
\label{fig:manylambda}
\end{figure}

The numerical analysis uncovers the following qualitative features: as a function of $z$, $\mathcal{F}(z,\lambda)$ grows to a local maximum, then decreases and becomes negative before rising sharply again. To illustrate how this behavior changes as one varies the coupling $\lambda$, we plot $\mathcal{F}(z,\lambda)$ for different values of $\lambda$ in Figure \ref{fig:manylambda}. The interacting part of the EEC goes from being featureless at weak coupling to develop sharp extrema at strong coupling. Numerically one finds as $\lambda\to\infty$ 
\begin{equation}
\begin{aligned}
    &z_{\text{min}}\sim {1\over 3 \lambda}, ~  z_{\text{max}} \sim {2\over 3\lambda} \,, \\
    &\mathcal{F}(z_{\text{min}},\lambda) \simeq -0.3 \, \lambda^2  \,, \\
    &\mathcal{F}(z_{\text{max}},\lambda) \simeq 0.08 \lambda^2 \,,
\end{aligned}
\end{equation}
consistently with what we found in Section \ref{sec:crossover}.

\begin{figure}[t]
\centering 
\includegraphics[width=0.7\linewidth]{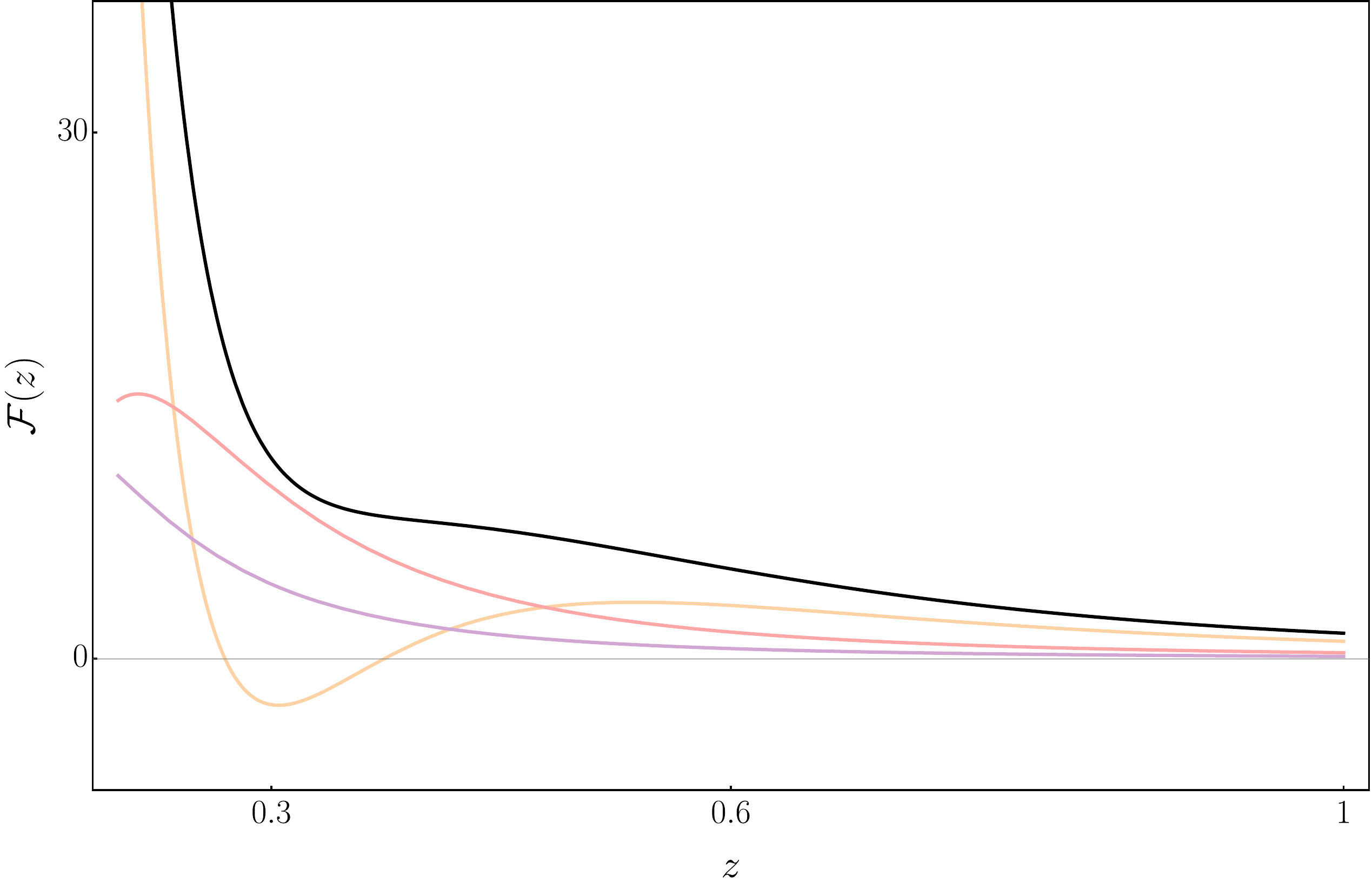}
\caption{$\mathcal{F}(z,\lambda)$ for $N=K=6$ and $\lambda=3$ as a function of $z$. \textcolor{Peach}{Orange}, \textcolor{Maroon}{red} and \textcolor{Thistle}{purple} are respectively the first, second and third summands in equation \eqref{eq:tosu6}. Adding them up reconstructs the black line. 
Since extrema of the summands occur in different places, features in the resulting EEC are destroyed by interference.}
\label{fig:su6}
\end{figure}

We conclude this section by discussing how $\mathcal{F}(z,\lambda)$ generalizes to $K,N>3$. 
In this case nontrivial cancellations can occur among the
$s$-sectors in \eqref{eq:Ffc}. Each individual term in the $s$-sum has a global minimum and a local maximum at
finite $z$, but these features can be smeared out by the sum as discussed in Section \ref{sec:crossover}.  Depending on $N$ and $K$, the minimum can
be lifted and survive only as a positive local minimum
or both the minimum and the maximum can disappear altogether.  The latter
behavior occurs for a set of canonical operators
starting at $N=6$.  Let us show
this explicitly for $N=K=6$.  Here $s$ runs from $1,\dots ,5$ and
\begin{equation}
    M_1^2=M_5^2=\lambda \,, \quad M_2^2=M_4^2=3\lambda \,, \quad M_3^2=4\lambda \,.
\end{equation}
Using the multiplicities in \eqref{eq:masses}, one obtains 
\begin{equation}\label{eq:tosu6}
    \mathcal{F}^{N=K=6}(z,\lambda)= 2\mathcal{F}^{N=K=3}\left(z,\frac{\lambda}{3}\right)+2\mathcal{F}^{N=K=3}\left(z,\lambda\right)+\mathcal{F}^{N=K=3}\left(z,\frac{4\lambda}{3}\right) \,.
\end{equation}
As shown in  Figure \ref{fig:su6}, the extrema of the individual terms disappear from the full EEC.

\subsection{Legendre Moments at Finite Coupling}
We now study the angular moments of the interaction-dependent EEC
\begin{equation}
    a^{\rm int}_\ell(\lambda) = \int_{0}^1dz P_\ell(1-2z)\CF_{\rm int}(z,\lambda)~,
\end{equation}
with $\CF_{\rm int}(z,\lambda)$ defined in \eqref{eq:ECC_final}. The homogeneous sum rules imply $a^{\rm int}_0=a^{\rm int}_1=0$. As discussed in the positivity analysis of Section \ref{sec:positivityprop}, the interaction moments need not be positive: positivity only constrains the complete moments $a_\ell=a_\ell^{\mathrm{free}}+{1\over \Delta_H^2}a_\ell^{\mathrm{int}}$.

\begin{figure}[t]
    \centering
    \includegraphics[height=0.37\textwidth]
        {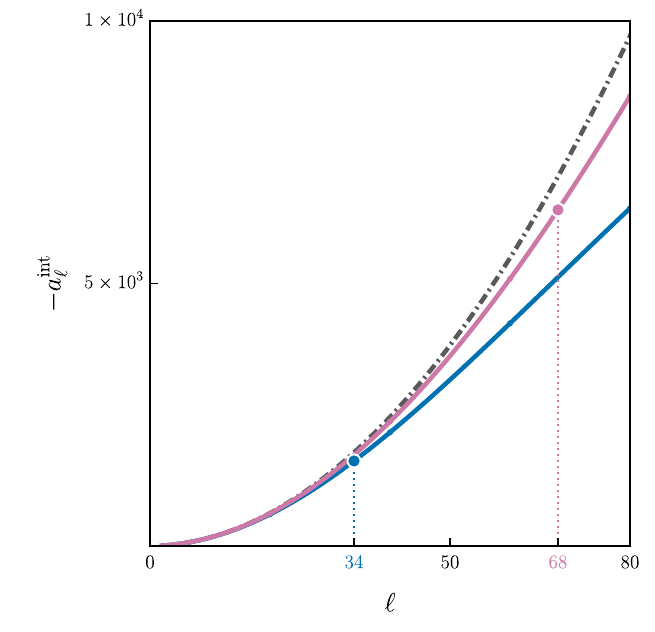}
    \hfill
    \includegraphics[height=0.37\textwidth]
        {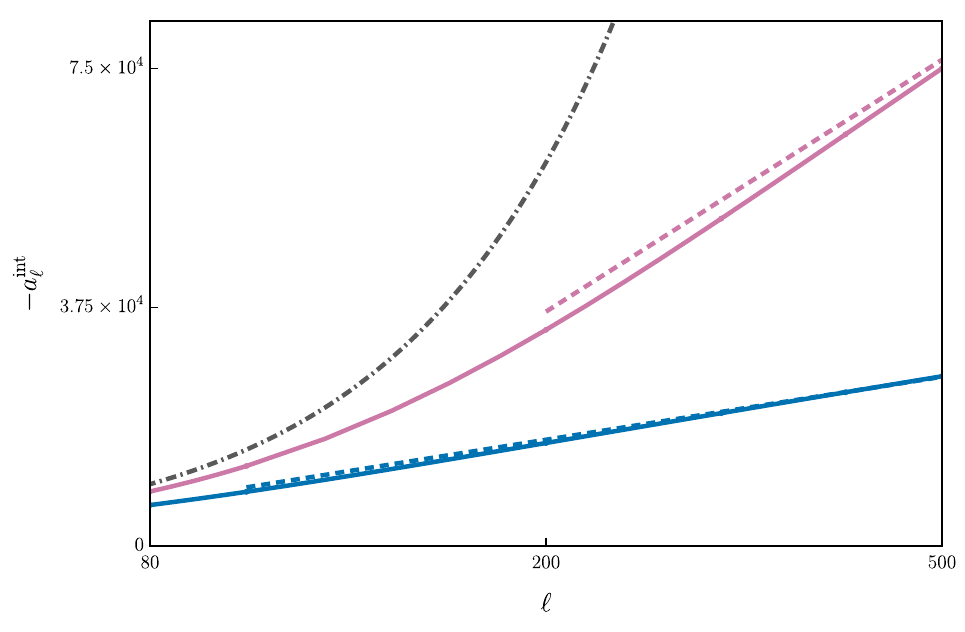}
    \caption{Legendre moments $- a^{\rm int}_\ell$, for $\lambda=100$ (\textcolor{blue}{blue}) and
$\lambda=400$ (\textcolor{purple}{purple}). \textbf{Left panel:} linear horizontal axis.
\textbf{Right panel:} logarithmic horizontal axis. The gray dashed
curve in both panels shows the quadratic behavior $\frac{3}2(\ell-1)(\ell+2)$.
The dashed blue and purple curves show the large-$\ell$ logarithmic
behavior $108\lambda\log\ell+C_\lambda$, with only the additive
constants $C_\lambda$ fitted to the numerical data.
The markers at $\textcolor{blue}{\ell_*}\simeq34$ and $\textcolor{purple}{\ell_*}\simeq68$ indicate an
approximately $10\%$ deviation from the quadratic behavior for
$\lambda=100$ and $\lambda=400$, respectively, consistent with
the expected change of behavior at $\ell_*\sim 3.4\sqrt{\lambda}$.}
    \label{fig:legendre-moments}
\end{figure}

For concreteness, here we continue to keep $K=N=3$. The two nonzero massive sectors are identical. Combining their contributions, the source-detector representation gives
\be
a_\ell^{\mathrm{int}}(\lambda)
=
\frac34
\left[
\sum_{J,n,r}
C_{J,n,r}\,
a_{\ell,J}(\gamma_{J,n,r})
-
2(\ell-1)(\ell+2)
\right],
\qquad \ell\geq2
~.
\ee
Here the sector label has been suppressed. The subtraction includes the endpoint distributions of the zero-coupling superblock sum. To evaluate numerically the above expression at finite coupling, we used and compared two different strategies. The first one, as done in the previous subsection, is to consider a truncation of the OPE sum up to a given $(J_{\rm max},n_{\rm max})$. The second more convenient strategy is to use the direct evaluation of the EEC on the cylinder as discussed at length in Section \ref{sec:semiclassics}. This leads to the formula:
\be\label{eq:aintjacobi}
\begin{aligned}
a_\ell^{\mathrm{int}}(\lambda)
={}&
\frac34(\ell-1)\ell(\ell+1)(\ell+2)
\int_0^1 dz\,z^2P_{\ell-2}^{(4,0)}(1-2z)
\left[
\frac1{4\pi^2}
\int_{-\pi}^{\pi}dt_1^-
\int_{-\pi}^{\pi}dt_2^-\,(t_s^2
-1)
\right]
~,
\end{aligned}
\ee
where $P_{\ell-2}^{(4,0)}$ is a Jacobi polynomial and either of the identical massive sectors may be used.\footnote{Integrating by parts moves the four derivatives in $\mathcal{D}$ from formula \eqref{eq:semiEEC} onto the Legendre polynomial. The resulting polynomial is evaluated using the standard differentiation identities for the Jacobi polynomials:
\be
\frac{d^2}{dz^2}
\left[(1-z)^2\frac{d^2}{dz^2}P_\ell(1-2z)\right]
=
(\ell+1)(\ell+2)\frac{d^2}{dz^2}
\left[(1-z)^2P_{\ell-2}^{(2,2)}(1-2z)\right]
=
(\ell-1)\ell(\ell+1)(\ell+2)
P_{\ell-2}^{(4,0)}(1-2z)~.
\ee}
The large-$\ell$ behavior of the moments at fixed finite coupling is controlled by the collinear limit of Section \ref{sec:smallz}. For $K=N=3$, the coefficient of the leading $1/z$ singularity in $\CF_{\rm int}$ is $54\lambda$. Using the exact overlap
\begin{equation}
\int_0^1 dz\,
P_\ell(1-2z)\left[\frac1z\right]_+
=
\int_0^1 dz\,
\frac{P_\ell(1-2z)-1}{z}
=
-2(\psi(\ell+1)+\gamma_E)~,
\end{equation}
where $\psi$ is the digamma function and $\gamma_E$ the Euler gamma constant. At large $\ell$ and fixed $\lambda$, we therefore obtain
\begin{equation}
    a_\ell ^{\rm int}(\lambda) = -108\lambda\log\ell + O(1)~,\quad \ell\to\infty~.
\end{equation}
For example, at $\lambda=1$, numerically evaluating the moments up to $\ell_{\rm max}=512$ supports this prediction and gives the large-$\ell$ estimate
\be
a^{\rm int}_\ell(1)
\simeq -108\log\ell+154.78~.
\ee
Here the additive constant is estimated with the logarithmic coefficient fixed to its analytic value.  The logarithmic growth at fixed coupling should be distinguished from the strong-coupling limit at fixed angular momentum. For each fixed $\ell$, one has
\be
\lim_{\lambda\to\infty}
a_\ell^{\mathrm{int}}(\lambda)
=
-\frac32(\ell-1)(\ell+2)
~.
\ee
At large but finite coupling, Legendre moments with small $\ell$ can therefore lie close to this quadratic expression even though the eventual large-$\ell$ growth is logarithmic. Comparing the angular resolution $z\sim\ell^{-2}$ with the crossover scale $z\sim\lambda^{-1}$ suggests that the change of behavior is associated parametrically with $\ell\sim\sqrt{\lambda}$. We have tested this phenomenon by evaluating $a^{\rm int}$ for different values of $\lambda$, see Figure \ref{fig:legendre-moments}.

The strong-coupling expansion obtained in Section \ref{sec:semiclassics} describes the EEC at fixed nonzero angular separation. The expansion breaks down near $z=0$, so we cannot integrate its terms separately to compute the Legendre moments. In particular, the terms $z^{-3}$ appearing in \eqref{eq:cubicl} and $z^{-4}$ \eqref{eq:subleading_strong_coupling_EEC} have divergent integrals. They must be defined as distributions, including contact terms at $z=0$. These contact terms are needed to satisfy the homogeneous sum rules and obtain the correct strong-coupling moments.

Using the formula \eqref{eq:aintjacobi} we introduced above, we can obtain the following strong-coupling expansion for the first three non-trivial moments up to order $1/\lambda^2$:
\be
\begin{aligned}
a_2^{\mathrm{int}}(\lambda)
={}&
-6+\frac{L_\lambda-1/3}{2\pi^2\lambda}+
\frac1{2\pi^4\lambda^2}
\left[
\frac{\pi^2+6}{15}
\left(L_\lambda-\frac{53}{30}\right)
+\frac{7\pi^2}{20}
-\frac{21+2\kappa_{\mathrm m}}5
\right]
+\cdots
~,
\\[2mm]
a_3^{\mathrm{int}}(\lambda)
={}&
-15+\frac{5(L_\lambda-1)}{2\pi^2\lambda} +
\frac1{2\pi^4\lambda^2}
\left[
\frac{2(\pi^2+6)}3
\left(L_\lambda-\frac{53}{30}\right)
-\frac{133\pi^2}{12}
+\frac{277}{3}
-4\kappa_{\mathrm m}
\right]
+\cdots
~,
\\[2mm]
a_4^{\rm int}(\lambda)={}&-27+\frac{15(L_\lambda-\frac32)}{2\pi^2\lambda}+\frac{1}{2\pi^4\lambda^2}
\left[
\frac{10(\pi^2+6)}3\left(L_\lambda-\frac{53}{30}\right)
+\frac{2975\pi^2}{36}-\frac{5593}{6}-20\kappa_{\rm m}
\right]+\cdots~.
\end{aligned}
\ee
where we defined
\begin{equation}
    L_\lambda \equiv \log(12\pi^2\lambda) +2\gamma_E~,\quad \kappa_{\rm m} \equiv \log 2 +\sum^{\infty}_{n=2} \frac{\log(1-n^{-2})}{n^2} \approx 0.600482~.
\end{equation}
We now use these moments to fix the distributional terms that should appear in the fixed-$z$ results derived in Section \ref{sec:semiclassics}. The idea is as follows, at order $1/\lambda$ we expect from our strong-coupling calculation that
\be\label{eq:fintdistansatz}
\CF_{\rm int}(z,\lambda)|_{\lambda^{-1}} = \frac{1}{\lambda}\left(\#_1\left[\frac{1}{z^3}\right]_{+++}+\#_2\delta''(z)\right)~,
\ee
where the coefficients $\#_{1,2}$ can be fixed by using the following overlaps:\footnote{Similarly, to footnote \ref{footnote}, our definition of $\left[\frac{1}{z^3}\right]_{+++}$ is given by: \begin{equation}\nonumber
   \int_{0}^1 dz~\left[\frac{1}{z^3}\right]_{+++}\varphi(z) \equiv \int_{0}^1~\frac{dz}{z^3}\left[\varphi(z)-\varphi(0) -z\varphi'(0)-\frac{z^2}{2}\varphi''(0)\right]~. 
\end{equation}}
\begin{equation}
\int_0^1 dz\,P_\ell(1-2z)
\left[\frac1{z^3}\right]_{+++}
=
\begin{cases}
0 \,, & \ell=2 \,,\\
-20 \,, & \ell=3 \,,\\
-105 \,, & \ell=4
\end{cases}
\quad
\int_0^1 dz\,P_\ell(1-2z)\,\delta''(z)
=
\begin{cases}
12 \,, & \ell=2 \,,\\
60 \,, & \ell=3 \,,\\
180 \,, & \ell=4
\end{cases}
~.
\end{equation}
Thus $\ell=2$ fixes the coefficient $\#_2$ and similarly $\ell=3$ fixes $\#_1$:
\begin{equation}
   \#_1= \frac{1}{12\pi^2}~,\quad \#_2 = \frac{L_\lambda-1/3}{24\pi^2}~,\quad 
\end{equation}
We can also use $\ell=4$ to give an independent consistency check:
\be
180\#_2 - 105\#_1 = \frac{15(L_\lambda-3/2)}{2\pi^2}~.
\ee
exactly matching our result. Therefore, the final result for the EEC at order $1/\lambda$ is given by
\begin{equation}
    \left.{\mathcal F_{\mathrm{int}}}(z,\lambda)\right|_{\lambda^{-1}}
=
\frac1{12\pi^2\lambda}
\left(
\left[\frac{1}{z^3}\right]_{+++}
+
\frac{L_\lambda-1/3}{2}\delta''(z)
\right)
~.
\end{equation}
For completeness, we now extend this result also to the next order in $\lambda$. The fixed-$z$ result at this order reported in \eqref{eq:subleading_strong_coupling_EEC} has been derived in appendix \ref{app:subleading_strong_coupling}. Note that, in addition to the inverse powers of $z$, the result contains an additional function
\begin{equation}
    R(z) = \frac{3(1-z^2)+(1+4z+z^2)\log z}{(1-z)^5}~.
\end{equation}
As $z\to0$, this function behaves like $R(z)=\log z +3 + O(z\log z)$, and for $z\to 1$ we have $R(1)=-1/30$. While $R(z)$ poses no problem for the convergence of the integral in $z$, it still needs a distribution subtraction to satisfy the homogeneous sum rules coming from $a^{\rm int}_0$ and $a^{\rm int}_1$. Explicitly we have:
\begin{equation}
    [R(z)]_{++} = R(z) + \frac{5}{24}\delta(z) -\frac{1}{24}\delta'(z)~.
\end{equation}
Now, similarly to what has been done for \eqref{eq:fintdistansatz}, to fix the EEC at order $1/\lambda^2$ one has to introduce the regulated distributions $[z^{-3}]_{+++}$, ${[z^{-4}]_{++++}}$ and $[R(z)]_{++}$ together with two contact terms respectively proportional to $\delta''(z)$ and $\delta'''(z)$ and then proceed to fix all the unknown constants by matching the relevant Legendre overlaps. Omitting the intermediate algebra, we obtain 
\begin{equation}
\begin{aligned}
\left.{\mathcal F_{\mathrm{int}}}(z,\lambda)\right|_{\lambda^{-2}}
=
\frac1{2\pi^4\lambda^2}
\Bigg\{
&
-\frac{\pi^2+6}{60}[z^{-4}]_{++++}
+\frac{\pi^2+6}{90}[z^{-3}]_{+++}
-\frac25[R(z)]_{++}
\\
&+
\frac{
(\pi^2+6)(L_\lambda-151/60)-6\kappa_{\mathrm m}
}{180}
\,\delta''(z)
\\
&+
\frac{
(\pi^2+6)(L_\lambda-18/5)-6\kappa_{\mathrm m}
}{360}
\,\delta'''(z)
\Bigg\}
~.
\end{aligned}
\end{equation}
We also checked that pushing the calculation $a^{\rm int}_{\ell}(\lambda)$ up to $\ell=10$ agree with our calculation and provide further consistency checks.

\section{Conclusions}\label{sec:conclusions}

In this paper we have studied the energy-energy correlator in the large charge state created by a canonical operator in $SU(N)$ ${\cal N}=4$ SYM at finite $N$ and arbitrary values of the large charge 't Hooft coupling $\lambda \sim \Delta_H g_{\rm YM}^2$. Starting from the finite-coupling HHLL four-point function obtained
in \cite{Brown:2025cbz}, we calculated the EEC by mapping the OPE
expansion derived there to the source-detector OPE representation
of the energy correlator \cite{Kologlu:2019bco}. At small angles, $z \ll \lambda^{-1}$, the relevant physics is effectively the one of collinear emission in the conformal phase $\langle \Phi \rangle =0$. At larger angles, $z \gtrsim \lambda^{-1}$, the relevant degrees of freedom are excitations on top of a nontrivial scalar profile $\langle \Phi \rangle \neq 0$. Perhaps the most striking feature of the EEC studied in the present paper is that it exhibits a large charge jet regime in the range $\lambda^{-1} \lesssim z \lesssim \lambda^{-1/3}$, where the EEC $\sim {1 \over \lambda z^3}$, see Figure \ref{fig:intro}. Notice that these regimes \emph{can only} be meaningfully resolved at strong coupling $\lambda \gtrsim 1$.

Supersymmetry has played an important role in our analysis, allowing us to relate the four-point function of scalar operators to the one involving stress tensors \cite{Chicherin:2023gxt}. The same relation is not expected to hold beyond two detectors. However, our results suggest a more general picture. For any $n$-point energy correlator we expect that the physics at small angles is dominated by radiation in the vacuum, whereas for larger angles we expect that the leading behavior is captured by an EFT for massive W-bosons on top of a nontrivial scalar profile $\langle \Phi \rangle \neq 0$. Understanding what precisely this EFT is and how to systematically perform calculations using it is an open problem. 

Our strong-coupling calculation suggests that the connected coupling-dependent energy correlators at strong coupling are related to \emph{the half light transform on the Coulomb branch} of ${\cal N}=4$ SYM, where  we place all detectors at the origin of Minkowski space and integrate over the outgoing light-rays to null infinity.\footnote{The origin of this prescription is due to the fact that at strong coupling the energy correlators is dominated by the kinematics where all detectors approach the same endpoint.} 
On the Coulomb branch, the dependence on $z$ follows simply from transformation properties of stress-tensor half light transforms under boosts leading to $\vev{\CE(\vn_1)\CE(\vn_2)} \sim {1 \over \lambda z^3}$. The same argument---if correct---would imply that the leading strong coupling behavior of the connected three-point energy correlator is $\vev{\CE(\vn_1)\CE(\vn_2)\CE(\vn_3)} \sim {1 \over \lambda^{3/2} \left( z_{12} z_{13} z_{23} \right)^{3/2}}$,\footnote{ We have checked that it is the case for the three-point correlation function of scalar detectors.} whereas we will have nontrivial cross ratios starting from the four-point energy correlator. It would be very interesting to explore this connection further. We expect that the regime, where this relationship approximately holds generalizes the large charge jet regime that we observed at the level of the two-point energy correlator to higher-point functions. 

It would be also very interesting to study the behavior of other detectors, such as the charge operators as well as more general detector operators \cite{Kravchuk:2018htv,Caron-Huot:2022eqs}. The OPE between these detectors in vacuum obey the simple selection rules \cite{Kologlu:2019mfz,Chang:2020qpj}. It would be very interesting to understand what are the properties of the light-ray OPE in the symmetry-broken phase studied here.

 One motivation for this work was to study energy correlators in
a controlled many-particle setting, where the effects of interactions
can be calculated systematically. It would be interesting to
understand whether the phenomena identified here have counterparts
in energy correlators in heavy-ion collisions, where these
observables have been proposed as probes of medium scales and
collective dynamics \cite{Andres:2022ovj,Barata:2023zqg,Yang:2023dwc,Barata:2023bhh,Bossi:2024qho,Duan:2026icj,Barata:2026pgh}. It would be also interesting to calculate the EEC in the $O(2)$ Wilson--Fisher model in the double-scaling limit $n\to\infty$, $\epsilon\to0$ with $\lambda = \epsilon n$ held fixed, and compare its behavior as a function of $\lambda$ with our results. Similarly, perhaps the same analysis of the energy correlators can be performed for the multi-particle production of massive particles in $\phi^4$ theory \cite{Demidov:2021rjp,Demidov:2022ljh}.

Within the Standard Model, a possible analogue of the angular
sensitivity to symmetry breaking studied here is provided by
electroweak showers in collisions far above the electroweak scale. At transverse momenta well above the weak scale, the shower
approximately follows the dynamics of the unbroken electroweak
theory, while the Higgs expectation value controls the emergence
of masses and additional symmetry-breaking splittings near the
weak scale \cite{Chen:2016wkt,Capdevilla:2024amg}.
It would be interesting to investigate whether energy correlators
can resolve this crossover by varying the angular separation
between detectors, thereby scanning the characteristic transverse
scale of the shower.

\section*{Acknowledgments}

We thank Jo\~{a}o Barata, Gabriel Cuomo, and Robin Karlsson for discussions. 
We have used ChatGPT Pro to check and perform calculations, as well as for editing of the draft. This work is partially supported by the Swiss National Science Foundation Grant No.~218510, the NCCR SwissMAP, the European Research Council and the Swiss State Secretariat for Education, Research and Innovation through the consolidator grant ProbQuant.

\appendix

\section{Four-point Function and Mellin Space Detector Transform}
\label{app:fourpoint_mellin}
This appendix contains some technical details needed to derive \eqref{eq:Ffc}. In particular, following \cite{Costa:2012cb}, we compute the Mellin amplitude associated with the HHLL four-point function $\langle \mathcal O_H T_2 T_2 \overline{\mathcal O}_H\rangle$ and use it to derive the expansion of the EEC in detector superblocks.

\subsection{HHLL Correlator in the Detector Channel}
\label{app:fourpoint_channel}

Let
$\mathcal O_H\equiv\mathcal O_K^m$, with
$\Delta_H=mK$. We normalize the HHLL correlator by the protected
heavy two-point function and separate its coupling-independent and
interaction-dependent parts as
\begin{equation}
\label{eq:app_HHLL_split}
\frac{
\left\langle
\mathcal O_H(x_1,Y_1)T_2(x_2,Y_2)T_2(x_3,Y_3)
\overline{\mathcal O}_H(x_4,Y_4)
\right\rangle
}{
\left\langle
\mathcal O_H(x_1,Y_1)\overline{\mathcal O}_H(x_4,Y_4)
\right\rangle
}
=
\mathcal G_{\rm free}(x_i,Y_i)
+\mathcal I_4(x_i,Y_i)\,
\mathcal T_{m,K}(u,v;\lambda)~.
\end{equation}
Here $\mathcal I_4$ is fixed by the superconformal Ward identities
\cite{Eden:2000bk,Nirschl:2004pa}, while
$\mathcal T_{m,K}$ is independent of the R-symmetry cross-ratios.
Our convention is such that
$\mathcal T_{m,K}(u,v;0)=0$ at generic $u,v$. The light-ray transform of
$\mathcal G_{\rm free}$ is the free event shape discussed in the main
text and will not be included in the Mellin analysis below.

We take the sink polarization to be conjugate to the source
polarization, $Y_4=\overline Y_1$, and choose
\begin{equation}
\label{eq:app_detector_polarizations}
Y_2=Y_3=X~,
\qquad
X^2=0~,
\qquad
Y_1\cdot X\neq0~,
\qquad
Y_4\cdot X\neq0~.
\end{equation}
This projects the $T_2\times T_2$ OPE onto the
$[0,4,0]$ channel. The correlator with two stress tensors is then
obtained from this scalar channel by superconformal descent, as
reviewed in \cite{Chicherin:2023gxt}. In this channel the universal prefactor becomes
\begin{equation}
\label{eq:app_detector_prefactor}
\begin{aligned}
\mathcal I_4(x_i,Y_i)
=
\mathcal N_Y
\frac{x_{14}^6x_{23}^2}
{x_{12}^2x_{13}^4x_{24}^4x_{34}^2}
=
\frac{\mathcal N_Y}{x_{23}^4}\frac{v^3}{u}~,\qquad
\mathcal N_Y
=
\frac{1}{(4\pi^2)^2}
\left[
\frac{(Y_1\cdot X)(Y_4\cdot X)}
{Y_1\cdot Y_4}
\right]^2.
\end{aligned}
\end{equation}
We henceforth strip off the cross-ratio-independent factor
$\mathcal N_Y$.

It is important to distinguish the reduced function appearing in
\eqref{eq:app_HHLL_split} from the scalar function entering the
canonically normalized detector transform. Let
$\mathcal A_{\rm int}$ denote the interaction-dependent reduced
correlator written in the conventions of \cite{Costa:2012cb}; in
particular, it does not include the free contribution contained in the
full correlator of that reference. At the order in the large-charge
expansion considered here, our normalization dictionary is
\begin{equation}
\label{eq:app_normalization_dictionary}
u^{-\Delta_H/2}\mathcal A_{\rm int}(u,v;\lambda)
=
v\,\mathcal T_{m,K}(u,v;\lambda)
=
2\Delta_H\,\Phi_{\Delta_H}(u,v;\lambda).
\end{equation}
The scaling $1/\Delta_H$ converts the heavy-two-point-normalized
scalar correlator into the scalar function associated with
canonically normalized energy detectors. The remaining numerical
factor in \eqref{eq:app_normalization_dictionary} is fixed, in the
normalization of $T_2$ used here, by matching the known one-loop
event shape.

Multiplying the heavy-light expansion in the main text by
$u^{\Delta_H/2}$, one obtains
\begin{equation}
\label{eq:app_Aint_OPE}
\begin{aligned}
\mathcal A_{\rm int}(u,v;\lambda)
={}
\frac{v^2}{2}
\sum_{s=1}^{K}\nu_s
\sum_{J\geq0}\sum_{n\geq0}\sum_{r=1}^{J+1}
C_{J,n,r}^{(s)}
u^{(\Delta_H+\gamma_{J,n,r}^{(s)})/2}
\mathcal G_J^1(\cos\phi)
-
\frac{1}{2}
\sum_{s=1}^{K}\nu_s\,u^{\Delta_H/2}.
\end{aligned}
\end{equation}
The physical heavy-light superconformal primary appearing here has
dimension
\begin{equation}
\label{eq:app_physical_dimension}
\Delta_{J,n,r}^{(s)}
=
\Delta_H+\gamma_{J,n,r}^{(s)}+2~.
\end{equation}
By contrast, the ordinary conformal block entering
$\mathcal A_{\rm int}$, and hence the Mellin pole below, has dimension
\begin{equation}
\label{eq:app_ordinary_dimension}
\Delta_{\rm ord}
=
\Delta_{J,n,r}^{(s)}-2
=
\Delta_H+\gamma_{J,n,r}^{(s)}~.
\end{equation}
Note that the positive weights $C_{J,n,r}^{(s)}$ are most directly extracted
in the auxiliary conjugate-light polarization. Since all polarization
dependence has been isolated in $\mathcal I_4$, the same reduced
function can be continued to
\eqref{eq:app_detector_polarizations}. Ordinary OPE-coefficient
products in the detector polarization need not themselves be
positive.

\subsection{Mellin Space Derivation}
\label{app:mellin_calculation}

Let $M(j_1,j_2;\lambda)$ be the Mellin amplitude of
$\Phi_{\Delta_H}$ defined in the main text. We first relate it to the
reduced Mellin amplitude $M^{(R)}(s,t;\lambda)$ associated with
$\mathcal A_{\rm int}$. Specializing the representation of
\cite{Costa:2012cb} to external dimensions
$(\Delta_H,2,2,\Delta_H)$, one has
\begin{equation}
\label{eq:app_Costa_Mellin}
\begin{aligned}
\mathcal A_{\rm int}(u,v;\lambda)
={}&
\int\frac{ds\,dt}{(4\pi i)^2}\,
M^{(R)}(s,t;\lambda)
u^{t/2}v^{-(s+t)/2}
\\
&\times
\Gamma\left(\frac{\Delta_H+2-t}{2}\right)^2
\Gamma\left(\frac{2-\Delta_H-s}{2}\right)^2
\Gamma\left(\frac{s+t}{2}\right)
\Gamma\left(\frac{s+t+2\Delta_H-4}{2}\right).
\end{aligned}
\end{equation}

For an ordinary primary of spin $J$ and dimension
$\Delta_{\rm ord}=\Delta_H+\gamma$, the pole contribution to the
reduced Mellin amplitude is
\cite{Costa:2012cb}
\begin{equation}
\label{eq:app_MR_poles}
M^{(R)}(s,t;\lambda)\big|_{\Delta_{\rm ord},J}
\simeq
\sum_{a=0}^{\infty}
\frac{
C_{H_1 2_2\Delta_{\rm ord},J}
C_{2_3 H_4\Delta_{\rm ord},J}
\,\mathcal Q_{J,a}(s)
}{
t-\Delta_{\rm ord}+J-2a
}~.
\end{equation}
Here $a$ labels conformal descendants and, for $h=d/2=2$,
\begin{equation}
\label{eq:app_Mack_residue}
\begin{aligned}
\mathcal Q_{J,a}(s)
={}-
\frac{
2\Gamma(\Delta_{\rm ord}+J)(\Delta_{\rm ord}-1)_J
}{
4^J
\Gamma\left(
\frac{\Delta_{\rm ord}+J+\Delta_H-2}{2}
\right)^2
\Gamma\left(
\frac{\Delta_{\rm ord}+J-\Delta_H+2}{2}
\right)^2
}\times\frac{
Q_{J,a}(s)
}{
a!(\Delta_{\rm ord}-h+1)_a
\Gamma\left(
\frac{\Delta_H+2-\Delta_{\rm ord}+J}{2}-a
\right)^2
}~,
\end{aligned}
\end{equation}
where $Q_{J,a}$ is a Mack polynomial of degree $J$.

Matching the powers of $U=u/v$ and $V=1/v$ gives
\begin{equation}
\label{eq:app_j_st_map}
j_1=\frac{t-\Delta_H}{2}~,
\qquad
j_2=\frac{s+\Delta_H}{2}~.
\end{equation}
The Jacobian $ds\,dt=4\,dj_1\,dj_2$ converts the measure in
\eqref{eq:app_Costa_Mellin} exactly into
$dj_1\,dj_2/(2\pi i)^2$. Equations
\eqref{eq:app_normalization_dictionary} and
\eqref{eq:app_j_st_map} therefore imply
\begin{equation}
\label{eq:app_M_to_MR}
\begin{aligned}
M(j_1,j_2;\lambda)
={}&
\frac{1}{2\Delta_H}
M^{(R)}(2j_2-\Delta_H,2j_1+\Delta_H;\lambda)
\\
&\times
\Gamma(1-j_1)^2\Gamma(1-j_2)^2
\Gamma(j_1+j_2)
\Gamma(\Delta_H-2+j_1+j_2).
\end{aligned}
\end{equation}
This Mellin amplitude must be paired with the universal kernel in
Eq.~\eqref{eq:large_charge_kernel}.

Let us now set $w=j_1+j_2$. Substituting \eqref{eq:app_M_to_MR} into the
detector transform \eqref{eq:EEC_Mellin_master} gives
\begin{equation}
\label{eq:app_before_j1_residue}
\begin{aligned}
 {\mathcal F_{\mathrm{int}}}(z,\lambda)
={}&
\frac{1}{4}
\int\frac{dw\,dj_1}{(2\pi i)^2}\,
\Gamma(1-w)^2\Gamma(w)\Gamma(\Delta_H-2+w)
\kappa(w,z)
\\
&\times
M^{(R)}
(2w-2j_1-\Delta_H,2j_1+\Delta_H;\lambda)
+ O(\Delta_H^{-1})~.
\end{aligned}
\end{equation}
For the pole in \eqref{eq:app_MR_poles},
\begin{equation}
\label{eq:app_j1_residue}
\underset{j_1=(\gamma-J+2a)/2}{\operatorname{Res}}
\frac{1}{2j_1-\gamma+J-2a}
=
\frac{1}{2}~.
\end{equation}
The $j_1$ contour is closed to the right. Its clockwise orientation,
together with \eqref{eq:app_j1_residue}, yields
\begin{equation}
\label{eq:app_after_j1_residue}
\begin{aligned}
 {\mathcal F_{\mathrm{int}}}(z,\lambda)
\big|_{\Delta_{\rm ord},J}
={}&-
\frac{
C_{H_1 2_2\Delta_{\rm ord},J}
C_{2_3 H_4\Delta_{\rm ord},J}
}{
8
}
\sum_{a\geq0}
\int\frac{dw}{2\pi i}\,
\Gamma(1-w)^2\Gamma(w)
\\
&\times
\Gamma(\Delta_H-2+w)\kappa(w,z)
\mathcal Q_{J,a}
(2w+J-2a-\Delta_{\rm ord})
+ O(\Delta_H^{-1})~.
\end{aligned}
\end{equation}
The pole at $w=0$ is excluded by the Mellin prescription of
\cite{Chicherin:2023gxt}. To evaluate the remaining integral, we take
$\Delta_H\to\infty$ at fixed $J,a,\gamma,w$. In this regime,
\begin{equation}
\label{eq:app_large_Delta_residue}
\begin{aligned}
&\Gamma(\Delta_H-2+w)
\mathcal Q_{J,a}
(2w+J-2a-\Delta_{\rm ord})
\\
&\qquad\simeq-
\frac{
2^{1-2J}\Delta_H^{w+J-a}
}{
a!\,
\Gamma\left(\frac{2+J-2a-\gamma}{2}\right)^2
\Gamma\left(\frac{2+J+\gamma}{2}\right)^2
}
Q_{J,a}(2w+J-2a-\Delta_{\rm ord})~.
\end{aligned}
\end{equation}
We expand the Mack polynomial as
\begin{equation}
\label{eq:app_Mack_expansion}
Q_{J,a}(2w+J-2a-\Delta_{\rm ord})
=
\sum_{p\geq0}
\Delta_H^{-p}q^{(p)}_{J,a}(w,\gamma)~.
\end{equation}
Expanding the recursion relation of
\cite{Costa:2012cb} at large $\Delta_H$ gives
\begin{align}
q^{(p)}_{J,a}(-\ell,\gamma)&=0~,
\qquad\ell<J-a-p~,
\label{eq:app_Mack_zeros}
\\
q^{(p)}_{J,a}(-\ell,\gamma)
\big|_{\ell=J-a-p}
&=
(-1)^J2^J(J-p)!
\binom{J-a}{p}
\left(-\frac{J+\gamma}{2}\right)_p^2~.
\label{eq:app_Mack_survivor}
\end{align}
The first identity removes all poles that would generate positive
powers of $\Delta_H$. The $w$ contour can therefore be closed to
the left, and only the poles of $\Gamma(w)$ at
\begin{equation}
\label{eq:app_w_poles}
w=-\ell~,
\qquad
\ell=J-a-p\geq1
\end{equation}
contribute at this order. Equivalently,
\begin{equation}
a=0,\ldots,J-1~,
\qquad
p=0,\ldots,J-a-1~.
\end{equation}
Combining these residues reproduces the ordinary detector block
$\mathcal I_{J,\gamma}(z)$ displayed in
Eq.~\eqref{eq:ordinary_detector_block}:
\begin{equation}
\label{eq:app_ordinary_result}
 {\mathcal F_{\mathrm{int}}}(z,\lambda)
\big|_{\Delta_{\rm ord},J}
=
\frac{1}{8}
C_{H_1 2_2\Delta_{\rm ord},J}
C_{2_3 H_4\Delta_{\rm ord},J}
\mathcal I_{J,\gamma}(z)
+ O(\Delta_H^{-1}).
\end{equation}
For the last pole, $\ell=1$, the detector polynomial is understood
with the endpoint completion
\begin{equation}
\label{eq:app_contact_completion}
\kappa(-1,z)
\longrightarrow
\kappa(-1,z)+6\delta(z).
\end{equation}
This is the contact term appearing in
$\mathcal I_{J,\gamma}$. Its coefficient is fixed by the homogeneous
event-shape sum rules in the endpoint convention used in the main
text. Since the pole at $w=0$ is excluded, one also has
\begin{equation}
\label{eq:app_I_zero_spin}
\mathcal I_{0,\gamma}(z)=0.
\end{equation}
With the spinning three-point tensor structures normalized by the
Gegenbauer convention used here, reversing the two legs in the second
three-point function gives
\begin{equation}
\label{eq:app_OPE_sign}
C_{H_1 2_2\Delta_{\rm ord},J}
C_{2_3 H_4\Delta_{\rm ord},J}
=
(-1)^J
C_{H_1 2_2\Delta_{\rm ord},J}
C_{H_4 2_3\Delta_{\rm ord},J}~.
\end{equation}
Consequently, after removing the common factor
$1/(8\Delta_H^2)$, an ordinary block of the form
\begin{equation}
C_{H_1 2_2\Delta_{\rm ord},J}
C_{H_4 2_3\Delta_{\rm ord},J}
2^{-J}u^{\Delta_{\rm ord}/2}
\mathcal G_J^1(\cos\phi)
\end{equation}
is mapped to
\begin{equation}
(-1)^J
C_{H_1 2_2\Delta_{\rm ord},J}
C_{H_4 2_3\Delta_{\rm ord},J}
\mathcal I_{J,\gamma}(z)~.
\end{equation}

Finally, it remains to recombine the ordinary blocks into the superconformal combination in \eqref{eq:app_Aint_OPE}. For $J\geq1$, the
Gegenbauer recursion gives
\begin{equation}
\label{eq:app_v2_Gegenbauer}
\begin{aligned}
v^2\mathcal G_J^1(x)
={}&
(1+4u+u^2)\mathcal G_J^1(x)
\\
&-
2(u^{1/2}+u^{3/2})
\left[
\mathcal G_{J+1}^1(x)+\mathcal G_{J-1}^1(x)
\right]
\\
&+
u\left[
\mathcal G_{J+2}^1(x)+\mathcal G_{J-2}^1(x)
\right],
\qquad
x=\frac{u+1-v}{2\sqrt u},
\end{aligned}
\end{equation}
where negative-spin polynomials are set to zero. At $J=0$, the exact
identity in this convention is instead
\begin{equation}
\label{eq:app_v2_Gegenbauer_J0}
v^2\mathcal G_0^1(x)
=
(1+3u+u^2)\mathcal G_0^1(x)
-2(u^{1/2}+u^{3/2})\mathcal G_1^1(x)
+u\mathcal G_2^1(x).
\end{equation}
The difference between
\eqref{eq:app_v2_Gegenbauer} and
\eqref{eq:app_v2_Gegenbauer_J0} is an ordinary spin-zero block. Its
leading detector transform vanishes by
\eqref{eq:app_I_zero_spin}. The detector recombination therefore gives
the superblock $\mathcal S_{J,\gamma}(z)$ defined in
Eq.~\eqref{eq:detector_superblock} for every $J\geq0$. Direct
substitution also shows that the endpoint terms in
\eqref{eq:app_contact_completion} cancel in this superconformal
combination.

The first line of \eqref{eq:app_Aint_OPE} consequently produces the
superblock sum $\mathcal F(z,\lambda)$. Since
$t_s(u,v;0)^2=1$, linearity of the detector transform identifies the
second line with the zero-coupling sum
$-\mathcal F(z,0)$. This completes the Mellin-space derivation of ~\eqref{eq:ECC_final}.

\section{The EEC Superblock}\label{app:eecs}
In this appendix we derive various representations for the detector superblock useful for weak coupling and at large spin at strong and finite coupling that have been used extensively in the main text. We begin with the scalar detector block $B_{j,b}$ on which the differential $\mathcal{D}$ operator \eqref{eq:calD} acts on
\begin{equation}\label{eq:scalar_b}
    \mathcal{I}_{j,j+2b}(z)=(-1)^j 2^{-j}\mathcal{D} \, B_{j,b}(z) \,,
\end{equation}
where
\begin{equation}
    B_{j,b}(z)=\frac{2 \sin^2\pi b}{\pi^2} \sum_{\ell=1}^j \sum_{m=0}^{j-\ell} (-1)^\ell z^{\ell+1} \frac{(m+1)_\ell (j-m-\ell+1)_\ell}{((b+m)_{\ell+1})^2} \,.
\end{equation}
We have omitted contact terms since they will cancel in the superconformal combination. The strategy is to obtain a representation for $B_{j,b}$ in the desired limit and to reconstruct the detector superblock afterwards.

\subsection{Weak Coupling}
Before moving to large spin, let us derive expressions that will enter the weak coupling analysis. At weak coupling in fact the relevant objects are $\partial_\gamma^n\mathcal{S}_{J,\gamma}|_{\gamma=J}$. To compute them we expand the scalar block at small $b$. Since
\begin{equation}
    \frac{\sin^2 \pi b}{((b+m)_{\ell+1})^2} \sim \begin{cases} O(b^2) \,, \quad m\ge1 \,, \\ O(1) \,, \quad m=0 \,, \end{cases}
\end{equation}
up to order $O(b^2)$ only the $m=0$ term in the sum contributes. Simplifying the resulting expression we get
\begin{equation}
    B_{j,b}(z) \simeq 2z((1-z)^j-1)-4b\,z \sum_{\ell=1}^j \binom{j}{\ell} (-z)^\ell H_\ell \,.
\end{equation}
Substituting 
\begin{equation}
    H_\ell = \int_0^1 \frac{1-t^\ell}{1-t}dt
\end{equation}
we can perform the binomial sum 
\begin{equation}
    B_{j,b}(z) \simeq 2z((1-z)^j-1)+4b\,z \int_0^1 \frac{(1-tz)^j-(1-z)^j}{1-t}dt \,.
\end{equation}
The superblock recombination involves different factors of $B_{j,b}$ with one or both argument shifted. Notice however that
\begin{equation}
    B_{j,b+k}(z) = \frac{2 \sin^2\pi b}{\pi^2} \sum_{\ell=1}^j \sum_{m=0}^{j-\ell} (-1)^\ell z^{\ell+1} \frac{(m+1)_\ell (j-m-\ell+1)_\ell}{((b+m+k)_{\ell+1})^2} = O{(b^2)}
\end{equation}
since the denominator in the summand is of order $O(1)$ also for $m=0$ now. Therefore only terms with shifted $j$ and unshifted $b$ will contribute to the superblock at this order, and we get
\begin{equation}\label{eq:trublS}
    \mathcal{S}_{J,J+2b}(z) \simeq \frac{1}{2} \mathcal{D} \left(B_{J,b}(z)-2B_{J+1,b}(z)+B_{J+2,b}(z)\right) \,.
\end{equation}
Simplifying the resulting expressions we get
\begin{equation}
    \mathcal{S}_{J,J+2b}(z) \simeq \mathcal{D} \left[z^3(1-z)^J\right]+2b \mathcal{D} \left[z^3 \int_0^1 \frac{t^2(1-tz)^J-(1-z)^J}{1-t}\right] \,,
\end{equation}
and we can read off
\begin{equation}\label{eq:1LSS}
    \mathcal{S}_{J,J}(z)=\mathcal{D} \left[z^3(1-z)^J\right] \,, \quad \partial_\gamma \mathcal{S}_{J,\gamma}(z)|_{\gamma=J}= \mathcal{D} \left[z^3 \int_0^1 \frac{t^2(1-tz)^J-(1-z)^J}{1-t}\right] \,.
\end{equation}
We remark that since shifting $b$ by an integer produces terms of order $O(b^2)$ one has
\begin{equation}\label{eq:sublargJS}
    \mathcal{S}_{J,J+2n+{2b}}(z) = O(b^2) \,, \quad n \in \mathbb{N}_{>0}
\end{equation}
which gives
\begin{equation}\label{eq:sublargJSn0}
\mathcal S_{J,J+2n}(z)=\partial_\gamma\mathcal S_{J,\gamma}(z)|_{\gamma=J+2n} = 0 \,.
\end{equation}

In order to get an expression for the second derivative of the detector superblock with respect to $\gamma$ we have to expand previous expressions up to $O(b^2)$. At this order \eqref{eq:trublS} has to be modified to include terms with shifted $b$ as well, and blocks with $n\ne0$ will be nonvanishing as well. We report the final result here. For $n=0$ we get
\begin{equation}
\begin{aligned}
\left.\partial_\gamma^2\mathcal S_{J,\gamma}(z)\right|_{\gamma=J}=\frac12\mathcal D\Bigg[z^3\sum_{k=0}^{J}(-z)^k\Bigg\{
&\binom{J}{k}\left(2H_{k+2}^2+H_{k+2}^{(2)}-\frac{\pi^2}{3}\right)+\\
&+(k+1)^2(k+2)^2\sum_{m=1}^{J-k}\frac{(m+1)_k(J-m-k+1)_k}{(m)_{k+3}^{\,2}}\Bigg\}\Bigg] \,,
\end{aligned}
\end{equation}
while for $n\ne0$
\begin{equation}
\begin{aligned}
\left.\partial_\gamma^2\mathcal S_{J,\gamma}(z)\right|_{\gamma=J+2n}=\frac12\mathcal D\Bigg[z^3\sum_{k=0}^{J}&(-z)^k(k+1)^2(k+2)^2
\sum_{m=0}^{J-k}\frac{(m+1)_k(J-m-k+1)_k}{(n+m)_{k+3}^{\,2}}\Bigg] \,.
\end{aligned}
\end{equation}

\subsection{Large Spin, Small $z$}\label{app:largesapp}
In this subsection we derive the large-spin form of the detector superblock
$\mathcal S_{J,\gamma_{J,n,r}^{(s)}}$ in the collinear scaling regime
\be
    J\to\infty\,,\qquad z\to0\,,\qquad x=Jz\quad\text{fixed}\,.
\ee
It is useful to organize the derivation in two steps. We first extract the large-$j$ behavior of the scalar block $B_{j,b}(z)$. We then reconstruct the full superblock. 
In order to extract the large $j$ behavior of the scalar block it is convenient to introduce a generating function. Applying twice the identity
\begin{equation}
    \frac{1}{(b+m)_{\ell+1}}=\frac{1}{\Gamma(\ell+1)}\int_0^1 t^{b+m-1}(1-t)^\ell dt \,.
\end{equation}
to the two Pochhammer factors in the denominator of the summand in $B_{j,b}(z)$  gives
\begin{equation}
    B_{j,b}(z)=\frac{2 \sin^2\pi b}{\pi^2}\int_0^1 dt\int_0^1 du \sum_{\ell=1}^j \sum_{m=0}^{j-\ell} \frac{(-1)^{\ell}z^{\ell+1}}{\Gamma(\ell+1)^2} (m+1)_\ell (j-m-\ell+1)_\ell (t u)^{b+m-1}(1-t)^{\ell}(1-u)^{\ell} \,.
\end{equation}
The sums are now easily performed after introducing the generating function
\be
    \mathcal{K}_b(w)=\sum_{j\ge0}w^j B_{j,b}(z)\,.
\ee
One finds
\begin{equation}
    \mathcal{K}_b(w)=\frac{2 \sin^2\pi b}{\pi^2}\int_0^1dt\int_0^1du\, \frac{-wz^2(1-t)(1-u)(t u)^{b-1}}{(1-w)(1-tuw)\left((1-w)(1-tuw)+(1-t)(1-u)wz\right)} \,.
\end{equation}
The coefficient $B_{j,b}$ is then recovered from
\begin{equation}
B_{j,b}(z)=\frac{1}{2\pi i}\oint dw \, \frac{\mathcal{K}_b(w)}{w^{j+1}} \,.
\end{equation}
This representation is useful to take the large $j$ limit. To do so we blow up the $w$ contour, pick up the poles closest to the origin and then take $j\to\infty$ with $x=jz$ fixed. Poles at $w=w^*$ with $|w^*|>1$ are exponentially suppressed by $w^{-j}$. Keeping the three relevant contributions gives
\begin{equation}
    B_{j,b}(z)\simeq -\int_0^1 dt \int_0^1 du \,\, z\,\,\frac{2\sin^2 b \pi}{\pi^2} \frac{(t u)^{b-1}}{1-t u}\left(1-e^{-\frac{(1-t)(1-u)}{1-tu}x}+(tu)^{j}e^{\frac{(1-t)(1-u)}{1-tu}x}\right) \,.
\end{equation}
The last term is subleading at large $j$, thus
\begin{equation}
    B_{j,b}(z)\simeq-\frac{1}{j} \frac{2\sin^2 b \pi}{\pi^2}\int_0^1 dt \int_0^1 du \,\,  x\,\, \frac{(t u)^{b-1}}{1-t u}\left(1-e^{-\frac{(1-t)(1-u)}{1-tu}x}\right) \,.
\end{equation}
For future reference we define the large-spin block $\Psi_b(x)$
\begin{equation}\label{eq:B_large_spin_scaling}
\begin{aligned}
    &B_{j,b}(z)\simeq \frac{1}{j}\Psi_b(x) \,, \qquad x=jz \,, \\
    &\Psi_b(x)=-\frac{2\sin^2 b \pi}{\pi^2}\int_0^1 dt \int_0^1 du \,\,  x\,\, \frac{(t u)^{b-1}}{1-t u}\left(1-e^{-\frac{(1-t)(1-u)}{1-tu}x}\right) \,.
\end{aligned}
\end{equation}
We now assemble the superconformal combination. Setting $\gamma_{HL}=J+2\beta$ from equation \eqref{eq:detector_superblock} we have
\begin{equation}
\begin{aligned}
    \mathcal S_{J,J+2\beta}
    &=\mathcal{D}\left(\frac{1}{2}B_{J,\beta}-B_{J+1,\beta}+\frac{1}{2}B_{J+2,\beta}\right)\\
    &\quad+\mathcal{D}\left(2B_{J,\beta+1}-B_{J+1,\beta+1}-B_{J-1,\beta+1}\right)\\
    &\quad+\mathcal{D}\left(\frac{1}{2}B_{J,\beta+2}-B_{J-1,\beta+2}+\frac{1}{2}B_{J-2,\beta+2}\right)\,.
\end{aligned}
\end{equation}
Substituting \eqref{eq:B_large_spin_scaling} and expanding the
finite shifts of $J$ at fixed $x=Jz$ yields
\begin{equation}\label{eq:B14}
\mathcal S_{J,J+2\beta}(z)\simeq \frac{1}{J^3} \mathcal{D}\left(1-x\partial_x+\frac{x^2}{2}\partial_x^2\right)\left(\Psi_\beta(x)-2\Psi_{\beta+1}(x)+\Psi_{\beta+2}(x)\right) \,.
\end{equation}
It is useful to introduce
\begin{equation}
    \Phi_b(x)=\left(1-x\partial_x+\frac{x^2}{2}\partial_x^2\right)\left(\Psi_b(x)-2\Psi_{b+1}(x)+\Psi_{b+2}(x)\right) \,,
\end{equation}
which can be simplified to
\begin{equation}\label{eq:Phi_large_spin_integral}
\begin{aligned}
    \Phi_b(x)
   =\frac{\sin^2\pi b}{\pi^2}x^3\int_0^1 dt\int_0^1du\,\frac{(tu)^{b-1}(1-t)^2(1-u)^2}{1-tu}e^{-\frac{(1-t)(1-u)}{1-tu}x}\,.
\end{aligned}
\end{equation}
In the same scaling limit the detector differential operator reduces to
\begin{equation}\label{eq:Dx_large_spin_operator}
    \mathcal{D}\simeq J^3 \frac{1}{x}\partial_x x^2 \partial_x \frac{1}{x^2}\partial_x x^2\partial_x = J^3 \left(4\,\partial_x^3+x\,\partial_x^4\right)=J^3 \partial_x^4 \,x \,.
\end{equation}
Notice that the $J^{-3}$ factor in \eqref{eq:B14} cancels against the
$J^3$ from $\mathcal D$. All in all the large-spin superblock takes the compact form
\begin{equation}\label{eq:S_large_spin_scaling}
\mathcal S_{J,J+2\beta}(z)\simeq\partial_x^4 \left(x\,\Phi_\beta(x) \right) \,.
\end{equation}

For both the loop expansion and the collinear limit it is useful to compute the superblock and its derivatives at $b=0$. To do so we make the change of variables
\begin{equation}
    p=\frac{(1-t)(1-u)}{1-tu} \,, \qquad
    y= \frac{u-t}{t+u-2tu} \,.
\end{equation}
The $y$ integral can be performed exactly, yielding
\begin{equation}
\begin{aligned}
&\Phi_b(x)=x^3\int_0^1 2b(1-p)^{2b-1}\rho_b(p)e^{-px} dp \,,\\ 
&\rho_b(p)=\frac{\sin^2\pi b}{\pi^2}\frac{B\left(\frac{1}{2},b\right)}{b}p^2 \frac{1}{(1+p)^{2b}} \,{}_2F_1\left(-\frac{3}{2},b,b+\frac{1}{2},\left(\frac{1-p}{1+p}\right)^2\right) \,.
\end{aligned}
\end{equation}
Notice that since
\begin{equation}
\int_0^\infty \partial_x^4 \left(x^4 e^{-p x}\right) dx = 0 
\end{equation}
one readily obtains
\begin{equation}\label{eq:x0int}
    \int_0^\infty \partial^4_x \left(x\Phi_\beta(x)\right) dx = 0 \,.
\end{equation}

The limit $b\to0$ must be treated distributionally. The factor $\rho_b(p)$ is
smooth both as $b\to0$ and as $p\to1$, with $\rho_b(1)=1 -\pi^2 b^2/2+O(b^3)$ and $\rho_0(p)=p^2$.
Using
\begin{equation}
2b(1-p)^{2b-1}=\delta(1-p)+2b \left[\frac{1}{1-p}\right]_+ + 4b^2 \left[ \frac{\log(1-p)}{1-p} \right]_++\dots
\end{equation}
and integrating against a test function gives
\begin{equation}
    \int_0^1 f(p) \rho_0(p)\,\,2b(1-p)^{2b-1}\sim f(1) +2b\int_0^1 \frac{p^2 f(p)-f(1)}{1-p}+\dots\,.
\end{equation}
Consequently,
\begin{equation}\label{eq:Phi_b_zero}
\Phi_0(x)=x^3 e^{-x} \,, \quad \frac{1}{2}\partial_b \Phi_b(x)|_{b=0}= x^3\int_0^1dp \,\frac{p^2 e^{-px}-e^{-x}}{1-p} \,.
\end{equation}
This gives for the superblock
\begin{equation}\label{eq:xderi}
\mathcal S_{J,J}(z)\simeq \partial^4_x \, \left(x^4 e^{-x}\right) \,, \quad \partial_\gamma\mathcal S_{J,\gamma}(z)|_{\gamma=J}\simeq \partial_x^4 \left(x^4 \int_0^1 dt\, \frac{t^2e^{-tx}-e^{-x}}{1-t}\right)
\end{equation}
These formulas can be also obtained by setting $z=x/J$ in equation \eqref{eq:sublargJS}. We remark that again consistently with previous results one finds
\begin{equation}\label{eq:no0n}
\mathcal S_{J,J+2n}=\partial_\gamma\mathcal S_{J,\gamma}|_{\gamma=J+2n} = 0 \,.
\end{equation}

The second derivative is slightly more delicate because differentiating
twice with respect to $b$ removes the quadratic zero from $\sin^2\pi b$ at
integer $b$. Around $b=0$,
\begin{equation}
    \rho_b(p)\simeq p^2\left(1-b\frac{(1-p)^2(1+p)+2p^2(3+p)\log p}{(1+p)^3}+\dots\right) \,.
\end{equation}
This gives
\begin{equation}
\begin{aligned}
    \frac{1}{4}\partial_b^2\Phi_b(x)|_{b=0}
    &=2x^3\int_0^1 \frac{p^2e^{-px}-e^{-x}}{1-p}\log(1-p)\\
    &\quad-x^3\int_0^1 \frac{p^2e^{-px}}{1-p}\frac{(1-p)^2(1+p)+2p^2(3+p)\log p}{(1+p)^3}
    -\frac{\pi^2}{4}x^3 e^{-x} \,.
\end{aligned}
\end{equation}

It is also useful to compute the second derivative at $b=n$ positive integer. In this case there are no endpoint subtleties. The second derivative simply removes the quadratic zero from $\sin^2\pi b$, and one finds
\begin{equation}\label{eq:Phi_second_derivative_integer_b}
    \frac{1}{4}\partial_b^2 \Phi_b(x)|_{b=n\ge1}=x^3\int_0^1dp\,B\left(\frac{1}{2},n\right)p^2 \frac{(1-p)^{2n-1}}{(1+p)^{2n}} {}_2F_1\left(-\frac{3}{2},n,n+\frac{1}{2},\left(\frac{1-p}{1+p}\right)^2\right) e^{-px} \,.
\end{equation}
Evaluating these expressions at $x=0$ one gets
\begin{equation}\label{eq:S0laJ}
\begin{aligned}
  &\lim_{J\to\infty}\partial_\gamma^2\mathcal S_{J,\gamma}(0)\big|_{\gamma=J}= 8(\pi^2-6) \,,\\& \lim_{J\to\infty}\partial_\gamma^2\mathcal S_{J,\gamma}(0)\big|_{\gamma=J+2n}=24\left(\frac{\pi^2}{2}-3H_n^{(2)}-\frac{(2n+1)(3n^2+3n-1)}{2n^2(n+1)^2}\right)\,.
\end{aligned}
\end{equation}

\subsection{Large Spin, Strong Coupling}\label{app:strong}
At strong coupling we get contributions from $J,r,n \sim M_s$. Let us expand the superblocks in this region. Let us start again from $\mathcal{I}$ with $J=M_sk$ and $\gamma_{HL}=M_s \alpha$. Then proceeding as before we have
\begin{equation}
    \mathcal{I}_{j,\gamma}(z) =(-1)^j2^{-j}\mathcal{D}\frac{2 \sin^2\pi \frac{\gamma-j}{2}}{\pi^2} \sum_{\ell=1}^j \sum_{m=0}^{j-\ell} (-1)^\ell z^{\ell+1} \frac{(m+1)_\ell (j-m-\ell+1)_\ell}{((\frac{\gamma-j}{2}+m)_{\ell+1})^2} \,.
\end{equation}
Provided $z<1$ we do not need to scale $\ell$ with $j$, since large $\ell\sim j$ is suppressed by $\sim z^{j}$. We take
\begin{equation}
    j=M_s k \,, \quad \gamma=M_s \alpha \,, \quad m=M_s k u \,, u\in[0,1] \,.
\end{equation}
Expanding the Pochhammer symbols in the sum we get
\begin{equation}
    B_{j,\frac{\gamma-j}{2}}(z)\simeq\frac{2 \sin^2\left(M_s\pi \frac{k-\alpha}{2}\right)}{\pi^2}  \frac{k}{M_s} \int_0^1 du\,\sum_{\ell\ge1} (-1)^\ell z^{\ell+1}\frac{k^{2\ell}u^\ell(1-u)^\ell}{(\frac{\alpha-k}{2}+ku)^{2\ell+2}} \,.
\end{equation}
The sum is geometric and the integral can be performed exactly. The result is
\begin{equation}
    B_{j,\frac{\gamma-j}{2}}(z)\simeq \frac{2 \sin^2\left(M_s\pi \frac{k-\alpha}{2}\right)}{M_s \pi^2} \left(\frac{-4zk}{\alpha^2-k^2}+\frac{2\sqrt{z}}{\sqrt{\alpha^2-(1-z)k^2}} \text{arctanh}\left(\frac{2k\sqrt{z(\alpha^2-(1-z)k^2)}}{\alpha^2-(1-2z)k^2}\right)\right) \,.
\end{equation}
Let us now reconstruct the superblock $\mathcal S_{J,\gamma}$. All $J$ and $\gamma$ shifts are $1/M_s$ suppressed. Expanding at large $M_s$ one finds 
\begin{equation}
\begin{aligned}
    &\mathcal S_{J,\gamma_{J,n,r}^{(s)}}(z)\simeq \frac{\sin^2\left(M_s\pi \frac{k-\alpha}{2}\right)}{\pi^2 M_s^5} \mathcal{D}_z\left(\partial_k^2-\partial_\alpha^2\right)^2 h(\alpha,k,z) \,, \\ &h(\alpha,k,z)=\left(\frac{-4zk}{\alpha^2-k^2}+\frac{2\sqrt{z}\,\text{arctanh}\left(\frac{2k\sqrt{z(\alpha^2-(1-z)k^2)}}{\alpha^2-(1-2z)k^2}\right)}{\sqrt{\alpha^2-(1-z)k^2}} \right) \,.
\end{aligned}
\end{equation}

\section{Subleading Strong Coupling Expansion}
\label{app:subleading_strong_coupling}

In this appendix we derive the first subleading correction to the fixed $0<z<1$, strong coupling result. 
Our starting point is the semiclassical representation
\eqref{eq:semiEEC},
\begin{equation}
 {\mathcal F_{\mathrm{int}}}(z,\lambda)
=
\frac{1}{32\pi^2}
\sum_{s=1}^{K}\nu_s\,
\left.
\chi^3\mathcal D_\chi G_s(\chi)
\right|_{\chi=1/z}
~,
\end{equation}
where
\begin{equation}
G_s(\chi)
=
\frac1\chi
\int_{-\pi}^{\pi}dt_1^-
\int_{-\pi}^{\pi}dt_2^-\,
(t_s^2-1)
~,
\qquad
\mathcal D_\chi
=
\frac{d^2}{d\chi^2}
(1-\chi)^2\chi^2
\frac{d^2}{d\chi^2}
~.
\end{equation}
Through order $M_s^{-4}$, it is sufficient to retain the two shortest images,
\begin{equation}
t_s^2
\simeq
\frac{v^2}{u}
\left[
W_s \left(\phi,\frac{i t_{12}^-}{2}\right)
+
W_s \left(2\pi-\phi,\frac{i t_{12}^-}{2}\right)
\right]^2
~.
\end{equation}
There are two relevant integration regions. Near the equal
endpoints, the distances from the endpoints scale as $M_s^{-1}$,
and we must expand the leading calculation one order further.
Near the opposite endpoints, these distances instead scale as
$M_s^{-2}$; both propagator images contribute, including their
interference.

We first consider the equal endpoint $(\pi,\pi)$, using the
coordinates introduced in the main text,
\begin{equation}
t_i^-=\pi-\frac{2\vartheta_i}{M_s}
~.
\end{equation}
Expanding the relation determining $\phi$ gives
\begin{equation}
\phi^2-\left(\frac{t_{12}^-}{2}\right)^2
=
\frac{4\vartheta_1\vartheta_2}{\chi M_s^2}
+
\frac{4(1-\chi)\vartheta_1^2\vartheta_2^2}
{3\chi^2M_s^4}
+
 O(M_s^{-6})
~,
\end{equation}
at fixed $\vartheta_i$. With
$\rho=4\sqrt{\vartheta_1\vartheta_2/\chi}$, the resulting
propagator expansion is
\begin{equation}
\begin{aligned}
t_s^2
={}&
\rho^2\Bigg[
K_1(\rho)^2
+\frac1{M_s^2}
\left(
\frac{(3-\chi)\rho^2}{48}K_1(\rho)^2
+
\frac{(1-\chi)\rho^3}{96}
\frac{d}{d\rho}K_1(\rho)^2
\right)
\Bigg]
+\mathcal O(M_s^{-4})
~.
\end{aligned}
\end{equation}
The other equal endpoint gives the same contribution.
Performing the integral over $y$ in
$\vartheta_{1,2}=\sqrt{\chi}\rho e^{\pm y}/4$, with the same
matching cutoff $\vartheta_0$ as in the main text, gives
\begin{equation}
\left.G_s^{\mathrm{same}}(\chi)\right|_{M_s^{-4}}
=
\frac1{M_s^4}
\int_0^\infty d\rho\,
\Bigg[
\frac{3-\chi}{48}\rho^5K_1(\rho)^2
+
\frac{1-\chi}{96}\rho^6
\frac{d}{d\rho}K_1(\rho)^2
\Bigg]
\log\frac{16\vartheta_0^2}{\chi\rho^2}
~.
\end{equation}
Using
\begin{equation}
\int_0^\infty d\rho\,\rho^5K_1(\rho)^2=\frac85
~,
\qquad
\int_0^\infty d\rho\,\rho^6
\frac{d}{d\rho}K_1(\rho)^2=-\frac{48}{5}
~,
\end{equation}
the coefficient multiplying the logarithm is $\chi/15$.
The remaining terms are
annihilated by the differential operator $\mathcal D_\chi$. Consequently,
\begin{equation}
\left.G_s^{\mathrm{same}}(\chi)\right|_{M_s^{-4}}
=
-\frac{\chi}{15M_s^4}\log\chi
+
\text{terms annihilated by }\mathcal D_\chi
~.
\end{equation}
Since
\begin{equation}
\mathcal D_\chi(\chi\log\chi)=6\chi-4
~,
\end{equation}
the equal endpoints contribute
\begin{equation}
\left.
 {\mathcal F_{\mathrm{int}}}(z,\lambda)
\right|_{\mathrm{same},\,M_s^{-4}}
=
\frac1{8}
\sum_{s=1}^{K}\frac{\nu_s}{M_s^4}
\left(
-\frac1{10\pi^2z^4}
+\frac1{15\pi^2z^3}
\right)
~.
\end{equation}
We next consider the opposite endpoint $(\pi,-\pi)$.
A convenient parametrization is
\begin{equation}
t_1^-=\pi-\frac{\rho^2\eta}{4\pi M_s^2}
~,
\qquad
t_2^-=-\pi+\frac{\rho^2(1-\eta)}{4\pi M_s^2}
~,
\qquad
0<\eta<1
~.
\end{equation}
Defining
\begin{equation}
a=\sqrt{1-\frac{4\eta(1-\eta)}{\chi}}
~,
\end{equation}
the two Bessel arguments become
\begin{equation}
\begin{aligned}
2M_s\sqrt{\phi^2-\left(t_{12}^-/2\right)^2}
&\simeq \rho\sqrt{1-a}
~,
\\
2M_s\sqrt{(2\pi-\phi)^2-\left(t_{12}^-/2\right)^2}
&\simeq \rho\sqrt{1+a}
~.
\end{aligned}
\end{equation}
Both remain finite in this region. The measure and propagator
reduce to
\begin{equation}
\frac1\chi\,dt_1^-dt_2^-
=
\frac{\rho^3}{8\pi^2\chi M_s^4}\,d\rho\,d\eta
~,
\end{equation}
and
\begin{equation}
t_s^2
\simeq
\frac{\rho^2}{4a^2}
\left[
(1+a)\sqrt{1-a}\,K_1(\rho\sqrt{1-a})
-
(1-a)\sqrt{1+a}\,K_1(\rho\sqrt{1+a})
\right]^2
~.
\end{equation}
The relative minus sign follows from
$\sin(2\pi-\phi)=-\sin\phi$~.
The $M_s^{-4}$ suppression is now supplied directly by the
integration measure.

The radial integral can be evaluated using Bessel integrals.
In addition to the integral above, the interference term uses
\begin{equation}
\int_0^\infty d\rho\,\rho^5
K_1(\rho\sqrt{1-a})K_1(\rho\sqrt{1+a})
=
\frac{4}{a^5\sqrt{1-a^2}}
\left[
3a-2a^3+3(a^2-1)\operatorname{arctanh}a
\right]
~.
\end{equation}
Including both opposite endpoints and using the symmetry
$\eta\leftrightarrow1-\eta$, we therefore obtain
\begin{equation}
\begin{aligned}
G_s^{\mathrm{opp}}(\chi)
=
\frac{1}{8\pi^2\chi M_s^4}
\int_{\epsilon}^{1/2}\frac{d\eta}{a^2}
\Bigg[
&\frac85
\left(
\frac{(1+a)^2}{(1-a)^2}
+
\frac{(1-a)^2}{(1+a)^2}
\right)
\\
&-
\frac{8(1-a^2)}{a^5}
\left(
3a-2a^3+3(a^2-1)\operatorname{arctanh}a
\right)
\Bigg]
~,
\end{aligned}
\end{equation}
The regulator $\epsilon$ introduced above is necessary because the integral is divergent for $\eta \to 0$.
To see this, impose an upper cutoff $\vartheta_0$ on each of
the scaled endpoint distances
$M_s^2(\pi-t_1^-)/2$ and $M_s^2(\pi+t_2^-)/2$.
On $0<\eta<1/2$, this gives
\begin{equation}
\rho_{\max}^2=\frac{8\pi\vartheta_0}{1-\eta}
~,
\qquad
\rho_{\max}^2(1-a)
\simeq
\frac{16\pi\vartheta_0\eta}{\chi}
\qquad (\eta\to0)
~.
\end{equation}
We accordingly split the integration region $[0,1/2]$ in $[0,\epsilon]~\cup~[\epsilon,1/2]$ with
\begin{equation}
\epsilon=\frac{C\chi}{16\pi\vartheta_0}
~,
\qquad
C\gg1
~,
\qquad
\epsilon\ll1
~,
\end{equation}
retaining the finite radial cutoff below $\epsilon$. Furthermore, it can be shown that the integral in the region $[0,\epsilon]$ produces terms which are all annihilated by $\mathcal{D}_\chi$.

Evaluating the remaining finite integral and
applying the $\mathcal{D}_\chi$ operator gives 
\begin{equation}
\begin{aligned}
\left. {\mathcal F_{\mathrm{int}}}(z,\lambda)\right|_{\mathrm{opp}}
=
\frac{1}{8\pi^4}
\sum_{s=1}^{K}\frac{\nu_s}{M_s^4}
\Bigg[
-\frac{3}{5z^4}
+\frac{2}{5z^3}
-\frac{12}{5}
\frac{
3(1-z^2)+(1+4z+z^2)\log z
}{
(1-z)^5
}
\Bigg]
~.
\end{aligned}
\end{equation}
Combining the equal- and opposite-endpoint contributions,
and recalling that $M_s^2\propto\lambda$, we obtain the full
correction at order $\lambda^{-2}$, at leading order
$\Delta_H^{-2}$ appearing in \eqref{eq:subleading_strong_coupling_EEC}.

\section{Asymptotics of the Double-scaling Function}
\label{app:double_scaling_asymptotics}

In this appendix we derive the two asymptotic expansions in
\eqref{eq:double_scaling_asymptotics}.  

\subsection{Mellin Representation}

Define the Mellin transform
\begin{equation}\label{eq:app_ds_Mellin_definition}
\mathcal M(\sigma;m_s)
=\int_0^\infty\rd\xi\,\xi^{\sigma-1} m_s^4 \, \mathfrak F(\xi m_s^2).
\end{equation}
Starting from \eqref{eq:double_scaling_integral}, set
$h=\eta(1-\eta)$ and change variables from $\xi$ to
$\delta_s=m_s^2\xi/(xh)$.  This gives
\begin{equation}\label{eq:app_ds_Mellin_before_integrals}
\begin{aligned}
\mathcal M(\sigma;m_s)
=\frac{m_s^{4-2\sigma}}{\pi^2}
&\int_0^\infty\rd x\,x^{\sigma-1}
\int_0^1\rd\eta\,h^{\sigma-2}
\int_0^1\frac{\rd t}{t}\int_0^1\frac{\rd u}{u}\,
p^2\mathcal D_x\left(x^3\re^{-px}\right)
\\
&\times\int_0^\infty\rd\delta_s\,
\delta_s^{\sigma-3}(tu)^{\delta_s}\sin^2(\pi\delta_s).
\end{aligned}
\end{equation}
Manipulating the expression above, we obtain
\begin{equation}\label{eq:app_ds_Mellin_compact}
\begin{aligned}
&\mathcal M(\sigma;m_s)
=\frac{m_s^{4-2\sigma}}{2\pi^2}G(\sigma)\mathcal J(\sigma),
\\
G(\sigma)
&=\frac{\Gamma(\sigma)^2\Gamma(\sigma-1)^2}
{\Gamma(2\sigma-2)}
(\sigma-3)(\sigma-4),
\\
\mathcal J(\sigma)
&=\int_0^\infty\rd L\,L\int_0^1\rd\rho\,
p^{2-\sigma}
\left[L^{2-\sigma}-\operatorname{Re}(L+2\pi\ri)^{2-\sigma}\right].\\
&=B(3-\sigma,3-\sigma)\,\mathcal K(\sigma)
\end{aligned}
\end{equation}
Here $B(a,b)=\Gamma(a)\Gamma(b)/\Gamma(a+b)$ denotes the Euler beta
function, while
\begin{align}
\mathcal K(\sigma)
&=\int_0^\infty\rd L\,A_\sigma(L)
\left[L^{2-\sigma}-\operatorname{Re}(L+2\pi\ri)^{2-\sigma}\right],
\label{eq:app_ds_K_definition}
\\
A_\sigma(L)
&=\left(1-\re^{-L}\right)^{3-\sigma}
{}_2F_1 \left(3-\sigma,3-\sigma;6-2\sigma;1-\re^{-L}\right).
\label{eq:app_ds_As_definition}
\end{align}
The inverse transform is
\begin{equation}\label{eq:app_ds_Mellin_inverse}
m_s^4\,\mathfrak F(\xi m_s)
=\int_{c-\ri\infty}^{c+\ri\infty}
\frac{\rd\sigma}{2\pi\ri}\,\xi^{-\sigma}
\mathcal M(\sigma;m_s),
\qquad 2<c<3.
\end{equation}
Indeed, as $L\to\infty$ one has $p\to1$ away from the endpoints and the
integrand of $\mathcal J(\sigma)$ behaves as $L^{1-\sigma}$, which requires
$\operatorname{Re}\sigma>2$.  As $L\to0$, instead,
$p\sim\rho(1-\rho)L$, and the leading integrand behaves as
$L^{5-2\sigma}[\rho(1-\rho)]^{2-\sigma}$.  Convergence of both the
$L$-integral and the $\rho$ endpoint integrals then requires
$\operatorname{Re}\sigma<3$.  Hence the common fundamental strip is
$2<\operatorname{Re}\sigma<3$.

\subsection{Large-\texorpdfstring{$\xi$}{xi} Expansion}
For large $\xi$ the contour in \eqref{eq:app_ds_Mellin_inverse} is closed to
the right.  The poles crossed in this deformation are located at the integers
\begin{equation}
\sigma=n,\qquad n=3,4,5,\ldots.
\end{equation}
The singular coefficients needed here are fixed by the small-$L$ expansion
of $\mathcal K(\sigma)$.  For the first three poles one finds
\begin{equation}\label{eq:app_ds_J_double_poles_alternative}
\mathcal J(\sigma)\sim
\begin{cases}
\displaystyle {1\over(\sigma-3)^2},&\sigma\to3,\\[2mm]
\displaystyle \left({1\over6}+{1\over\pi^2}\right)
{1\over(\sigma-4)^2},&\sigma\to4,\\[2mm]
\displaystyle \left({1\over40}+{9\over4\pi^4}\right)
{1\over(\sigma-5)^2},&\sigma\to5.
\end{cases}
\end{equation}
Since $G(\sigma)$ has simple zeros at $\sigma=3,4$ and is nonzero at
$\sigma=5$, equation~\eqref{eq:app_ds_Mellin_compact} gives
\begin{equation}
\mathcal M(\sigma;m_s)\sim
\begin{cases}
\displaystyle
-{1\over3\pi^2m_s^2}{1\over\sigma-3},&\sigma\to3,\\[2mm]
\displaystyle
{1\over\pi^2m_s^4}
\left({1\over10}+{3\over5\pi^2}\right){1\over\sigma-4},
&\sigma\to4,\\[2mm]
\displaystyle
{18(90+\pi^4)\over175\pi^6m_s^6}{1\over(\sigma-5)^2}
+ O\left({1\over\sigma-5}\right),&\sigma\to5.
\end{cases}
\end{equation}
The first two poles therefore give pure
powers, while the double pole at $\sigma=5$ gives a logarithm.  Thus
\begin{equation}
m_s^4 \, \mathfrak F(\xi m_s^2)
={1\over\pi^2}\left[
{1\over3m_s^2\xi^3}
-\left({1\over10}+{3\over5\pi^2}\right){1\over m_s^4\xi^4}
+{18(90+\pi^4)\over175\pi^4}
{\log(m_s^2\xi)\over m_s^6\xi^5}
\right]
+ O\left({1\over m_s^6\xi^5}\right),
\end{equation}
This reproduces the large-$\xi$ line of
\eqref{eq:double_scaling_asymptotics}.
More generally, the large-$\xi$ expansion has the structure
\begin{equation}\label{eq:app_ds_large_xi_all_order_alternative}
m_s^4 \mathfrak F(\xi m_s^2)\sim
\frac{b_3}{m_s^2\xi^3}+\frac{b_4}{m_s^4\xi^4}
+\sum_{n=5}^{\infty}\frac{m_s^{4-2n}}{\xi^n}
\left[a_n\log(m_s^2\xi)+b_n\right].
\end{equation}
The coefficients $a_n$ are determined by the local small-$L$ expansion of
$\mathcal K(\sigma)$ and can be computed explicitly to arbitrary order.  They
grow factorially,
$a_n\sim(1024\sqrt{3}/\pi)\,\Gamma(2n)/8^{2n}$, so the resulting expansion is
asymptotic.
\subsection{Small-\texorpdfstring{$\xi$}{xi} Expansion}
For small $\xi$ the contour in \eqref{eq:app_ds_Mellin_inverse} is closed to
the left.  The relevant poles are the simple pole at $\sigma=1$ and the
triple pole at $\sigma=0$; poles at negative integers give terms suppressed
by additional powers of $\xi$.  The two values of $\mathcal J$ needed for
the leading terms are
\begin{equation}
\mathcal J(1)=4\pi^2,
\qquad
\mathcal J(0)=24\pi^2-\frac{4\pi^4}{3}.
\end{equation}
 Substitution into
\eqref{eq:app_ds_Mellin_compact} gives
\begin{equation}
\mathcal M(\sigma;m_s)\sim
\begin{cases}
\displaystyle {24m_s^2\over\sigma-1}+\mathcal O(1),&\sigma\to1,\\[2mm]
\displaystyle {32(18-\pi^2)m_s^4\over\sigma^3}
+ O(\sigma^{-2}),&\sigma\to0.
\end{cases}
\end{equation}
The triple pole generates the leading double logarithm, and hence
\begin{equation}
m_s^4 \mathfrak F(\xi m_s^2)
={24m_s^2\over\xi}
+16(18-\pi^2)m_s^4\log^2(m_s^2\xi)
+ O\left(m_s^4\log(m_s^2\xi)\right),
\end{equation}
which reproduces the small-$\xi$ line of
\eqref{eq:double_scaling_asymptotics}.
More generally, the small-$\xi$ expansion has the following power-logarithmic structure
\begin{equation}\label{eq:app_ds_small_xi_general_structure}
m_s^4 \mathfrak F(\xi m_s^2)\sim
\frac{c_{-1}m_s^2}{\xi}
+\sum_{n=0}^{\infty}m_s^{2n+4}\xi^n
\left[A_n\log^2(m_s^2\xi)+B_n\log(m_s^2\xi)+C_n\right].
\end{equation}

\bibliographystyle{JHEP}

\bibliography{biblio.bib}

\end{document}